\documentclass{aastex63}

\submitjournal{AJ}

\shorttitle{Overview of the Fiber System for the Dark Energy Spectroscopic Instrument}
\shortauthors{DESI Collaboration}

\graphicspath{{./}{figures/}}

\begin{document}

\title{Overview of the Fiber System for the Dark Energy Spectroscopic Instrument}

\correspondingauthor{Claire Poppett }
\email{clpoppett@berkeley.edu}

\author{Claire Poppett}
\affiliation{Space Sciences Laboratory, University of California, Berkeley, 7 Gauss Way, Berkeley, CA  94720, USA}

\author{Luke Tyas}
\affiliation{Lawrence Berkeley National Laboratory, 1 Cyclotron Road, Berkeley, CA 94720, USA}
\author{J.~Aguilar}
\affiliation{Lawrence Berkeley National Laboratory, 1 Cyclotron Road, Berkeley, CA 94720, USA}

\author{Christopher~Bebek}
\affiliation{Lawrence Berkeley National Laboratory, 1 Cyclotron Road, Berkeley, CA 94720, USA}

\author{D.~Bramall}
\affiliation{Institute for Computational Cosmology, Department of Physics, Durham University, South Road, Durham DH1 3LE, UK}

\author{T.~Claybaugh}
\affiliation{Lawrence Berkeley National Laboratory, 1 Cyclotron Road, Berkeley, CA 94720, USA}

\author{J.~Edelstein}
\affiliation{Space Sciences Laboratory, University of California, Berkeley, 7 Gauss Way, Berkeley, CA  94720, USA}

\author{P.~Fagrelius}
\affiliation{NSF's National Optical-Infrared Astronomy Research Laboratory, 950 N. Cherry Avenue, Tucson, AZ 85719, USA}

\author{H.~Heetderks}
\affiliation{Space Sciences Laboratory, University of California, Berkeley, 7 Gauss Way, Berkeley, CA  94720, USA}

\author{P.~Jelinsky}
\affiliation{Lawrence Berkeley National Laboratory, 1 Cyclotron Road, Berkeley, CA 94720, USA}
\affiliation{Space Sciences Laboratory, University of California, Berkeley, 7 Gauss Way, Berkeley, CA  94720, USA}

\author{S.~Jelinsky}
\affiliation{Space Sciences Laboratory, University of California, Berkeley, 7 Gauss Way, Berkeley, CA  94720, USA}

\author{Robin Lafever}\thanks{deceased}
\affiliation{Lawrence Berkeley National Laboratory, 1 Cyclotron Road, Berkeley, CA 94720, USA}

\author{A.~Lambert}
\affiliation{Lawrence Berkeley National Laboratory, 1 Cyclotron Road, Berkeley, CA 94720, USA}

\author{M.~Lampton}\thanks{deceased}
\affiliation{Space Sciences Laboratory, University of California, Berkeley, 7 Gauss Way, Berkeley, CA  94720, USA}

\author{Michael E.~Levi}
\affiliation{Lawrence Berkeley National Laboratory, 1 Cyclotron Road, Berkeley, CA 94720, USA}

\author{P.~Martini}
\affiliation{Center for Cosmology and AstroParticle Physics, The Ohio State University, 191 West Woodruff Avenue, Columbus, OH 43210, USA}
\affiliation{Department of Astronomy, The Ohio State University, 4055 McPherson Laboratory, 140 W 18th Avenue, Columbus, OH 43210, USA}

\author{C.~Rockosi}
\affiliation{Department of Astronomy and Astrophysics, University of California, Santa Cruz, 1156 High Street, Santa Cruz, CA 95065, USA}
\affiliation{University of California Observatories, 1156 High Street, Sana Cruz, CA 95065, USA}

\author{J.~Schmoll}
\affiliation{Institute for Computational Cosmology, Department of Physics, Durham University, South Road, Durham DH1 3LE, UK}

\author{Ray~M.~Sharples}
\affiliation{Centre for Advanced Instrumentation, Department of Physics, Durham University, South Road, Durham DH1 3LE, UK}
\affiliation{Institute for Computational Cosmology, Department of Physics, Durham University, South Road, Durham DH1 3LE, UK}

\author{Martin ~Sirk}
\affiliation{Space Sciences Laboratory, University of California, Berkeley, 7 Gauss Way, Berkeley, CA  94720, USA}

\author{Edward ~Wishnow}
\affiliation{Space Sciences Laboratory, University of California, Berkeley, 7 Gauss Way, Berkeley, CA  94720, USA}

\author{Jiaxi ~Yu}
\affiliation{Ecole Polytechnique F\'{e}d\'{e}rale de Lausanne, CH-1015 Lausanne, Switzerland}


\author{S.~Ahlen}
\affiliation{Physics Dept., Boston University, 590 Commonwealth Avenue, Boston, MA 02215, USA}

\author{A.~Bault}
\affiliation{Department of Physics and Astronomy, University of California, Irvine, 92697, USA}

\author{S.~BenZvi}
\affiliation{Department of Physics \& Astronomy, University of Rochester, 206 Bausch and Lomb Hall, P.O. Box 270171, Rochester, NY 14627-0171, USA}

\author{D.~Brooks}
\affiliation{Department of Physics \& Astronomy, University College London, Gower Street, London, WC1E 6BT, UK}

\author{S.~Cole}
\affiliation{Institute for Computational Cosmology, Department of Physics, Durham University, South Road, Durham DH1 3LE, UK}

\author{A.~de la Macorra}
\affiliation{Instituto de F\'{\i}sica, Universidad Nacional Aut\'{o}noma de M\'{e}xico,  Cd. de M\'{e}xico  C.P. 04510,  M\'{e}xico}

\author[0000-0002-4928-4003]{Arjun Dey}
\affiliation{NSF's National Optical-Infrared Astronomy Research Laboratory, 950 N. Cherry Avenue, Tucson, AZ 85719, USA}

\author{P.~Doel}
\affiliation{Department of Physics \& Astronomy, University College London, Gower Street, London, WC1E 6BT, UK}

\author{K.~Fanning}
\affiliation{Kavli Institute for Particle Astrophysics and Cosmology, Stanford University, Menlo Park, CA 94305, USA}
\affiliation{SLAC National Accelerator Laboratory, Menlo Park, CA 94305, USA}

\author{A.~Font-Ribera}
\affiliation{Department of Physics \& Astronomy, University College London, Gower Street, London, WC1E 6BT, UK}
\affiliation{Institut de F\'{i}sica d’Altes Energies (IFAE), The Barcelona Institute of Science and Technology, Campus UAB, 08193 Bellaterra Barcelona, Spain}

\author{J.~E.~Forero-Romero}
\affiliation{Departamento de F\'isica, Universidad de los Andes, Cra. 1 No. 18A-10, Edificio Ip, CP 111711, Bogot\'a, Colombia}
\affiliation{Observatorio Astron\'omico, Universidad de los Andes, Cra. 1 No. 18A-10, Edificio H, CP 111711 Bogot\'a, Colombia}

\author{E.~Gaztañaga}
\affiliation{Institut d'Estudis Espacials de Catalunya (IEEC), 08034 Barcelona, Spain}
\affiliation{Institute of Cosmology and Gravitation, University of Portsmouth, Dennis Sciama Building, Portsmouth, PO1 3FX, UK}
\affiliation{Institute of Space Sciences, ICE-CSIC, Campus UAB, Carrer de Can Magrans s/n, 08913 Bellaterra, Barcelona, Spain}

\author{S.~Gontcho A Gontcho}
\affiliation{Lawrence Berkeley National Laboratory, 1 Cyclotron Road, Berkeley, CA 94720, USA}

\author{A.~X.~Gonzalez-Morales}
\affiliation{Consejo Nacional de Ciencia y Tecnolog\'{\i}a, Av. Insurgentes Sur 1582. Colonia Cr\'{e}dito Constructor, Del. Benito Ju\'{a}rez C.P. 03940, M\'{e}xico D.F. M\'{e}xico}
\affiliation{Departamento de F\'{i}sica, Universidad de Guanajuato - DCI, C.P. 37150, Leon, Guanajuato, M\'{e}xico}

\author{C.~Hahn}
\affiliation{Department of Astrophysical Sciences, Princeton University, Princeton NJ 08544, USA}

\author{K.~Honscheid}
\affiliation{Center for Cosmology and AstroParticle Physics, The Ohio State University, 191 West Woodruff Avenue, Columbus, OH 43210, USA}
\affiliation{Department of Physics, The Ohio State University, 191 West Woodruff Avenue, Columbus, OH 43210, USA}

\author{J.~Jimenez}
\affiliation{Institut de F\'{i}sica d’Altes Energies (IFAE), The Barcelona Institute of Science and Technology, Campus UAB, 08193 Bellaterra Barcelona, Spain}

\author{S.~Juneau}
\affiliation{NSF's National Optical-Infrared Astronomy Research Laboratory, 950 N. Cherry Avenue, Tucson, AZ 85719, USA}

\author{D.~Kirkby}
\affiliation{Department of Physics and Astronomy, University of California, Irvine, 92697, USA}

\author{A.~Kremin}
\affiliation{Lawrence Berkeley National Laboratory, 1 Cyclotron Road, Berkeley, CA 94720, USA}

\author{M.~Landriau}
\affiliation{Lawrence Berkeley National Laboratory, 1 Cyclotron Road, Berkeley, CA 94720, USA}

\author{L.~Le~Guillou}
\affiliation{Sorbonne Universit\'{e}, CNRS/IN2P3, Laboratoire de Physique Nucl\'{e}aire et de Hautes Energies (LPNHE), FR-75005 Paris, France}

\author{M.~Manera}
\affiliation{Departament de F\'{i}sica, Serra H\'{u}nter, Universitat Aut\`{o}noma de Barcelona, 08193 Bellaterra (Barcelona), Spain}
\affiliation{Institut de F\'{i}sica d’Altes Energies (IFAE), The Barcelona Institute of Science and Technology, Campus UAB, 08193 Bellaterra Barcelona, Spain}

\author{A.~Meisner}
\affiliation{NSF's National Optical-Infrared Astronomy Research Laboratory, 950 N. Cherry Avenue, Tucson, AZ 85719, USA}

\author{R.~Miquel}
\affiliation{Instituci\'{o} Catalana de Recerca i Estudis Avan\c{c}ats, Passeig de Llu\'{\i}s Companys, 23, 08010 Barcelona, Spain}
\affiliation{Institut de F\'{i}sica d’Altes Energies (IFAE), The Barcelona Institute of Science and Technology, Campus UAB, 08193 Bellaterra Barcelona, Spain}

\author{J.~Moustakas}
\affiliation{Department of Physics and Astronomy, Siena College, 515 Loudon Road, Loudonville, NY 12211, USA}

\author{E.~Mueller}
\affiliation{Department of Physics and Astronomy, University of Sussex, Brighton BN1 9QH, U.K}

\author{A.~Muñoz-Gutiérrez}
\affiliation{Instituto de F\'{\i}sica, Universidad Nacional Aut\'{o}noma de M\'{e}xico,  Cd. de M\'{e}xico  C.P. 04510,  M\'{e}xico}

\author{A.~D.~Myers}
\affiliation{Department of Physics \& Astronomy, University  of Wyoming, 1000 E. University, Dept.~3905, Laramie, WY 82071, USA}

\author{J.~Nie}
\affiliation{National Astronomical Observatories, Chinese Academy of Sciences, A20 Datun Rd., Chaoyang District, Beijing, 100012, P.R. China}

\author{G.~Niz}
\affiliation{Departamento de F\'{i}sica, Universidad de Guanajuato - DCI, C.P. 37150, Leon, Guanajuato, M\'{e}xico}
\affiliation{Instituto Avanzado de Cosmolog\'{\i}a A.~C., San Marcos 11 - Atenas 202. Magdalena Contreras, 10720. Ciudad de M\'{e}xico, M\'{e}xico}

\author{N.~Palanque-Delabrouille}
\affiliation{IRFU, CEA, Universit\'{e} Paris-Saclay, F-91191 Gif-sur-Yvette, France}
\affiliation{Lawrence Berkeley National Laboratory, 1 Cyclotron Road, Berkeley, CA 94720, USA}

\author{W.~J.~Percival}
\affiliation{Department of Physics and Astronomy, University of Waterloo, 200 University Ave W, Waterloo, ON N2L 3G1, Canada}
\affiliation{Perimeter Institute for Theoretical Physics, 31 Caroline St. North, Waterloo, ON N2L 2Y5, Canada}
\affiliation{Waterloo Centre for Astrophysics, University of Waterloo, 200 University Ave W, Waterloo, ON N2L 3G1, Canada}

\author{F.~Prada}
\affiliation{Instituto de Astrof\'{i}sica de Andaluc\'{i}a (CSIC), Glorieta de la Astronom\'{i}a, s/n, E-18008 Granada, Spain}

\author{D.~Rabinowitz}
\affiliation{Physics Department, Yale University, P.O. Box 208120, New Haven, CT 06511, USA}

\author{M.~Rezaie}
\affiliation{Department of Physics, Kansas State University, 116 Cardwell Hall, Manhattan, KS 66506, USA}

\author{G.~Rossi}
\affiliation{Department of Physics and Astronomy, Sejong University, Seoul, 143-747, Korea}

\author{E.~Sanchez}
\affiliation{CIEMAT, Avenida Complutense 40, E-28040 Madrid, Spain}

\author{.~F.~Schlafly}
\affiliation{Space Telescope Science Institute, 3700 San Martin Drive, Baltimore, MD 21218, USA}

\author{D.~Schlegel}
\affiliation{Lawrence Berkeley National Laboratory, 1 Cyclotron Road, Berkeley, CA 94720, USA}

\author{M.~Schubnell}
\affiliation{Department of Physics, University of Michigan, Ann Arbor, MI 48109, USA}
\affiliation{University of Michigan, Ann Arbor, MI 48109, USA}

\author{H.~Seo}
\affiliation{Department of Physics \& Astronomy, Ohio University, Athens, OH 45701, USA}

\author{D.~Sprayberry}
\affiliation{NSF's National Optical-Infrared Astronomy Research Laboratory, 950 N. Cherry Avenue, Tucson, AZ 85719, USA}

\author{G.~Tarl\'{e}}
\affiliation{University of Michigan, Ann Arbor, MI 48109, USA}

\author{M.~Vargas-Maga\~na}
\affiliation{Instituto de F\'{\i}sica, Universidad Nacional Aut\'{o}noma de M\'{e}xico,  Cd. de M\'{e}xico  C.P. 04510,  M\'{e}xico}

\author{B.~A.~Weaver}
\affiliation{NSF's National Optical-Infrared Astronomy Research Laboratory, 950 N. Cherry Avenue, Tucson, AZ 85719, USA}

\author{R.~Zhou}
\affiliation{Lawrence Berkeley National Laboratory, 1 Cyclotron Road, Berkeley, CA 94720, USA}



\begin{abstract}
The Dark Energy Spectroscopic Instrument (DESI) is a revolutionary instrument designed for precise measurements of cosmic distances and the investigation of dark energy. DESI utilizes 5,000 optical fibers to simultaneously measure the spectra of distant objects and aims to measure 40 million galaxies and quasars in a five-year survey. One of the critical challenges to DESI's success was ensuring that the fiber system was not only highly efficient, but also delivered a highly stable beam enabling more reliable sky subtraction for measurements of faint objects. We achieved this stability by minimizing the stress on the fiber system during manufacture and operation of the telescope and fiber positioning robots. We installed the DESI fiber system on the 4m-Mayall Telescope with $\geq$ 99\% of fibers intact and the instrument has delivered superb optical performance throughout the initial years of the DESI survey, including $\geq$ 90\% average throughput when injected with a focal ratio of $\sim$f/3.9 as delivered by the primary focus corrector, excluding fiber absorption losses. The design of DESI required multiple innovations to achieve these requirements, such as cleaved fibers bonded with a UV-curing epoxy to glass ferrules in the focal plane and fusion splicing instead of physical connectors. In this paper, we describe the development, delivery and installation of the fiber system, the innovations that made the state-of-the-art performance possible, and the key lessons learned that could benefit future projects.
\end{abstract}

\section{Introduction}
\label{sec:intro} 
The Dark Energy Spectroscopic Instrument (DESI) is a new multi-object, fiber-fed spectrograph that was designed to study the dark energy causing the mysterious accelerating expansion of the Universe \citep{levi2013desi}. DESI can obtain 5,000 spectra in the wavelength range of $0.36-0.98~\micron$ in a $3^\circ$ diameter field of view in a single exposure \citep{desi16a,desi16b,desi22}.  DESI was constructed by a global collaboration of hundreds of researchers and installed on the 4m-Mayall Telescope, which is located at the Kitt Peak National Observatory (KPNO) in Arizona. DESI is a project of the U.S. Department of Energy (DOE) Office of Science, and the project used DOE funds combined with contributions from private foundations and partners to build substantial new instrumentation on the Mayall-4m telescope. The DOE  selected Lawrence Berkeley National Laboratory (LBNL) as the lead laboratory for DESI and appointed the LBNL Project Director.

DESI will measure the expansion history or distance–redshift relationship from the local universe to redshift 3.5 through precise measurements of the baryon acoustic oscillation (BAO) scale \citep{1970ApJ...162..815P,desi16a}. In addition to the expansion history and dark energy, DESI will also measure the growth of cosmic structure, provide new information on the sum of the neutrino masses, study the scale dependence of primordial density fluctuations from inflation, and test potential modifications to the general theory of relativity. DESI will make these measurements with a five-year survey  of 40 million galaxies and quasars in a 14,000 deg$^2$ survey footprint, and thereby construct the largest 3D map of the Universe to date \citep{desi23}.

Many instruments have been constructed specifically for galaxy redshift surveys \citep{1982ApJ...253..423D, 1996ApJ...470..172S, 2018ApJ...858...77H, 2000AJ....120.1579Y,2df2002}, and their multiplex capability, both for fiber-fed and slit spectroscopy, has increased substantially over the last four decades. 

The idea of using fibers for astronomical spectroscopy was first proposed in the 1970s \citep[see][for a historical overview]{1988ASPC....3...77H}. In 1977 Roger Angel led a group that measured the optical properties of a Valtec wet silica fiber and found that the transmission in the optical band was high enough to be considered for use in astronomical instrumentation \citep{1977ApJ...218..776A}. The same group at the Steward Observatory then constructed the MEDUSA instrument \citep{1980ApJ...242L..69H}, which was the first fiber-fed multi-object system built for astronomical research. The fibers were glued into holes in a focal plane aperture plate and fed into the Cassegrain spectrograph on the 2.3\,m Bok telescope. The simplicity of the focal plane aperture plate was adopted by many future projects, such as FOCAP , OPTOPUS, FLAIR, NESSIE, MOFOCS, SILFID, and the hugely successful SDSS \citep{1998Ap&SS.263...91C}. However, the cumbersome nature of these plates also inspired groups to develop more efficient fiber positioning techniques. 

When choosing a fiber positioning technology, trade studies must be done to optimize the instrument for scientific goals. These trade studies include focal plane reconfiguration time, target density, and fiber performance. 

The first such robotic fiber positioning system was the MX spectrometer, which was commissioned in 1985 with 32 individual robot arms that were able to reconfigure the 32 fibers within 45 seconds \citep{1984PhDT.........4H}. Autofib \citep{1988ASPC....3...93P}, designed by Durham University, was an alternative to the MX positioning system and used a single, very fast robot to position focal plane fibers bonded to magnetic "buttons". Autofib was replaced by 2dF which offered a larger field of view, more fibers, advanced fiber positioning technology, better spectral resolution and range, and greater operational efficiency \citep{1998ASPC..152...71L}. The future WEAVE instrument \citep{2012SPIE.8446E..0PD} utilizes a similar fiber positioning technology but increases the number of fibers to 1000. These instruments all use a robot to position fiber "buttons" on the focal plane. This system is effective and the cost is largely independent to the number of fibers on the focal plane, but the size of the magnets make the system more suitable to telescopes with a larger plate scale and focal plane reconfiguration times are limited by the speed of the robot. 

To increase focal plane reconfiguration times, the field was again led by the AAO group, who developed a fiber positioning system that used individual robots to position fibers. This allowed the entire focal plane to be reconfigured simultaneously via "Echidna" fiber positioners.The Echidna fiber positioning system enables each fiber to be positioned via its own "tilting spine" robot fiber positioner. This technology was first used on the FMOS instrument on the Subaru telescope\citep{10.1093/pasj/62.5.1135}. 4MOST on the VISTA telescope, with an expected start of operations in 2025, will also use Echidna fiber positioners to reconfigure $\sim2400$ fibers with fiber separations as small as 20 arcsec \citep{2019Msngr.175....3D}. Echidna fiber positioners are extremely precise, and their compact design allows a dense high density of fibers in the focal plane; however, as the fibers are positioned by tilting the spine, the angle of the light that is injected into the fiber is variable resulting in a less stable PSF and varying amounts of geometrically induced focal ratio degradation.

An alternative to tilting spine positioners is robots with rotating axis that keep the optical axis of fiber aligned with the chief ray. These robotic fiber positioners were first deployed on the LAMOST telescope \citep{2003SPIE.4837..548H} and use small step motors to drive the two rotating structures of the positioner. The LAMOST positioners have a 26mm center to center spacing with a 60\% overlap in area. An alternative, Cobra positioners, developed by NewScale Technologies, also use a two-stage rotary positioning system comprising two rotary piezoelectric SQUIGGLE micromotors. The robots have a 7.7 mm outer diameter and a 9.5 mm diameter patrol region, allowing 100\% coverage of the focal plane \citep{Cobra}. Cobra fiber positioners will be used on the Subaru Telescope PSF instrument \citep{2020SPIE11447E..7VW}. DESI elected to use this style of fiber positioners on the focal plane since the requirement for minimal and stable focal ratio degradation (FRD) \footnote{FRD is the effect whereby fibers degrade the incident focal ratio to faster output focal ratios and will be discussed in detail in section \ref{sec:FRD}} dominated over the requirement for the density of targets.  
DESI builds on the heritage of these many other projects and leverages numerous innovations to increase the number of objects measured within the 5-year survey by more than an order of magnitude relative to any previous project \citep{2023arXiv230606309S}. These innovations span all aspects of instrument design. Firstly, DESI is a dedicated survey instrument and so is the only instrument installed on the 4m-Mayall telescope. This allows the survey to take advantage of every useful night. Secondly, a new optical corrector delivers an extremely large, $3^\circ$ diameter field of view \citep{miller2023optical} that allows many targets to be observed in a single pointing. Finally, a highly multiplexed focal plane system uses 5,000 fiber positioners \citep{silber23} to rapidly reconfigure the focal plane and deliver light through the fiber system to ten highly optimized spectrographs \citep{2018SPIE10702E..7GE}. Although the high multiplexing is essential for DESI to meet its survey requirements, the power of DESI comes not just from the multiplexing of the fiber system, but also the high throughput and high stability of the fiber system.

Aside from the method of fiber positioning, as astronomers gained more experience with fiber-fed instruments, it became increasingly apparent that it was important to reduce FRD. FRD is the tendency of the fiber output to be lower $f$ / \# (`faster') than the input \citep{1988ASPC....3...26R}. The effects of FRD can be reduced by minimizing the amount of stress on the fiber system \citep{1989A&A...213..502C, 1988AJ.....96.2005B,2003PASP..115..854S,2008SPIE.7018E..2UH}. The method of end termination was the main focus of early R\&D into FRD and the development of these techniques is discussed in Section \ref{sec:PFA_RD}. Since DESI is a 5 year survey, the lifetime of the fiber system was a major area for study. We used the knowledge gained by the group of the VIRUS instrument\citep{2021AJ....162..298H} group who invested a large amount of R\&D to understand how $10+$ years of simulated wear on a VIRUS fiber bundle could affect both the transmission and the FRD of the optical fibers \citep{2012SPIE.8446E..5FM}. They reported that fiber transmission and FRD tests conducted before and after lifetime tests revealed that while total transmission values did not change over $10+$ years of simulated wear, there was a clear increase in FRD in all fibers tested. They posited that this increase in FRD was likely to be due to microfractures that developed over time from repeated flexure of the fiber bundle, and stands in contrast to the transient FRD events that stem from localized stress and subsequent modal diffusion of light within the fibers. DESI used the lessons learned from that study and FMOS \citep{2008SPIE.7014E..5LM} to build fiber cables with the same techniques used in undersea construction. This will be discussed in Section \ref{sec:RD_cables}. In addition to fiber-fed multiobject spectrographs, recent advances in high-resolution, fiber-fed instruments used in exoplanet searches have further increased our understanding of the importance of stability of the near- and far-field at the exit of the fiber \citep{2014SPIE.9147E..1FQ, 2018SPIE10702E..6QK}. The high stability of the fiber system enables DESI to address dark energy at higher redshifts by improving the accuracy of sky-subtraction on faint targets. 

Building upon the successes and understanding the limitations of these instruments was fundamental to the DESI fiber system design and our R\&D planning. The two main lessons were that careful end preparation and a robust, yet low-stress cabling system are essential. We also recognised that physical connectors were the main source of throughput loss for instruments without a continuous fiber run \citep{10.1117/12.856805}. 

\begin{figure}[ht]
\includegraphics[width=.95\textwidth]{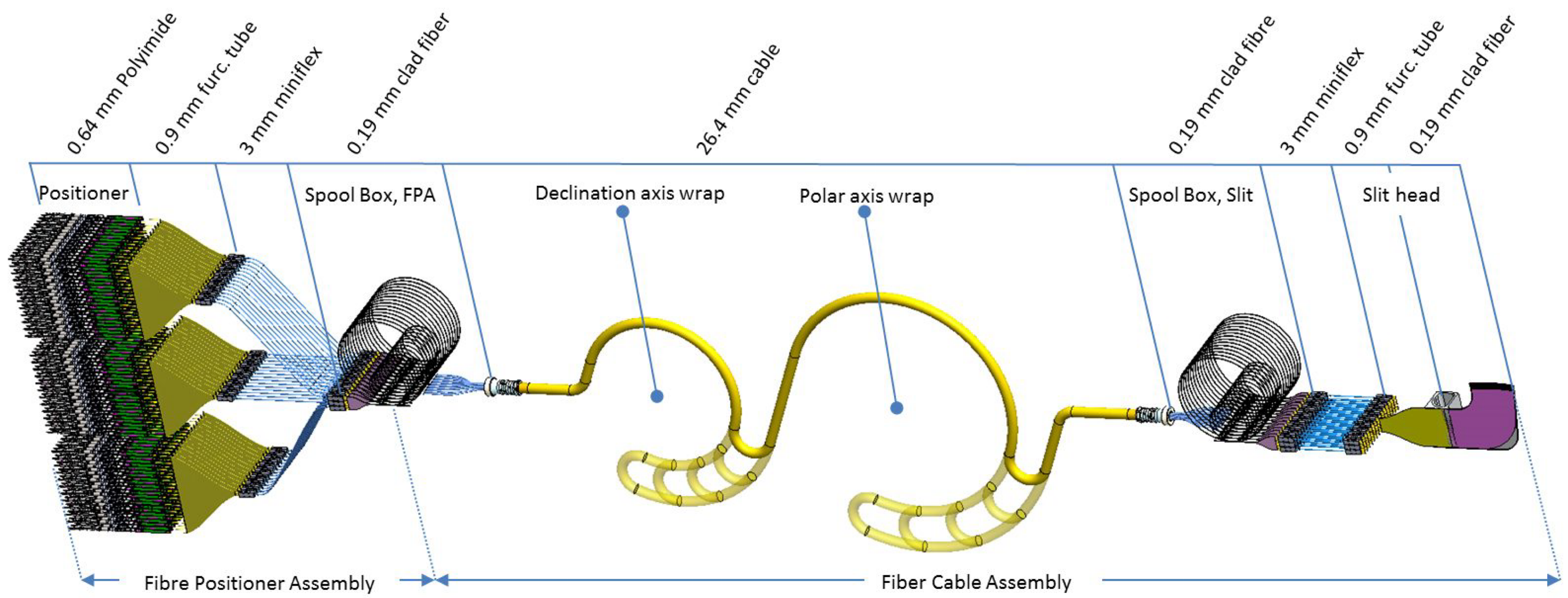}
\caption{Schematic of the DESI fiber system. The 5,000 fibers connecting to the spectrographs are divided into ten identical bundles of 500 fibers,
each of which is integrated with one focal plane unit, defined as a "petal". The prime focus corrector injects light into the focal plane fibers at an average focal ratio of f/3.9. A spool box at each end of the fiber cable allows for excess in the fiber system. Each 500 fiber cable is sub-furcated into ten smaller bundles and wrapped in a helical fashion about a central Kevlar strength member. The focal plane fibers are fused to the main fiber cables in the focal plane spool boxes. The output of each unit of the fiber system is terminated in a fiber slit. Each fiber slit feeds one of 10 spectrographs which accepts light within f/3.57. Values at the top of this figure give the outer diameter for each of the elements.}
\label{fig:fiber_schematic}
\end{figure}

A schematic of the DESI fiber system is shown in Figure~\ref{fig:fiber_schematic}. Light enters the 5,020 custom-built positioner fiber assemblies (PFAs) from the prime focus corrector, and the PFAs are installed in 5,020 robotic fiber positioners. The prime focus corrector design and the orientation of the fiber positioners ensures that the light is telecentric across the focal plane to within $\pm0.5^\circ$. 5,000 of these fibers are routed to the ten spectrographs for science observations and the remaining 20 feed the Sky Continuum Monitor System \citep{tie20} which is used as part of the dynamic Exposure Time Calculator to determine the exposure times \citep{expcalc}. The 5,000 fibers connecting to the spectrographs are divided into ten identical bundles of 500 fibers,
each of which is integrated with one focal plane unit, defined as a "petal". The 107~$\mu$m core fibers are precision cleaved and then bonded to fused silica ferrules that are in turn bonded to positioners. A polyimide tube is attached to the back of the ferrule to provide strain relief from the positioner as it moves between targets to reduce stress as the positioner moves \citep{2012SPIE.8446E..5FM}. Finally, the front face of the PFA has an antireflection (AR) coating to minimize Fresnel losses \citep{Poppett2018}. The PFA is installed into a positioner and the ferrule is bonded into the ferrule arm after being aligned in focus \citep{silber23}.

\begin{figure}[ht]
\includegraphics[width=\textwidth]{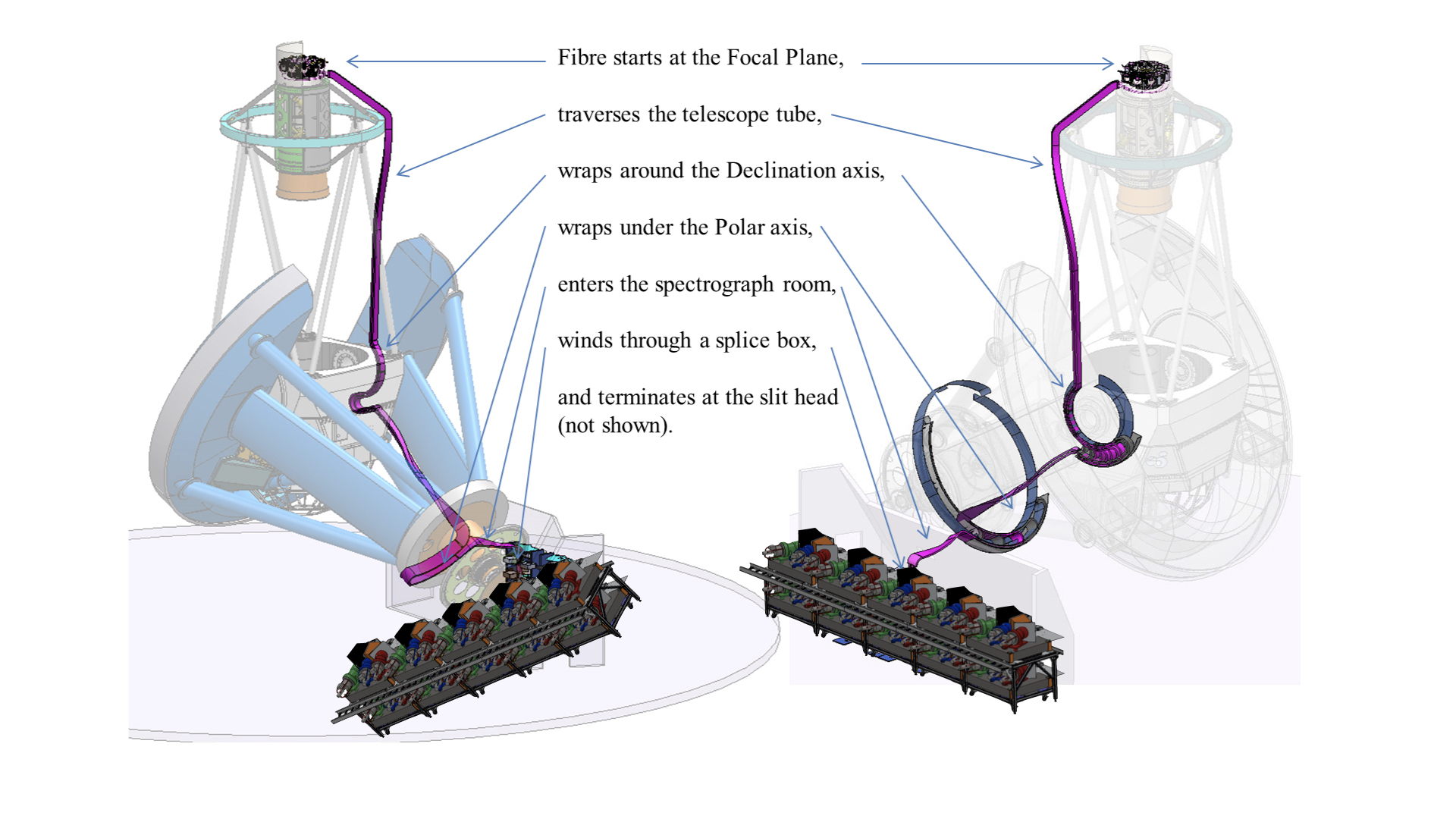}
\caption{Schematic of the DESI fiber system installed on the 4m-Mayall Telescope. The fiber system is installed on the outside of the telescope with no feed-throughs in order to facilitate maintenance.}
\label{fig:fiber_routing}
\end{figure}

Ten 47.5\,m long cables run from the focal plane enclosure (FPE) to the ten spectrographs, carrying 500 fibers (plus spares) per cable \citep{2018SPIE10702E..7NS}. The routing of the fiber system on the telescope is shown in Figure~\ref{fig:fiber_routing}. The routing was subject to a trade study between minimising the length of the fiber system in order to optimise throughput, and ensuring that the cable did not pass through any trusses in order to make future maintenance possible. Although we have not yet removed a full cable assembly from the telescope, it is likely that the survey will extend beyond the original lifetime. If this is achieved, it will probably become necessary to remove an assembly, and it is our expectation that we will be thankful for this design choice \citep{2018SPIE10702E..7PP}.  Each cable terminates in a slit assembly. Each slit assembly consists of 500 fibers arranged in 20 V-blocks (25 fibers per block), with the orientation of the blocks approximating the slit curvature. Fibers are routed through a spool box strain relief system before the cable and slit assemblages are connected to the focal plane via fusion splicing \citep{fagrelius16}. The fusion splice not only facilitates an effective fabrication, integration, and testing workflow but also maximizes the system throughput while minimizing FRD. 

This system was designed, tested, and built in collaboration with Durham University (DU) in the UK and some industry partners\footnote{M2FX installed the optical fiber into miniflex tubes and wound the fiber cables.}. The majority of the manufacture was performed in-house at LBNL and DU. In 2019 the fiber system was installed at the 4m-Mayall telescope and met all requirements. Highlights include that $\geq99\%$ of the fibers were intact, $\geq90\%$ fibers had collimated FRD $\leq1.8^\circ$, and the average throughput exceeded $\geq90\%$ from the prime focus corrector to the spectrograph. 

The first DESI spectrograph was delivered to the Mayall in May 2018, commissioning began in October 2019, survey validation began in December 2020 \citep{2023arXiv230606307D,2023ApJ...943...68L} and the survey officially began in May 2021 making it the first stage IV \citep{albrecht2006report} dark energy program to begin operations. The Early data release is composed of 496,128 stars;
1,125,635 galaxies; and 90,241 quasars which were spectroscopically classified and free of any known hardware,
observational, or redshift fitting issues \citep{2023arXiv230606308D,2023MNRAS.525.5406M}. In only the first seven months of science operations, more than 7.5 million galaxies were measured, making it larger than all previous 3D redshift surveys combined.

The high performance we achieved with the DESI fiber system is due to many innovations which will be discussed in detail in section \ref{sec:RandD}, but there are a few key elements that were particularly important. At the focal plane, precision cleaving of the front end fibers rather than the traditional method of polishing fibers eased the production flow since epoxies could be chosen that did not have a long cure time \citep{2016SPIE.9908E..9BP}. Rather than bonding fibers into the traditional ceramic ferrules, we instead opted to use glass ferrules. This allowed us to carefully monitor exactly where the glue was applied and cure it when it had reached the appropriate length along the ferrule. Fusion splicing of the fibers reduced the throughput loss of the connection from $\sim10$\% for a physical connector to around $\sim2$\%. Our adoption of helically wound cables as used in FMOS \citep{2008SPIE.7014E..5LM} has proved extremely robust to the movement of the telescope. Finally, rather than requiring a monolithic curved slit, we bonded 20 flat slit blocks of 25 fibers to a curved slit plate with an additional AR-coated window bonded to each slit block \citep{2016SPIE.9908E..8TS}.

In section~\ref{sec:Reqs} we describe the requirements of the Fiber System. Section \ref{sec:design_choices} discusses the key design choices that were made in order to achieve these requirements. In Section \ref{sec:FRD} we describe the implications of Focal Ratio Degradation (FRD), its causes, how we set performance requirements and how we ensured that the required performance was met during production. Then in section \ref{sec:RandD}, we discuss the R\&D that led to numerous important improvements over previous fiber systems. This R\&D began in 2012 and focused on areas of the fiber system that we had identified as either high risk or that required a novel approach to development. Section \ref{sec:Production} presents the production of the fiber system, including scheduling, quality assurance, and pre-ship performance measurements. The fiber system was only fully integrated on-site at the 4m-Mayall telescope, and we discuss this process in section~\ref{sec:installation}. DESI's fiber system has performed extraordinarily well on sky and we present results on some on sky performance and validation results from DESI commissioning in Section~\ref{sec:on_sky}. Finally, in section \ref{sec:sum} we include a summary and describe some lessons learned which may be helpful to future projects.

\section{Requirements}
\label{sec:Reqs}
The key goal driving the design of DESI was to conduct a spectroscopic survey that would meet the definition of a Stage-IV dark energy survey within only five years. Early estimates showed that a survey of on order 30 million objects would meet the Stage-IV definition \citep{schlegel2009bigboss} and identified one of the most challenging requirements to be the measurement of emission line galaxies at least as faint as $10^{-16}$ erg s$^{-1}$ cm$^{-2}$ to $z\sim1.6$.

In order to successfully measure these faint galaxies, it was essential that the fiber system met three key requirements. Firstly, it must have high throughput in order to optimize survey speed. Secondly, it must deliver a stable optical beam, both in the near field and in the far field, to enable superb subtraction of night sky emission and measurement of spectrograph resolution by the data reduction pipeline \citep{guy23}, which was designed to employ the spectroperfectionism algorithm \citep{Bolton_2010}. Finally, since the survey will run for five years, the fiber system must be reliable for the duration of the survey.

Achieving a high throughput is dependent on choosing a fiber with a high bulk throughput, and minimising the contributions from both inherent FRD and geometric FRD \footnote{Geometric FRD is FRD due to the misalignment between the chief ray and the optical axis of the fiber}. Increased FRD means that light exiting the fibers is at a larger focal ratio than the input and so if the spectrograph cannot accept these angles, light will be lost. For DESI, the acceptance focal speed of the spectrograph is f/3.57, so any light that is faster than this will be lost. If the speed of the spectrograph is increased to accommodate for these losses, the resolution will be reduced, and so a trade study that balances FRD losses with resolving power is essential. The absolute throughput of the fiber can be minimised by keeping the fiber length as short as possible which is especially important in the blue where the absorption by the glass is the highest. Minimizing contributions due to geometric FRD can be achieved in the design of the prime focus corrector and by choosing a fiber positioner that keeps the fiber normal to the optical axis. Minimising inherent FRD can be achieved by reducing stress in the fiber system. These design choices are discussed in more detail in Section \ref{sec:design_choices}. 

A stable optical beam was not a well understood requirement for early fiber-fed instruments, however, it is essential in order to achieve the sky subtraction stability and quantify the spectral resolution of each
spectrum. If the beam emitted from the fiber is not stable in the far-field, the aberrations of the spectrograph will vary, causing a change in the point spread function (PSF). If the beam emitted from the fiber is not stable in the near-field, the PSF at the detector will not be stable enough to perform accurate sky subtraction.

During survey operations, calibrations are performed by pointing the fibers at a white spot in the dome and illuminating the detector with a series of arc lamps. The stability of the PSF drives the calibration cadence for DESI. The observing strategy for DESI requires the spectrograph to not require calibrations between each observation tile. It is therefore essential that the PSF is stable over the course of 12 hours and is not affected by normal instrument operations such as fiber positioning or telescope movements. 

With these requirements in mind, we developed a key set of requirements that the fiber system had to meet. In order to meet the survey requirements of the required number targets within 5 years, the fiber system had to achieve a multiplexing of 5000 fibers and deliver 90\% of the light injected at the focal plane to the spectrograph. Since it was a survey requirement that the instrument had to run for 5 years, we required that the fiber system survive to twice the number of positioner moves and cable flexures that were expected over that time frame with a degradation rate of fewer than 100 fibers per year suffering a loss of more than 20\% throughput. The desired SNR and expected seeing at the site required that each fiber have a core diameter of 107 $\mu$m. The performance requirements for the Fiber System are given in Table~\ref{tab:fibreq}.

\begin{table} 
    \centering
\begin{tabular}{|c|c|c|} 
 \hline
 \bf{Title} & \bf{Value} & \bf{Statement} \\ 
 \hline
 Fiber core diameter& 107 $\pm 3 \mu$m  & Fiber average core diameter \\
 \hline
 Fiber System Bulk Throughput& $\lambda$(\micron)\hspace{0.2in}Efficiency&The bulk throughput of the fiber \\ 
 \cline{2-2}
  &   0.360\hspace{0.5in}0.45&Efficiencies are minima.  \\ 
  &   0.375\hspace{0.5in}0.51 &Contributors include glass \\ 
 &  0.400\hspace{0.5in}0.59 &absorption, AR coatings, slit \\ 
   &0.500\hspace{0.5in}0.89 & meniscus optical cement, splice \\ 
 &0.600\hspace{0.5in}0.89 & connection loss, fiber end-finishing \\ 
 &0.700\hspace{0.5in}0.90 &loss (roughness scatter), slit \\
     &0.800\hspace{0.5in}0.914 &output pupil alignment \\ 
 &0.900\hspace{0.5in}0.929 & \\
 &0.980\hspace{0.5in}0.937 & \\
 \hline
 Fiber System Throughput &$\geq90$~\% &Main contributors:\\
 due to FRD loss&from 360~nm - 980~nm&Fiber end polishing/cleaving/bonding, \\
 &&Fiber splice induced FRD, Actuator induced FRD,  \\
 &&Cable and guide bend-induced FRD \\ 
 \hline
 Single Fiber System & No more than 100   &The total throughput of any single fiber, \\ 
 Lifetime Throughput Loss&fibers per year to suffer&\\
 &a loss of $\geq$20\%&including both bulk throughput and FRD losses, \\
 && shall not degrade from the nominal average value \\
 &&by more than 20\% over a 5-year lifetime of use.\\
 \hline
 Number of Science fibers&$\geq$5,000&There shall be at least 5,000 science fibers\\
 &&at the focal plane \\ 
 \hline
 Fiber positioner flexure lifetime&376,000 positioner moves&The fiber shall maintain performance \\
 &&after 376K positioner moves \\
 \hline
 PFA traceability&PFA to slit&It shall be known which position in the slit \\
 &&corresponds to which positioner \\
 \hline
 Fiber Cable flexure &169K cycles&Fiber system performance shall be maintained\\
 &&over 169K cable pivot bends \\ 
 lifetime performance&&\\	
 \hline
 Number of fibers per slit&$\geq$500&There shall be at least 500 fibers per slit \\ 
 \hline
 Near field stability & uniform and stable &The near field pattern of the fiber shall be uniform and stable  \\ 
 &&in order to facilitate the calibration strategy\\
 \hline
   \end{tabular}
\caption{Fiber System Requirements}
\label{tab:fibreq}
 \end{table}

\section{Design choices}
\label{sec:design_choices}
The key to a highly efficient mutli-object fiber-fed spectrograph is both high throughput and high beam stability. In the subsequent sections we discuss the various design choices which were made in order to achieve these metrics. We divide these contributions into those that affect the bulk throughput in Section \ref{sec:DC_conts}, those that affect the FRD performance in Section \ref{sec:DC_FRD} and the design choices that affect the beam stability in Section \ref{sec:DC_beam}. 

\subsection{Contributions to throughput} \label{sec:DC_conts}
\subsubsection{Fiber Size}
Fiber feeds for spectrographs crucially determine the achievable signal-to-noise ratio and hence the potential survey rate. Too large a diameter fiber is immune to seeing but dilutes the galaxy spectrum with excessive night sky continuum and line emission; too small is immune to night sky brightness but is very sensitive to seeing which, when poor, will seriously decrease the fiber capture throughput.

Accordingly, the choice of fiber size must be made with realistic estimates for night sky brightness and (especially) the delivered galaxy image size. Besides seeing, other factors that increase the effective size of the image are the galaxy size distribution, the telescope pointing accuracy, the fiber location accuracy, and optical aberrations within the corrector and ADC.

At the simplest level, the PSF of the galaxy image is dominated by the seeing, which is commonly described by a Moffat function and whose encircled energy is given below; the signal-to-noise ratio is this divided by the diameter of the fiber core, $D_{\text{fiber}}$. 
\begin{eqnarray}
\rm{Signal}\,\rm{Photons} &\sim& EE_{\text{fiber}}=1-\left [  1+(2^{1/\beta}-1){\frac{D^2_{\text{fiber}}}{\rm{FWHM}^2_{\rm{seeing}}}}\right ] \\
\rm{Sky}\,\rm{Photons} &\sim& D^2_{\text{fiber}} \\
\rm{Noise} &\sim& \sqrt{\rm{SkyPhotons}}=D_{\text{fiber}}
\end{eqnarray}
where EE$_{\rm{fiber}}$ is the encircled energy fraction. 
By considering that the expected seeing at the 4m-Mayall is 1.1 arcsec \citep{Dey_2014} with the expected flux of our target galaxies \citep{Mostek:2012nc} we determined that a fiber core diameter of 107~$\mu$m was optimal for the DESI survey, translating to 1.4 arcsecs on the sky. The fiber cladding size (150~$\mu$m diameter) was chosen to eliminate significant evanescent surface loss \citep{1998ASPC..152...14B}. The fiber has a polyimid buffer that is 20~$\mu$m thick. The fiber is drawn from a stock preform and so the ratio of the core to clad is set, but the core diameter can be specified when the quantity required by DESI is drawn. 

\subsubsection{Fiber transmission}
When it was first proposed that optical fibers could be used for astronomical spectrographs, it was understood that the spectral transmission efficiency of fibers would be a primary concern \citep{1998ASPC..152...14B}.

Fiber bulk attenuation is dictated by the fiber glass type and its spectral transmittance, the fiber length, and the ability of the fiber's numerical aperture to fully accept the input ($f/\#$) from the corrector. The DESI fiber system was built using Polymicro FBP fiber, a low-OH step-index fused silica fiber, with numerical aperture (NA) = 0.22, that has been well characterized to have excellent broad-band transmittance without the complex and potentially confusing spectral absorption features found in high-OH, UV enhanced type fibers. This fiber is the gold standard in astronomical instrumentation since the diameter tolerances on the core /clad / buffer are extremely tight, resulting in minimal microbending, and this produces both low stress and excellent FRD.

\begin{figure}[tb]
\centering
\includegraphics[height=3in]{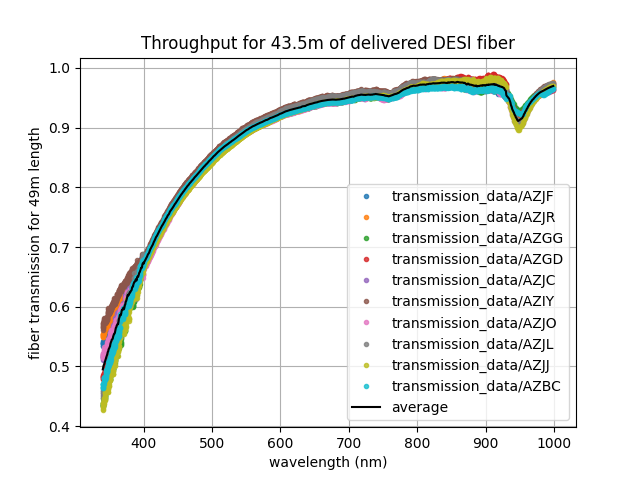}
\caption
{Bulk fiber transmission for 43.5m of DESI fiber. The attenuation of each batch of fiber was measured prior to shipment by the manufacturer. Although there is variation at blue wavelengths, it is assumed that this is a result of the measurement method and we see no dependence on throughput with fiber batch in the on sky data.}
\label{fig:fiber_length.jpg}
\end{figure}

When designing the fiber system, every effort was made to minimise the length in order to improve efficiency, which is especially important at wavelengths shorter than 400\,nm. The average length of the fiber system from the focal plane to the spectrograph slit is 43.5~m. The shortest unit is 42.7~m and the longest unit is 43.9~m. The DESI fiber was delivered in multiple batches and the spectral attenuation was measured prior to each shipment by the manufacturer. Figure \ref{fig:fiber_length.jpg} shows the bulk throughput of the delivered fiber for this length. In order to meet throughput requirements, the fiber length could not exceed 45~m, whilst a minimum length of 42.5~m was required in order to reach the spectrographs. In addition to the bulk throughput of the fiber, 94\% throughput was allocated to FRD losses on axis, 98\% throughput was allocated to splice losses, and 1-1.5\% was allocated to AR coating losses on each end of the fiber system. 

\subsection{Focal ratio degradation} \label{sec:DC_FRD}
As described earlier, FRD is the tendency of the fiber output to be at a lower $f$/\# (`faster') than the input. To optimally match the fiber to the downstream (spectrograph) optics, FRD should be minimized, stable, and repeatable. FRD is primarily caused by stress on the fiber in the form of end stress, microbending, and macrobending. There is no dependence of FRD with length  since end effects dominate \citep{2010MNRAS.404.1349P}. Additionally, `geometric' FRD is caused by alignment (telecentricity) errors: the output beam angle distribution is the combination of the input angle distribution and the angle of incidence of the chief ray.  In early fiber-based spectrographs, the downstream optical path often did not account for FRD \citep{1980ApJ...242L..69H}, leading to significant losses. DESI required that 90\% of the light incident on the focal plane, which is injected at an average focal ratio of f/3.9, should be output within the acceptance focal ratio of the spectrographs of f/3.57. Many models have been developed in an attempt to predict the optical performance of fibers \citep{Gloge:71, Marcuse1972DerivationOC, 1975ApOpt..14.1538G, 1994MNRAS.271....1C} , however, subtle effects often result in laboratory testing of the science fiber being required. The method we used to measure FRD during both R\&D and production is discussed in Section \ref{sec:FRD}.

\subsection{Beam stability and positioner design choice} \label{sec:DC_beam}
We considered a number of fiber positioner designs during the R\&D phase of DESI. The two main classes of design were a twirling post ($\theta\phi$) design and a tilting spine design.

Given the aforementioned requirement of the fiber system that 90\% of the light incident at the focal plane was injected into the spectrograph; full cone FRD testing at the input focal ratio of f/3.9 showed that the light incident on the fiber must be aligned to within $\pm 60'$ of the optical axis to meet this requirement. This placed an initial limitation on the tilting spines. These results are described in Section \ref{sec:FRD} in Figure \ref{fig:weight_FRD}.

The second requirement was that the PSF must be stable as the fiber moves around its patrol area. It was well understood how geometric FRD would affect the throughput, but it was less well understood how geometric FRD would affect the near term beam stability. By imaging the near field of the output end of the fiber we determined that any change to the shape of the PSF caused by stress at the end of the fiber was dominated by geometric FRD effects as the injection angle of the light changed. These results are described in more detail in Section \ref{sec:FRD} and ultimately led to a positioner down-select to a twirling post ($\theta\phi$) design of positioner.

\section{Focal Ratio Degradation}
\label{sec:FRD}
There are two standard tests used to characterize fiber FRD:  (1) Full-cone (or solid angle) tests \cite[e.g.,][]{coneFRD} measure the angular diffusion of a fiber by illuminating it with a filled cone of light with constant surface brightness over a specified angle. Cone tests therefore directly mimic how fibers behave in an instrument by measuring the energy enclosed within some $f$-ratio ($f/\#$) for a known input illumination geometry. (2) Collimated (or ring) tests \cite[e.g.,][]{collimatedFRD} measure FRD by injecting a collimated beam into the fiber at a fixed angle.  The fiber azimuthally scrambles the beam to form a ring illumination pattern.  The diameter of the ring gives a direct measurement of the incidence angle and FRD related to the thickness of the ring in the radial direction. Although less direct than the cone test, the ring test is very simple to perform and interpret since there are very few sources of measurement error. Both of these testing methods typically find that FRD and associated throughput losses are minimized by ensuring that the input beam $f/\#$ is modestly slower than the fiber numerical aperture.  Beyond this, FRD is unique to each fiber, its alignment, and its stress state when the experiments are performed. 

We used both the full cone and collimated tests to characterize the DESI fibers. The full cone test was primarily used during R\&D to establish requirements for the fiber system. The main benefit of the full cone test is that it can be designed to match the input illumination that the fiber will see at the telescope and so both the far field and near field illumination at the output of the fiber can be studied. The stability of the illumination is essential to understand and can only be studied with full cone illumination at the input. However, this test requires careful angular alignment in order to remove geometric effects from the dataset and for this reason it is not a good test to use for quality and assurance (QA) during manufacture. In contrast the collimated FRD test is an excellent tool to use during Q\&A, however, in order to set requirements based on collimated FRD performance, a relationship between full cone and collimated tests is necessary. In the following sections we first describe each of these tests, and then how we established a relationship between them both.

\subsection{Full cone FRD testing for throughput}
\begin{figure}
\centering
\gridline{\fig{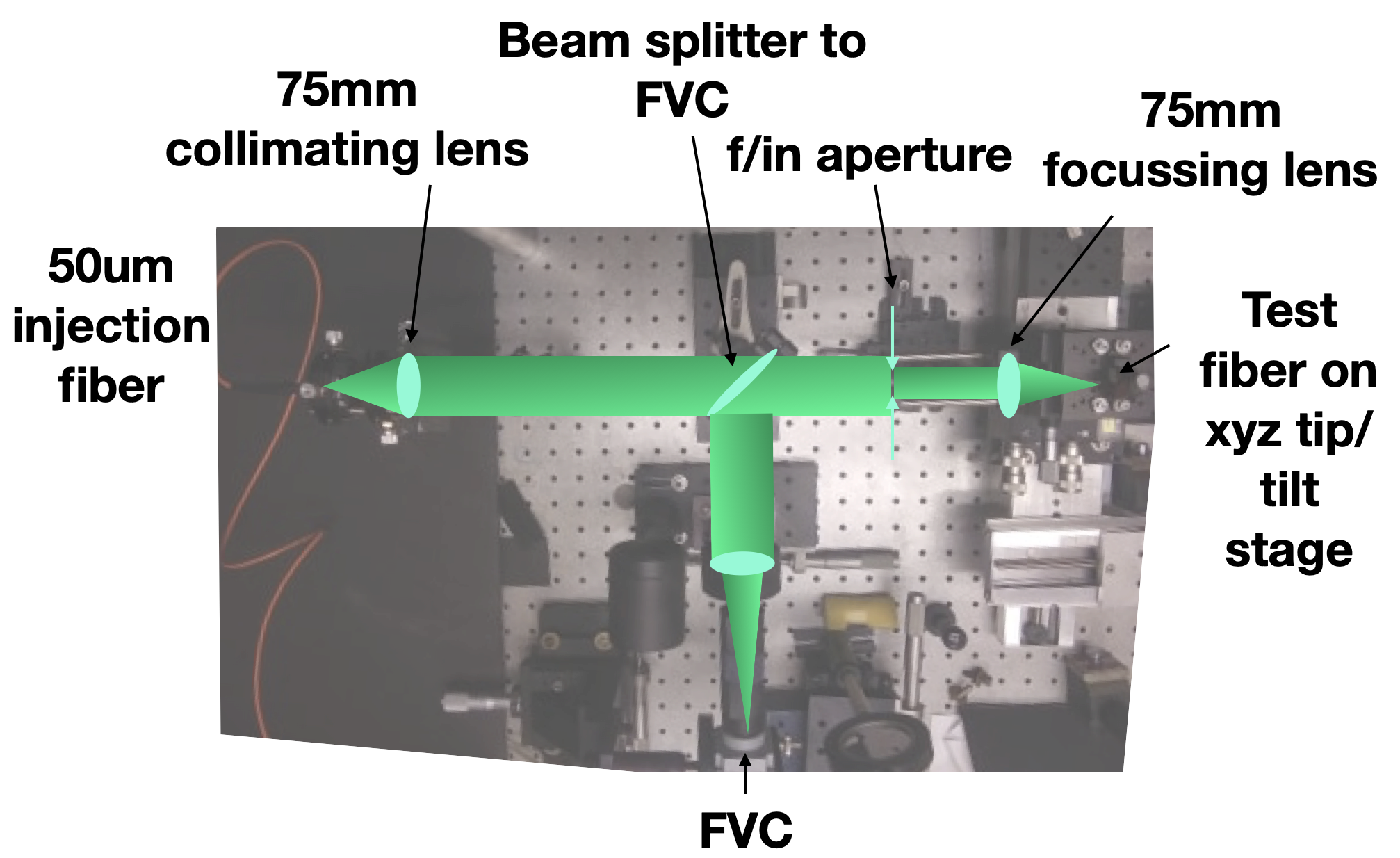}{0.48\textwidth}{(a)}
\fig{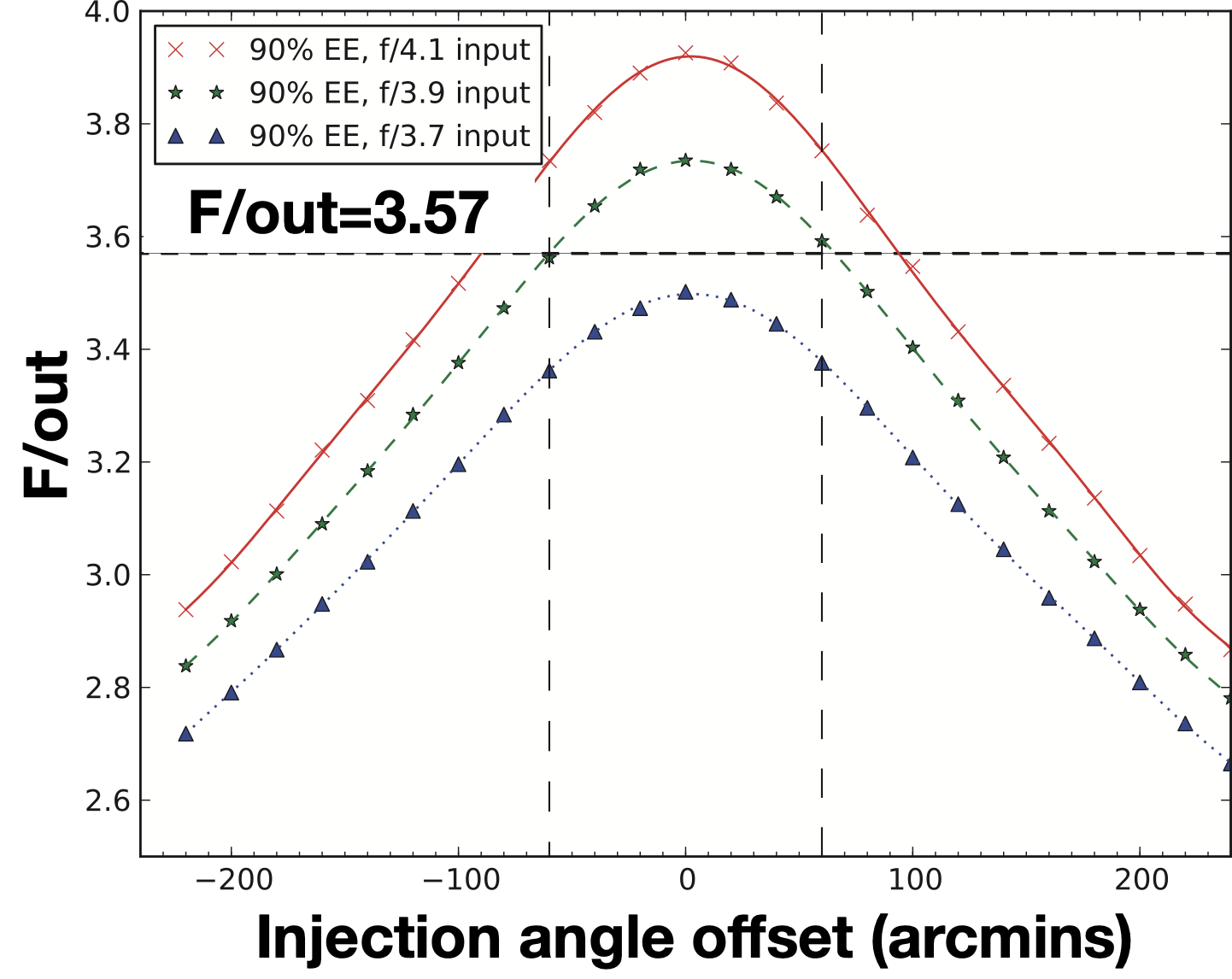}{0.4\textwidth}{(b)}}
\caption{Panel (a) shows the lab setup for full-cone FRD measurements. A $500\,\mu$m input spot is re-imaged onto a test fiber which is held on a $x$, $y$, $z$, tip, and tilt translation stage. An iris with obscuration is placed in the collimated beam to set the input focal ratio. Panel (b) shows the result of this test when measuring the energy enclosed in the far field at different input focal speeds of $f$/3.7, $f$/3.9, $f$/4.1 ({\it colored lines}) 
and with chief-ray injection angle tilts of $\pm240$\,arcmin.
The plot shows that 90\% of the input flux will be captured 
for the average DESI injection cone of $f$/3.9 ({\it green line})
and the DESI collimator acceptance speed of $f$/3.57 ({\it horizontal line})
if the injection angles are kept below about $\pm60$~arcmin ({\it vertical dashed lines}).}
\label{fig:FRD_rig}
\end{figure}

Our optical setup for full cone FRD testing is shown in panel (a) of Figure \ref{fig:FRD_rig}. A $500\,\mu$m core fiber is used as the injection spot in order to ensure that the test fiber is overfilled (since DESI is observing diffuse sources). This spot is collimated and a pupil mask is placed in the beam. The pupil mask sets the focal ratio of the beam and includes an obscuration scaled to the size of the prime focus corrector. A second lens then focuses this beam onto the test fiber. We aligned the test fiber in $x$, $y$, $z$, tip, and tilt and then  measured both the far field and near field of the test fiber as we changed the injection angle. These results are shown in panel (b) of Figure~\ref{fig:FRD_rig}.

The results of these measurements were used together with corrector, positioner and ferrule injection angle values to establish that the FRD throughput loss due to angular misalignment between the input beam and the optical axis of the fiber will not exceed 10\% given the $f/\#$ of the corrector injection and the spectrograph acceptance \citep{2016SPIE.9911E..0JB}. This result consequently set a requirement on the telecentricity of the beam entering the fiber. 

\subsection{Full cone FRD testing for beam stability}
The beam stability of a fiber system is effected by both modal noise, and changes in FRD performance. Modal noise can be minimised by increased scrambling \citep{2023AJ....166..105K} in the fiber system and FRD variation can be minimised by injecting at faster focal ratios \citep{10.1093/mnras/staa3904}, but the resulting `faster' beam of light then requires a faster collimator for the same size pupil, and a faster collimator might result in larger and more complicated optics to deliver the same efficiency and spectroscopic resolution. However, the penalty for generating FRD through injecting light into the fiber at too slow a FR is that the collimator and the rest of the instrument will then all grow in size if light is not to be discarded. 

\begin{figure}
\centering
  \includegraphics[height=.5\textwidth]{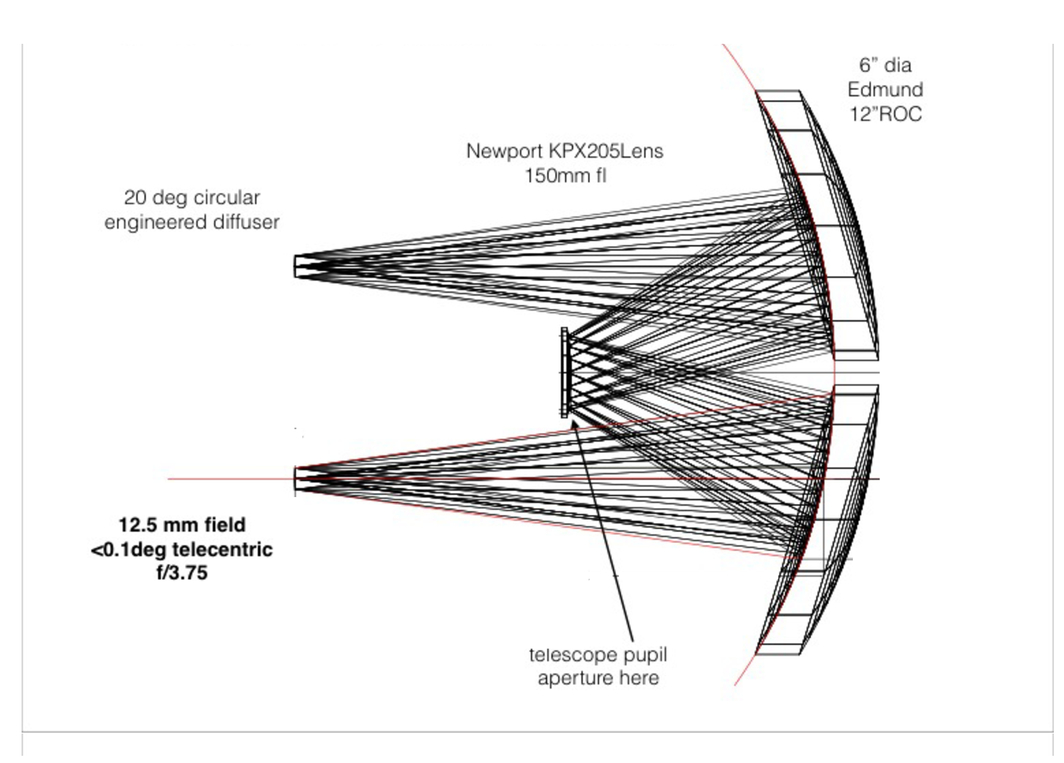}
\caption{An Offner relay was used in fiber testing since it allows a 12.5mm, telecentric planar field to be illuminated at $f$/3.9 which is the area weighted focal ratio of the DESI prime focus corrector.}
\label{fig:offner}
\end{figure}

The stability of the fiber system has a direct impact on the integrity of wavelength calibrations. The camera system inevitably has aberrations which lead to an aberrated image of the fiber. If the FRD is variable, the barycenter of the output of the fiber will also be variable resulting in a varying PSF and resulting aberrations. These variations mean that a calibrated measurement of the PSF can't be used for spectroperfectionism or generally to do sky subtraction.  For DESI, a particular concern is the distortions in the PSF due to movement of the fibers by the fiber positioners and movement of the telescope \citep{Poppett2014}. In order to test the performance of the fibers as they were moved around the patrol area of a fiber positioner, an Offner relay was used. The benefit of an Offner relay versus a traditional fiber full cone rig setup is that the relay can be designed to deliver an $f$/3.9 input beam (with scaled obscuration) over a 12.5mm, telecentric
planar field which is the patrol area of the DESI positioners. This setup is shown in Figure \ref{fig:offner}. In order to image the illumination distribution at the output of the fiber, a fast microscope objective was used to project onto a detector. A DESI test fiber with L$\sim$40m was placed in both a `twirling post` positioner and a `tilting spine' positioner and moved around the patrol area. 

\begin{figure}
\centering
  \includegraphics[width=.8\textwidth]{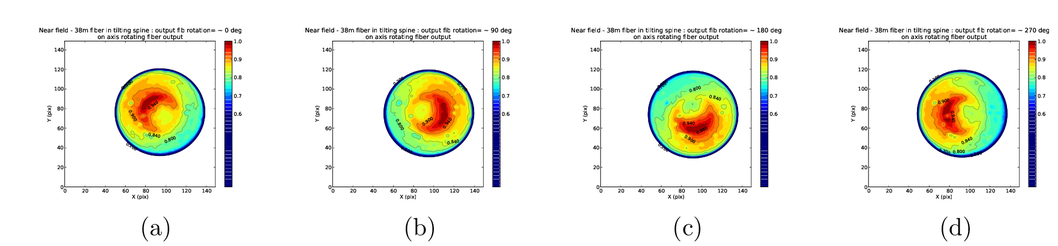}
\caption{The near field intensity distribution of the fiber output rotates as the fiber output is rotated and the input is held constant. This figure shows that the near field intensity distribution at the exit of the fiber is dominated by the stress at the end of the fiber caused by bonding the fiber into a  ferrule. The input (positioner end) of the fiber remained stationary during this test. The output of DESI fibers are different since they are bonded into slits, but the same intensity distribution was observed. The fact that it remains stable as the positioner moves is essential to the calibration cadence used during DESI operations.}
\label{fig:NF_rotate}
\end{figure}

Results of near field testing first showed that the illumination distribution at the end face of the fiber is strongly affected by stress at the output. The test fiber was bonded into a ferrule and polished and stress due to uneven glue distribution was seen in the illumination. This was verified by rotating the output ferrule and observing the illumination pattern follow this rotation. This result is shown in Figure \ref{fig:NF_rotate}.

\begin{figure}
\centering
  \includegraphics[width=.8\textwidth]{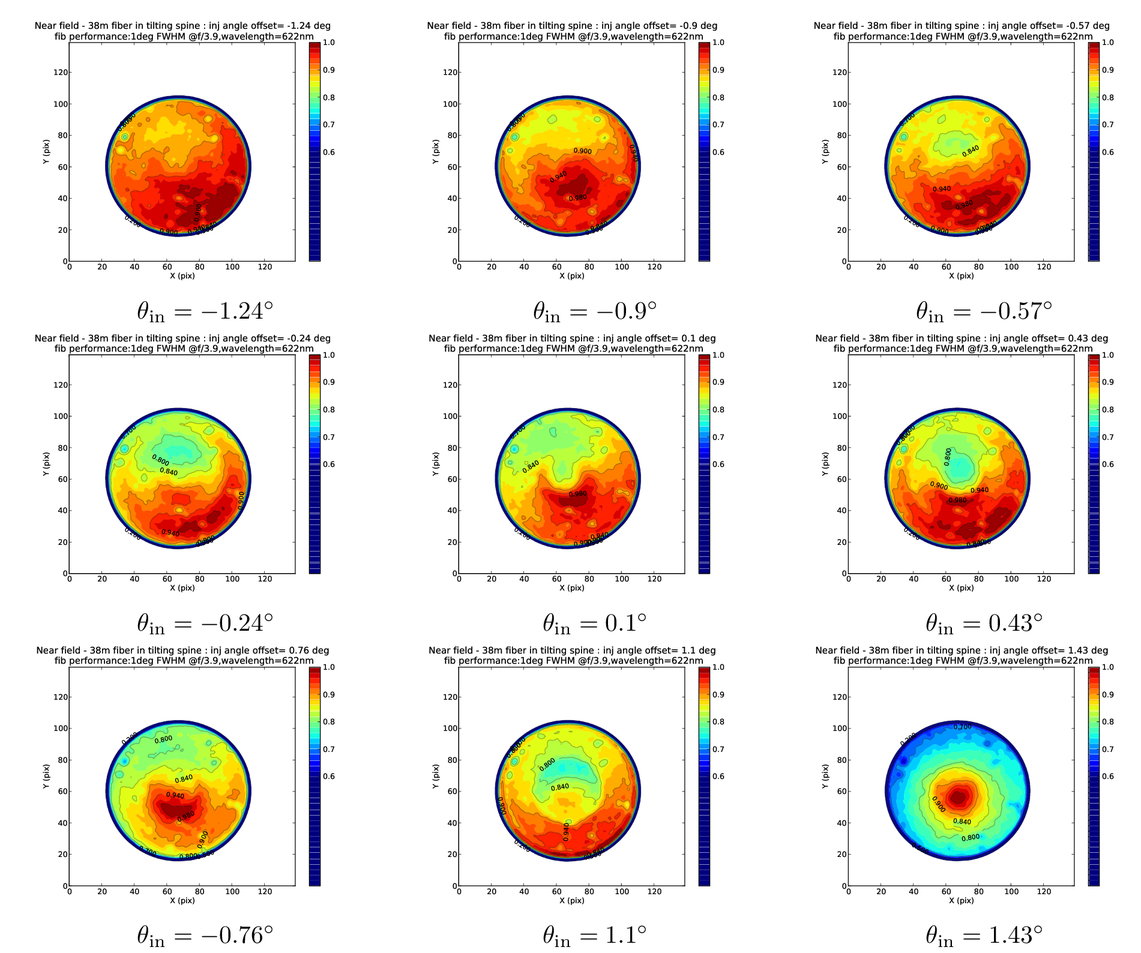}
\caption{Intensity distribution in the near field of the output of fiber changes as the fiber is misaligned with the optical axis. $\theta$in is the injection angle offset from the optical axis. The color scale is the same in all figures. The results for a twirling post positioner are not shown since the near field intensity distribution remains stable as the input moves. This is because the end effect dominates until geometric FRD is introduced.}
\label{fig:NF_angle}
\end{figure}

The second result of this test showed that geometric FRD dominated the illumination distribution. When the fiber was held in a `twirling post' positioner both the far field and the near field illumination distribution remained stable over the patrol area of the positioner. When the fiber was held in a `tilting spine' positioner, the FRD varied as a function of injection angle, as expected. However, an additional effect was seen in the near field illumination distribution whereby the near field intensity also varied as a function of injection angle, and this effect dominated over the effect of end stress observed in earlier testing. This result is shown in Figure \ref{fig:NF_angle} and is one of the reasons that a twirling post positioner design was chosen for DESI.

\subsection{Relationship between full cone and collimated FRD}
In order to establish a relationship between the full-cone and collimated FRD tests, a fiber must be measured on both test benches under the same conditions. However, this method will only provide the relationship for the specific fiber which was tested. In order to get the relationship for a range of fiber performances, we artificially stressed a 50\,m length test fiber in a repeatable way by adding weights to it in order to degrade the FRD performance. 

\begin{figure}
\centering
\gridline{\fig{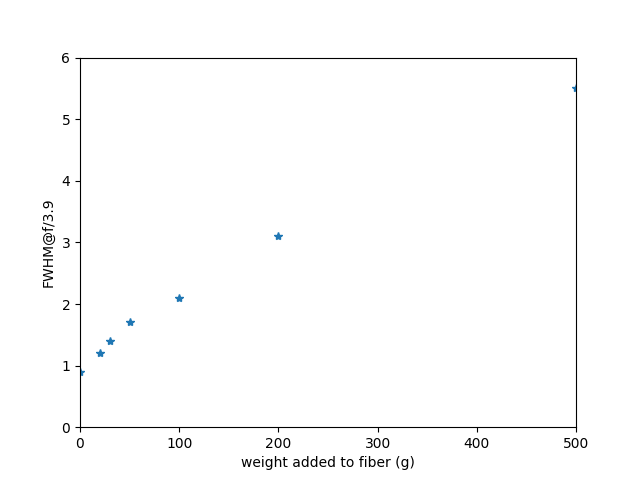}{0.45\textwidth}{(a)}
\fig{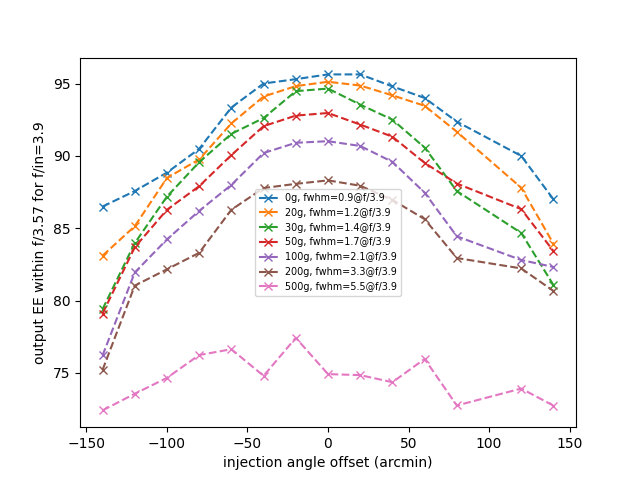}{0.45\textwidth}{(b)}}
\includegraphics[width=.95\textwidth]{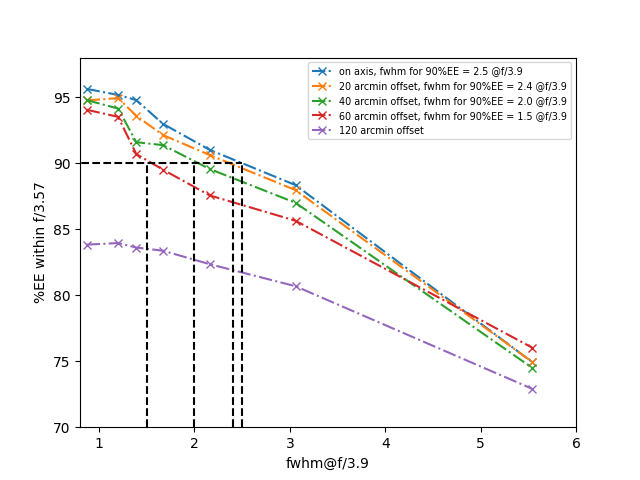}\\
{(c)}
\caption{The relationship between full cone performance to collimated FRD performance was established by measuring a fiber with artificially varying FRD performance on both a collimated FRD setup (panel (a)) and a full cone FRD setup (panel (b)). Since the focal plane  is designed so that the fiber is aligned to the chief ray to within 0.4 degrees (or 26 arcmins) and we require that losses due to FRD do not exceed 90\%,  we determine that the full fiber system must perform with a FWHM $\leq2.3^\circ$ for $f/in=3.9$ in the collimated FRD test. Panel (a) shows the Collimated FRD test data for a fiber with varying amounts of FRD performance. Panel (b) right panel shows the full cone test data for this same fiber with FRD artificially induced. Panel (c)shows the result of combining the results in these panels to establish an FRD requirement for various injection angle offsets.}
\label{fig:weight_FRD}
\end{figure}

In order to increase the FRD of the test fiber in a controlled manner, we sandwiched the test fiber  between two pieces of cardstock near the input end of the fiber and added weights. The weight varied from 0 to 500 grams, corresponding to a collimated FRD varying from $0.9^\circ$ to $5.6^\circ$ FWHM. The results are shown in panel (a) of Figure \ref{fig:weight_FRD}.

We placed this fiber into the full-cone test rig and applied weights in the same way to reproduce the degradation in FRD performance measured in the collimated test. By looking at the far field of the output of the fiber we were able to measure the enclosed energy for different amounts of artificial stress (and hence known FWHM performance). These results are shown in panel (b) of Figure~\ref{fig:weight_FRD}.

The results from these two plots was combined as shown in panel (c) of Figure \ref{fig:weight_FRD} to establish an FRD budget for the fiber system. The focal plane has an alignment budget that allocates $\pm0.4^\circ$ to the alignment of the fiber with respect to the chief ray. This requires that in order to meet 90\% throughput requirements, the full fiber system must perform with a FWHM $\leq2.3^\circ$ for $f/in=3.9$ in the collimated FRD test.

When the overall FRD requirement of the full fiber system was made, it was then possible for us to assign a budget to each subsystem.  In order to do this we needed to understand how the FRD performance of each subsystem would combine to contribute to the overall budget. This process is described in the following section.

\subsection{FRD budget} \label{sec:FRD_budget}
The main sources of FRD for the fiber system are the injection angle, performance of the PFAs, splicing performance, and cable and slit performance. We flow down this budget to the constituent parts of the fiber system. This budget was based on testing during the R\&D phase which established the performance that could be expected to be achieved for each for these items during manufacture.

By splicing together fibers with different FRD performances we were able to establish an equation for the final FRD performance \citep{2020SPIE11447E..11P}. This equation is:
\begin{equation}
FRD_{\text{splice}}=\sqrt{\rm{FRD}^2_{\text{PFA}} + \rm{FRD}^2_{\text{slit}}},
\end{equation}
where $\rm{FRD}_{\text{PFA}}$ is the collimated FRD performance of the PFA at f/3.9 input and  $\rm{FRD}_{\text{slit}}$ is the collimated FRD performance of the cable and slit assembly at f/3.9 input. $\rm{FRD}_{\text{splice}}$ is the FRD of the full system after splicing. The high quality of the fusion splices mean that they do not contribute significantly to the total FRD. Based on this equation we required that the PFA should be rejected if it has a FWHM $\geq1.5^\circ$, whilst the cable and slit should be rejected if it has a FWHM $\geq1.7^\circ$ in order to ensure that the final spliced system has a FWHM $\leq2.3^\circ$ at $f/in=3.9$ as summarized in table \ref{tbl:FRD_budget}.

\begin{table}[!ht]
\centering
\begin{tabular}{ |c|c| } 
 \hline
 Subsytem & Max FWHM @ f/3.9 input \\ 
 \hline \hline
 PFAs & 1.5 \\ 
 \hline
 Cable + Slit & 1.7 \\
 \hline
 Final Spliced system& 2.3 \\
 \hline
\end{tabular}
 \caption{FRD budget for the fiber system}
\label{tbl:FRD_budget}
\end{table}

\section{Fiber System Research and Development} \label{sec:RandD}
Research and development for the fiber system started in 2012 and focused on areas identified as being high risk or requiring a novel approach to their development. In the following sections we discuss each of these areas in the order in which they see photons. We begin with the development of the Positioner Fiber Assemblies (PFAs) in Section \ref{sec:PFA_RD}. PFAs help to ease production flow, and since the ferrulized fibers must be inserted into the positioner from the front, the fiber system must be built as discrete sections and then connected together. In Section \ref{sec:splice_RD} we discuss the R\&D that was necessary to optimise the connection of the PFA to the fiber cable and slit. In Section \ref{sec:RD_routing} we discuss the R\&D used to guide the design of the fiber routing around the telescope, and in Section \ref{sec:RD_cables} we discuss the fiber cable development effort. Finally, the R\&D required to develop the design and production of the fiber slit is discussed in Section \ref{sec:slit_RD}.

\subsection{Positioner Fiber Assembly (PFA) Development}\label{sec:PFA_RD}
PFAs are the section of fiber that is held by the fiber positioner. They are a single strand of optical fiber of length 3.1\,m and one end is optically terminated. It would be preferable to have a continuous fiber run, but PFAs were advantageous for a number of reasons. The primary reason for PFAs was to ease production flow and increase yield. By isolating the production of the fiber system from the cable and slits we reduced risk and increased production rate by being able to manufacture both items at the same time. Another advantage to building PFAs was that it is easier to apply anti-reflection (AR) coating to a 3\,m length of fiber than to a fully populated fiber cable and slit, since the coating is applied in a vacuum chamber. Finally, it reduced some requirements on the positioner design if the fiber could be inserted from the front of the positioner, rather than being fed through from the back.  

\begin{figure}[htbp]
\centering
\includegraphics[height=5cm]{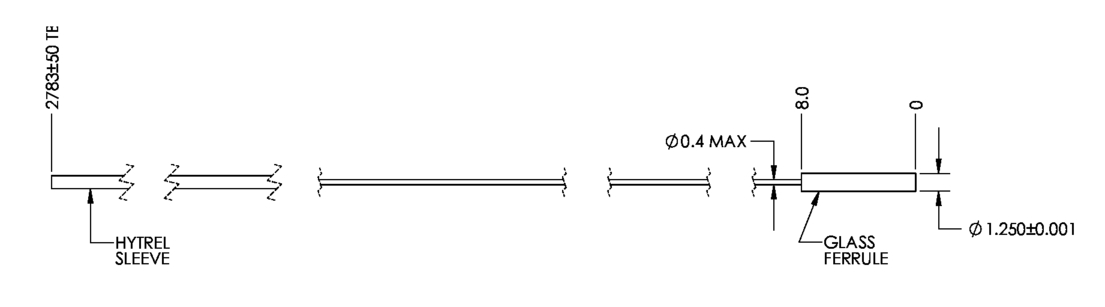} 
\caption{Drawing for the Positioner Fiber Assembly. A $\sim$3.1\,m length of optical fiber is cleaved and bonded into a glass ferrule with a UV curing epoxy (Dymax431). A tube polyimid tube is used to provide strain relief and is bonded to the back of the ferrule with a viscous UV curing epoxy (Loctite 3751) to protect the fiber as it routes through the positioner. The strain relief is shown with an OD of 0.4mm and fits within a pocket at the back of the ferrule. A hytrel sleeve is added to the fiber to further protect it at the back of the focal plane but is not bonded so that the fiber is free to rotate inside of it.}
\label{fig:PFA_drawing}
\end{figure}

The assembly begins at the focus of the focal plane and terminates in a fiber Spool Box where it is connected to the long run of fiber bound for a spectrograph. A drawing of the PFA is shown in Figure~\ref{fig:PFA_drawing}. We first cleaved a $\sim$3.1\,m length of optical fiber and bonded it into a glass ferrule with a UV curing epoxy (Dymax431). The fiber has an OD of $170\pm3\mu$ m and the ferrule has an ID of $176\pm2\mu$ m. These tolerances ensured that the fiber was properly centered within the ferrule and the length of 8mm ensured no tilt was introduced. A bond length of 3mm was required at the back of the ferrule. We then bonded strain relief to the back of the ferrule with a more viscous UV curing epoxy (Loctite 3751) to protect the fiber as it routes through the positioner. 

Essentially, all instruments with fibers at the focal plane must be terminated into some kind of focal plane assembly (regardless of positioning method, or single fiber versus Integral Field Unit or IFU) in order to place the fiber on target without inducing additional stress to the fiber system. Historically, this has been achieved by bonding the fiber into a ferrule, and then polishing the unit so that it is coplanar. Early instruments such as SDSS and MANGA bonded the fiber into a  stainless steel ferrule since the mechanical tolerances of these ferrules was high. However, the coefficient of thermal expansion (CTE) between the fiber and the ferrule resulted in high stress when the assembly experienced temperature changes. Later instruments such as LAMOST transitioned to ceramic ferrules, which have a better CTE match. However, the forces exerted on the fiber during polishing required a hard epoxy with a relatively long cure time, such as EpoTek-301, and this has a big impact on production work flow.

DESI took a different path due to many trade studies performed during initial R\&D. Rather than polishing the fiber, we determined that an optical cleave produced an extremely high quality optical surface that was both reproducible and reliable. DESI fibers were cleaved via the `tension-and-scribe' described by \citep{senior2009optical}, where tension is applied along the length of the fiber followed by an automatic scribing process with a diamond cleave blade. After the blade scribes the fiber, we maintain tension, causing the scribe to propagate across the fiber width and complete the cleave. We optimised the amount of tension applied to the fiber to maximise the mirror region on the face of the fiber and minimise hackle (a defect in the cleaved end face of an optical fiber, defined as multiple surface irregularities across the fiber surface) \citep{Poppett2014}.

\begin{figure}
\centering
\gridline{\fig{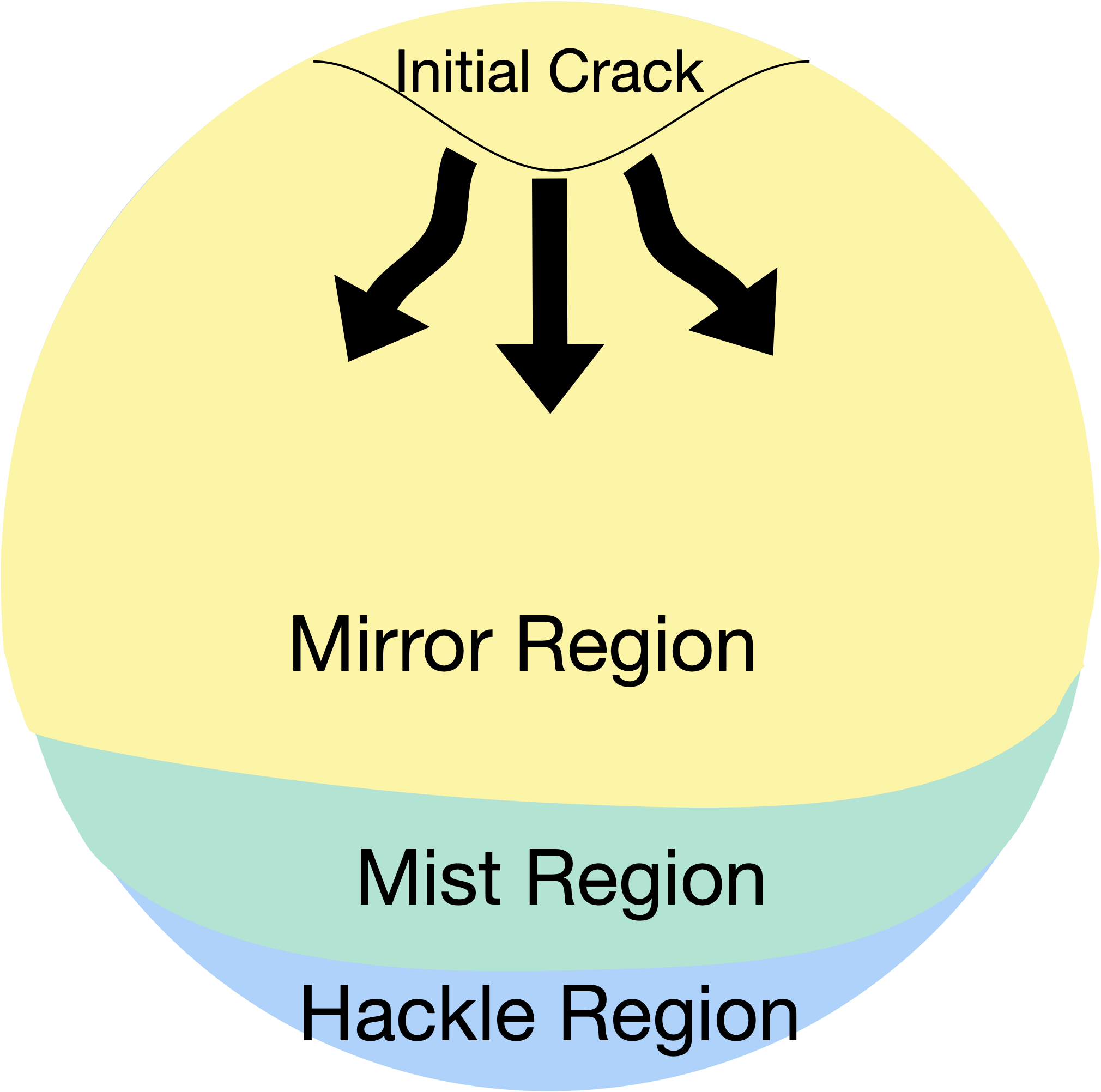}{.18\textwidth}{(a)}
\fig{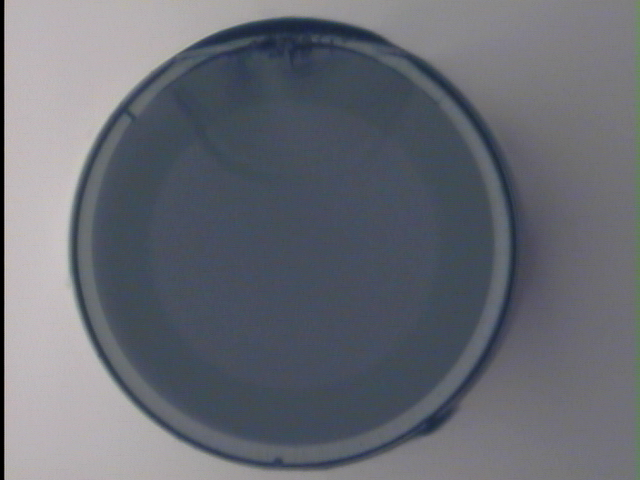}{.22\textwidth}{(b)}
\fig{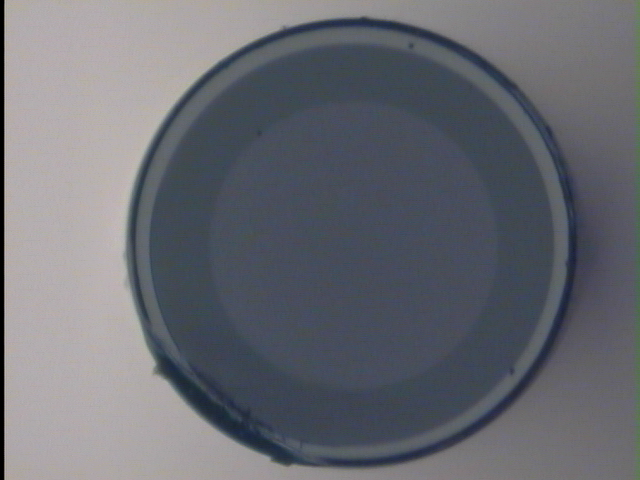}{.22\textwidth}{c)}
\fig{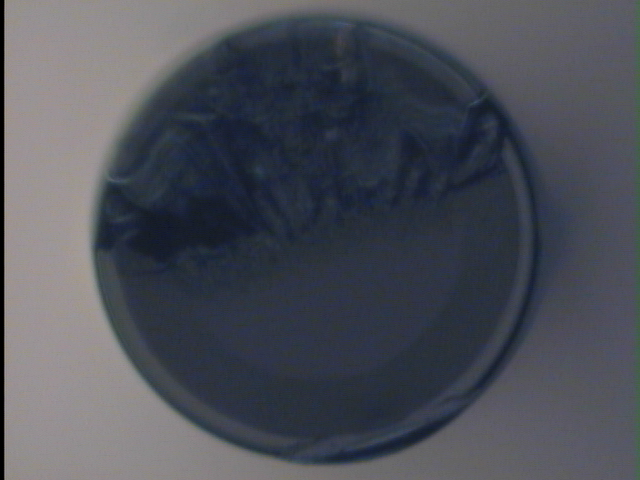}{.22\textwidth}{(d)}}
\caption{Cleave parameters for the DESI fiber were first optimised visually. The panel on the left shows a cleave schematic. The mist region is the transition area between the mirror region and the hackle region. Panels after this first show a DESI fiber cleaved under tension that is too low, then an optimal cleave, and finally a fiber cleaved under tension that is too high.}
\label{fig:cleave_images}
\end{figure}

Figure \ref{fig:cleave_images} shows microscope images of a DESI fiber that was cleaved with different amounts of tension. Too much tension will cause the cleave to propagate too fast, creating hackle on the cleaved end. If too little tension is used, the scoring edge needs to penetrate too deeply into the fiber to initiate the cleave, that results in a poor cleave.

We determined an approximate value for the optimum cleave tension by inspection with a microscope, as is shown in Figure \ref{fig:cleave_images}. We then refined this value by cleaving a fiber with various tensions close to the initial value from visual inspection and selecting the cleave tension that produced the smallest ring width in a collimated FRD test. 

Once the fiber is cleaved, it is bonded into a ferrule. Rather than using a traditional ceramic ferrule, we opted for a borosilicate ferrule. This has two main advantages. The first is that it allowed us to directly observe the amount of glue applied to the fiber and control where it went \citep{2018SPIE10702E..7PP}. The second is that it allowed UV curing epoxies to be used. This meant that there was a bigger range of glues to select from in order to find the properties we desired - easy to apply (viscosity somewhere between water and honey) and low stress at room temperature after curing with a CTE match to glass.

Our bonding rig setup allowed us to observe the effect on FRD of various glues as they were applied and as they cured. In all cases we applied a similar amount of glue and optimised the curing parameters. Testing showed that Dymax 431 was the optimal glue to bond the DESI fibers into ferrules based on both its performance and ease of handling \citep{2016Poppett}.

The final aspect to PFA termination is aligning the fiber to the front of the ferrule. This was achieved using a microscope with a fast objective in order to focus on both the fiber and the ferrule. We first focused the microscope onto the front of the ferrule, then translated the fiber until it was at the same focus, and finally bonded it into place. This process is described in more detail in Section~\ref{sec:ProdPFA}.

\subsection{Fiber connection development} \label{sec:splice_RD}

The most common way to connect fiber-optic cables is using mechanical connectors. A variety of connectors are widely used in many industries, including science applications. In the fall of 2012, the DESI team tested a number of commercially available connectors using the collimated FRD test and found that the performance of neither multiple nor single fiber connectors met our requirement of 90\% of the incident light being injected into the spectrograph \citep{fagrelius16}. The performance of these tests encouraged us to investigate other solutions and this proved to be one of the most important innovations for DESI.

Fusion splicing is a technique used extensively in the telecommunications industry.  With very low levels of loss, two uncoated fiber ends can be aligned, melted with an electric filament, and carefully pushed together to create a solid bond. However, fibers used in telecommunications are typically single mode. In order to optimise for multimode fibers and minimise focal ratio degradation, we had to work with vendors to improve performance.

\begin{figure}
\centering
\gridline{\fig{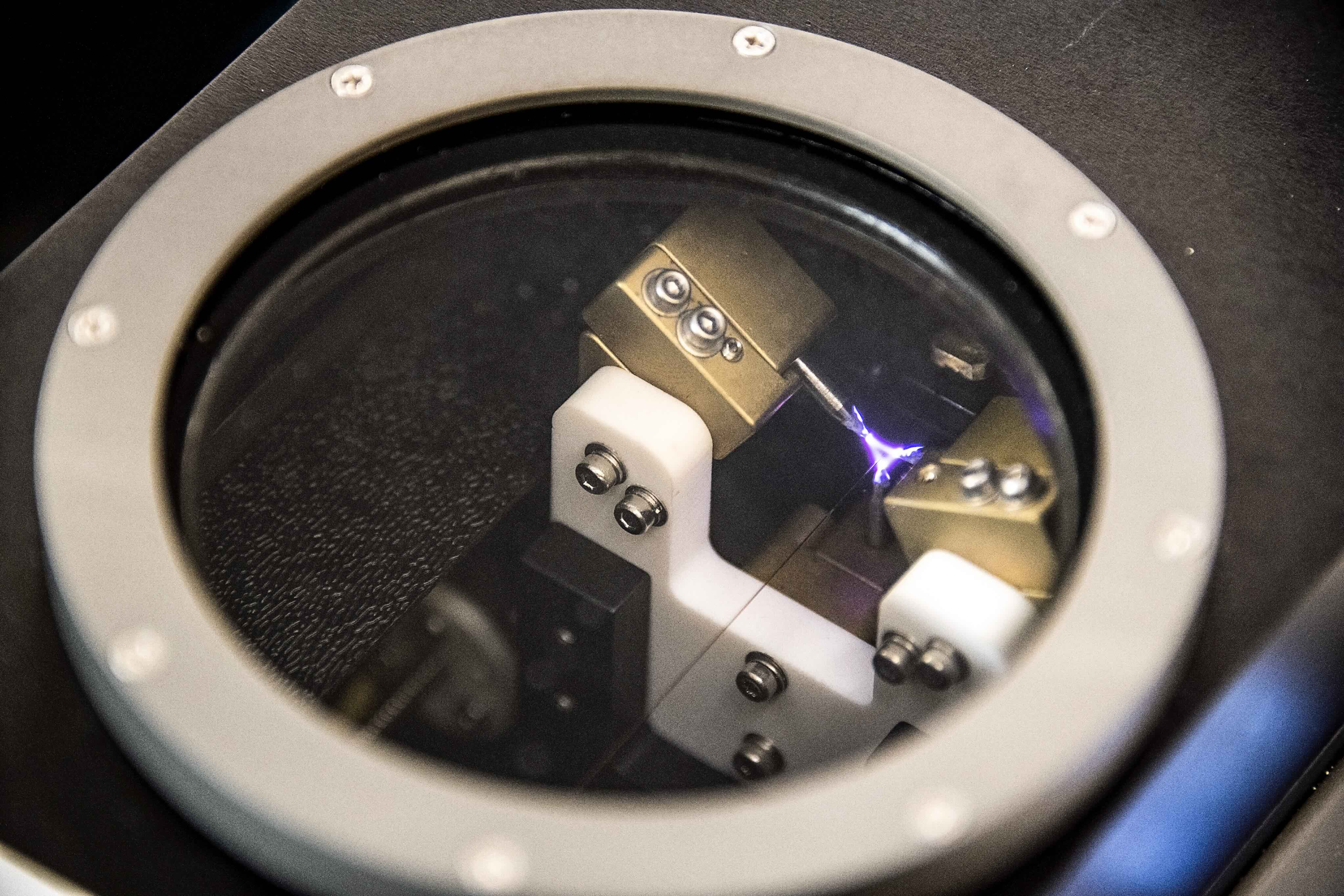}{0.48\textwidth}{(a)}
\fig{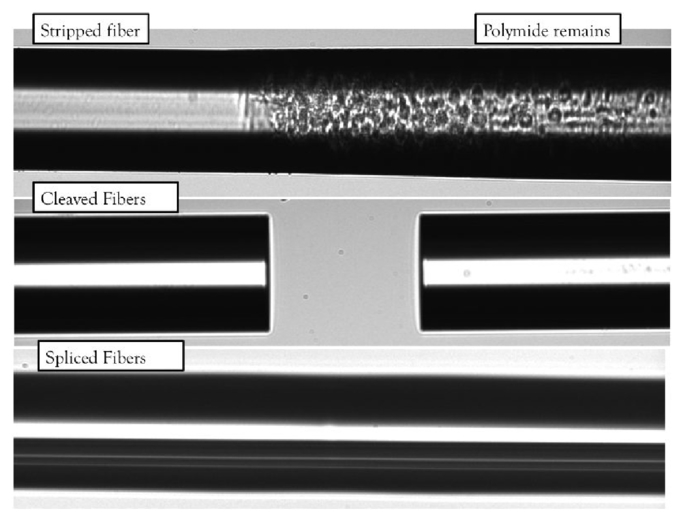}{0.45\textwidth}{(b)}}

\caption{The splicing process has four distinct steps. First, the polyimide buffer is removed from the fiber to expose the glass. Then each fiber is cleaved to obtain a high quality optical surface. Next, the two prepared fibers are aligned, heated, and pushed together to form a splice. Finally, the fiber is re-coated with acrylate to protect the splice. Panel (a) shows the `Ring of fire' stripper removing polyimid buffer from the fiber. Panel (b) shows images of the fiber during the splicing process. }
\label{fig:splice_process}
\end{figure}

Fusion splicing requires four steps, and each must be optimised to achieve the best performance. First, the polyimide buffer of the fiber must be removed in order to access the glass core. Multiple methods were tested to remove the buffer including mechanically via a sharp blade, ablation via a plasma, and chemically via a hot acid. We determined that a plasma stripper manufactured by 3SAE was both effective and efficient. This `ring-of-fire' fusion splicer is shown in action in the left panel of Figure~\ref{fig:splice_process}. The ion etching maintains the mechanical integrity
of the fiber by tapering the amount of coating on the fiber over a short length rather than having a sharp cutoff. Care must be taken to ensure that all coating is removed as even a thin layer of remaining polyimide can impact the performance of the splice. When the polyimide is fully removed from a short (10mm) length of fiber, the impact on the FRD is very small, but if the polyimide is heated during splicing, it can damage the fiber.

The next step is to splice the two fibers together. This requires clean optical surfaces. We achieved these via cleaving using the method we developed for the PFAs described in Section \ref{sec:PFA_RD}. We splice the fibers by carefully aligning the two ends and pushing them together with exactly the correct amount of force before heating them with a certain amount of power for a set amount of time. These parameters were the major source of development when optimising our splicing process. The heat is applied via an $\Omega$ shaped filament that surrounds the fiber and is heated by supplying power. Using the collimated FRD test, we found that minimal FRD is achieved when splicing with the lowest heat that would still create a good bond \citep{fagrelius16}. We designed a custom ergonomic splicing table to facilitate the scale of production required for the DESI focal plane. The production process is described and shown in Section~\ref{sec:prod_splicing}.

\subsection{Fiber routing} \label{sec:RD_routing}
The path the fibers take between the focal plane and the spectrograph is another area that required R\&D. The two main options were to feed the fiber system through the mount, or to route it around the outside of the structure. Whilst it was desirable to design the routing in a way that minimised the length of the fiber system in order to maximise efficiency, we also recognised that routing on the outside of the structure would simplify both installation and maintenance. 

The 4m-Mayall telescope's equatorial mounting has an inclination of 32 degrees with a `horseshoe' bearing (right ascension) and a perpendicular bearing (declination). The fiber system must be routed in such a way as to accommodate the movements around these two bearings without stressing the fiber system and a large amount of R\&D was required to minimise the length of the system and plan the installation to the telescope.

\begin{figure}
\centering
\includegraphics[height=.7\textwidth]{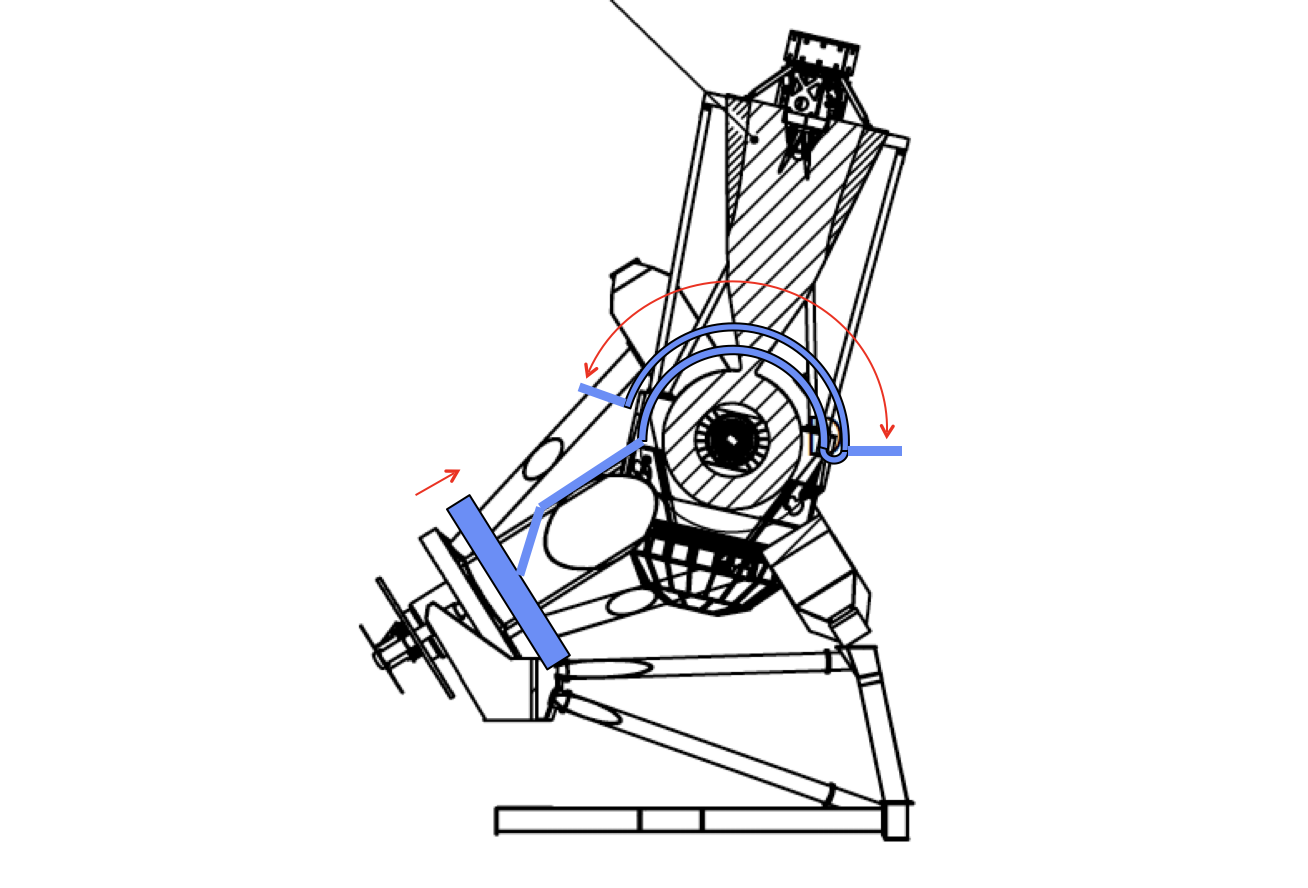}\\{(a)}\\
\gridline{\fig{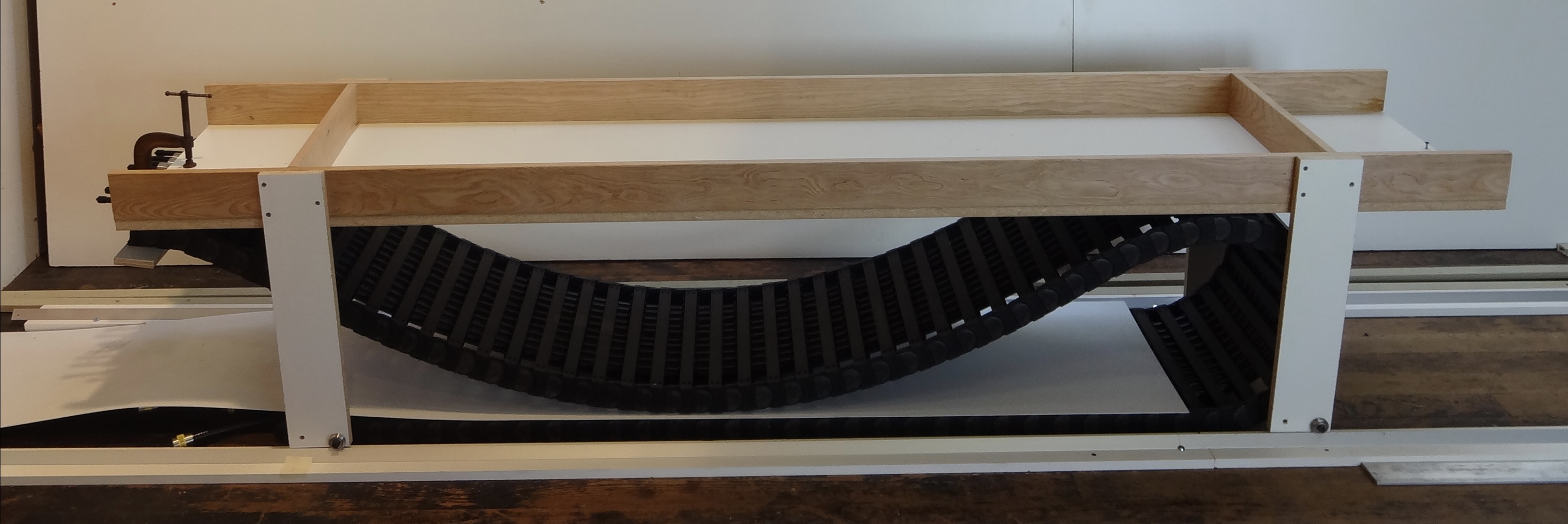}{0.6\textwidth}{(b)}
\fig{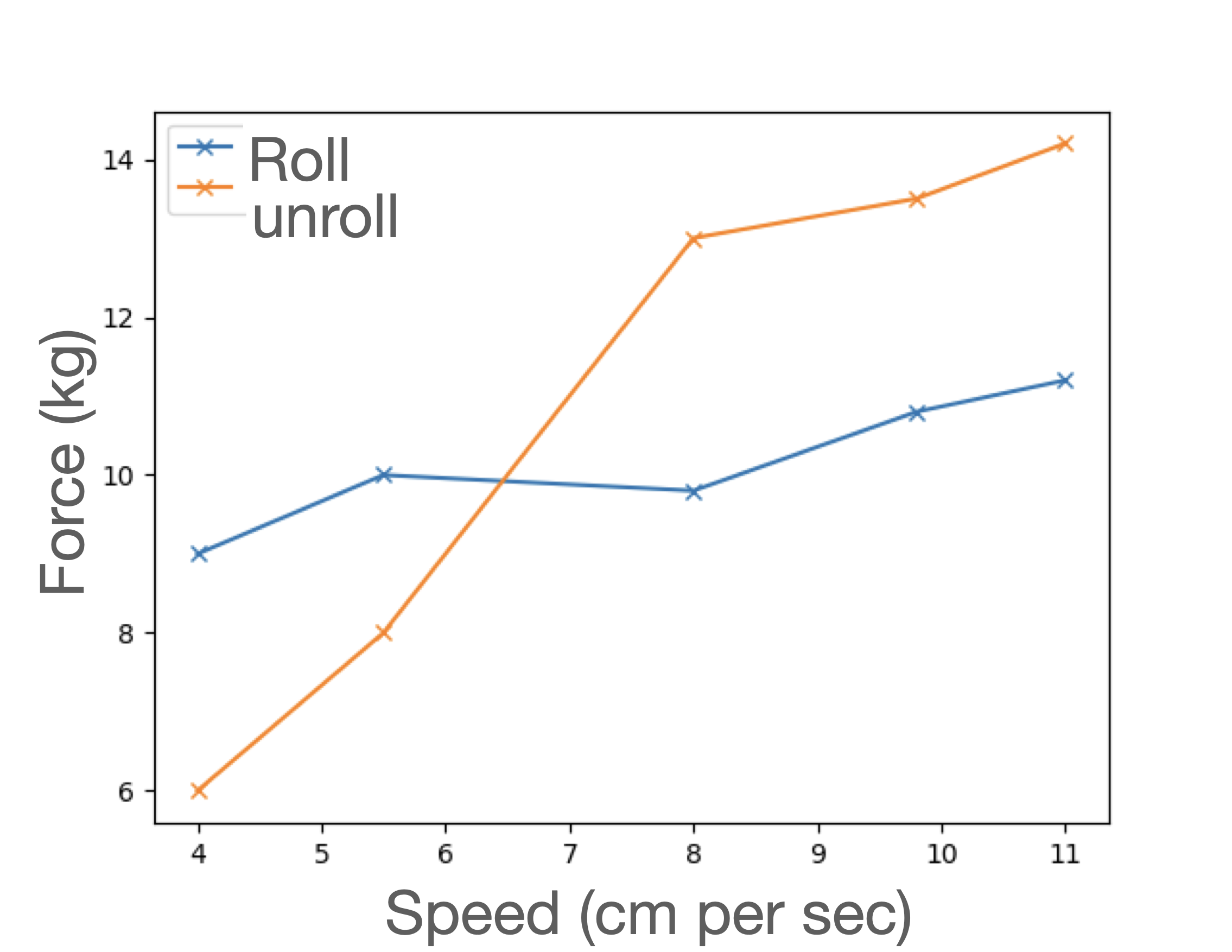}{0.3\textwidth}{(c)}}
\caption{Panel (a) shows the route of the fiber cables around the declination axis. The cables route around this axis in E-chain and to test the load that this would place on the telescope bearing we loaded ten short lengths of cable  into a section of E-Chain and measured the force required to roll and unroll the bundle. Panel (B) shows the lab setup. A force guage was attached to the top tray and the force as the cable was rolled and unrolled was measured. The tests were performed at speeds much faster than the slew rate of the telescope. The results are shown in panel (c) and the results reassured us that the movement of the telescope would not be impeded with this routing design.}
\label{fig:cables_RD}
\end{figure}

These fiber cables and the wraps were required to have some stiffness and so R\&D was required to ensure that the movement of the telescope would not be impeded around the declination bearing \footnote{This analysis was internal to DESI but Michael Warner at CTIO and the 150” telescope balancing manual contributed values for existing motor torques, telescope inertia, friction, and imbalance}.  In order to perform this test we built a lab setup that would mimic the fibers as they routed around the declination axis as shown in panel (a) of Figure~\ref{fig:cables_RD}. We loaded 10 short lengths of cable into a section of E-Chain cable wrap and measured the force required to roll and unroll the bundle as is shown in panel (b) of Figure~\ref{fig:cables_RD}. The tests were performed at speeds much faster than the slew rate of the telescope in order to stress the system and the results shown in panel (c) of this figure demonstrated that the movement of the telescope would not be impeded with this routing design.

\subsection{Fiber cable development}\label{sec:RD_cables}
The fiber cables couple the 5,000 PFA assemblies at the focal plane (500 per petal) to the ten spectrograph slits located in the Large Coud\'e Room. Each of the ten spectrograph-petal pairs have a single interlinking cable containing more than 500 bundled fibers (each cable contains a small number of spare and sky fibers). All cables are identical in construction and are $\sim44$\,m in total length.

\begin{figure}
\centering
\gridline{\fig{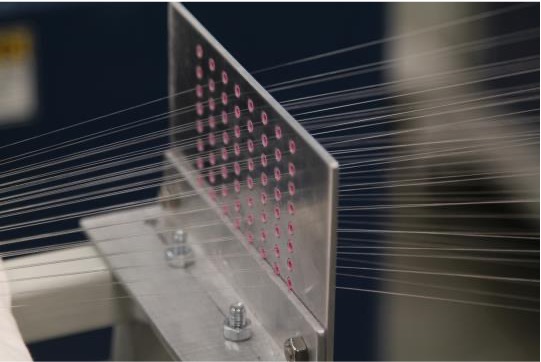}{0.48\textwidth}{(a)}
\fig{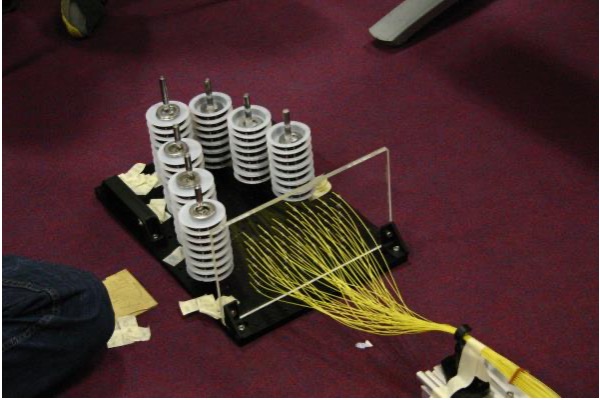}{0.48\textwidth}{(b)}}
\caption{Panel (a) - A bare fiber bundle inserted into a Miniflex conduit having passed through the mandrel with ceramic eyelets. Panel (b) - Experimental 50\,m fiber cable draw inserted into Miniflex conduit with individual fiber reels and a deterministic fiber feed using a mandrel of low-friction, low-static PTFE tubes plus a de-ionizer.}
\label{fig:fiber_mandrel_photo}
\end{figure}

The fiber cables were built by Durham University based on extensive design heritage from FMOS \citep{2008SPIE.7014E..5LM}. The main improvement from the FMOS fiber cable is in how we inserted the fibers into the Miniflex tubing which is wound into the fiber cables. Testing for DESI showed that using a mandrel system (see Figure~\ref{fig:fiber_mandrel_photo}) prevented static build up on the fibers and reduced tangling upon insertion. We also placed a de-ioniser in the vicinity of the bare fibers to reduce static, although this did not prevent fiber tangling. In conjunction with industry partner PPC Miniflex Ltd.\footnote{PPC Broadband Fiber Ltd, Unit1, Parham Airfield, Woodbridge, IP13 9EZ, United Kingdom}, a better method was developed. This saw clutch mechanisms added to the spools of bare fiber, a mandrel consisting of ceramic eyelets (rather than PTFE tubing) and using a fine powder of Boron Nitride as a lubricant (in a vibrating hopper through which the fiber was drawn) rather than a de-ioniser. Additionally, rather than attaching a bundle of fibers to a draw string and pulling them through furcation tubing, the tubing was in fact extruded around the fibers instead.

\subsection{Slit assembly development} \label{sec:slit_RD}
R\&D for the fiber slit focused on the trade study between three options. The first option was to use a monolithic slit such as in VIRUS. This option is superior in terms of optical performance, but is difficult to manufacture. The second option was an array of diverging slit blocks with flat optical windows bonded onto a curved plate. The final option was an array of parallel slit blocks with flat optical windows bonded onto a curved plate such as in MaNGA \citep{2015AJ....149...77D} and WEAVE \citep{10.1117/12.2560331}. 

\begin{figure}
\centering
\gridline{\fig{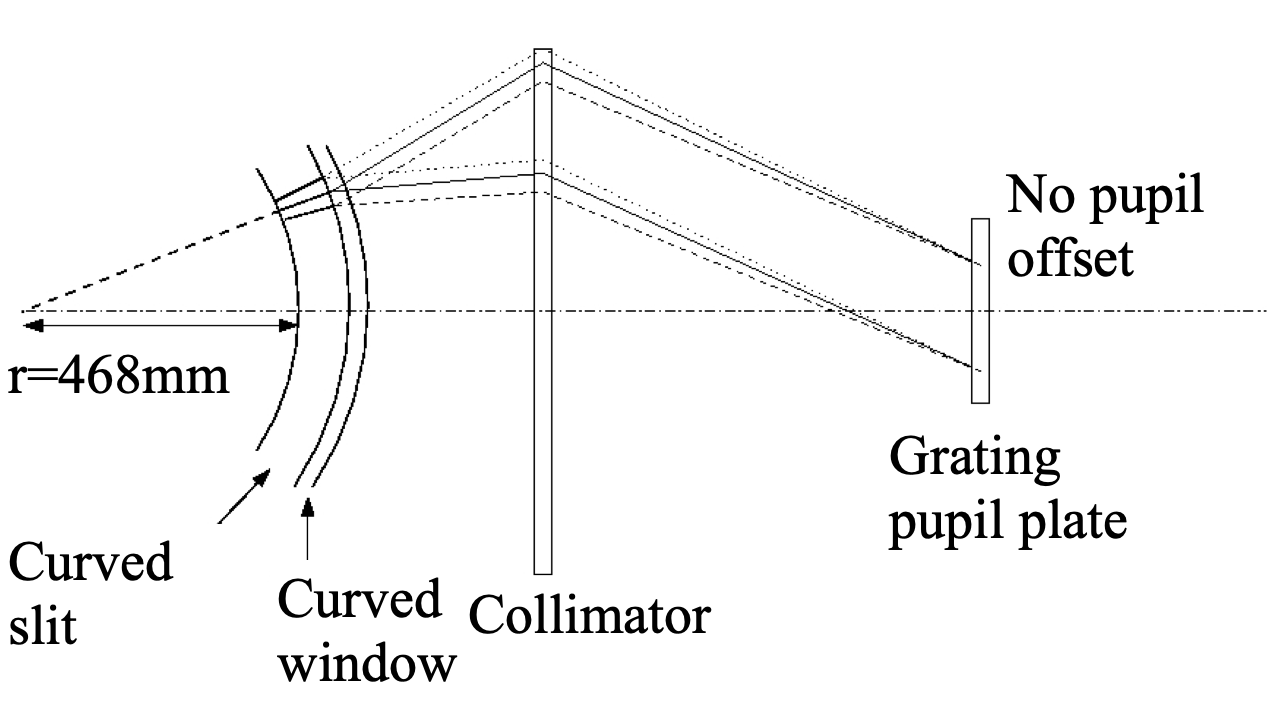}{0.48\textwidth}{(a)}
\fig{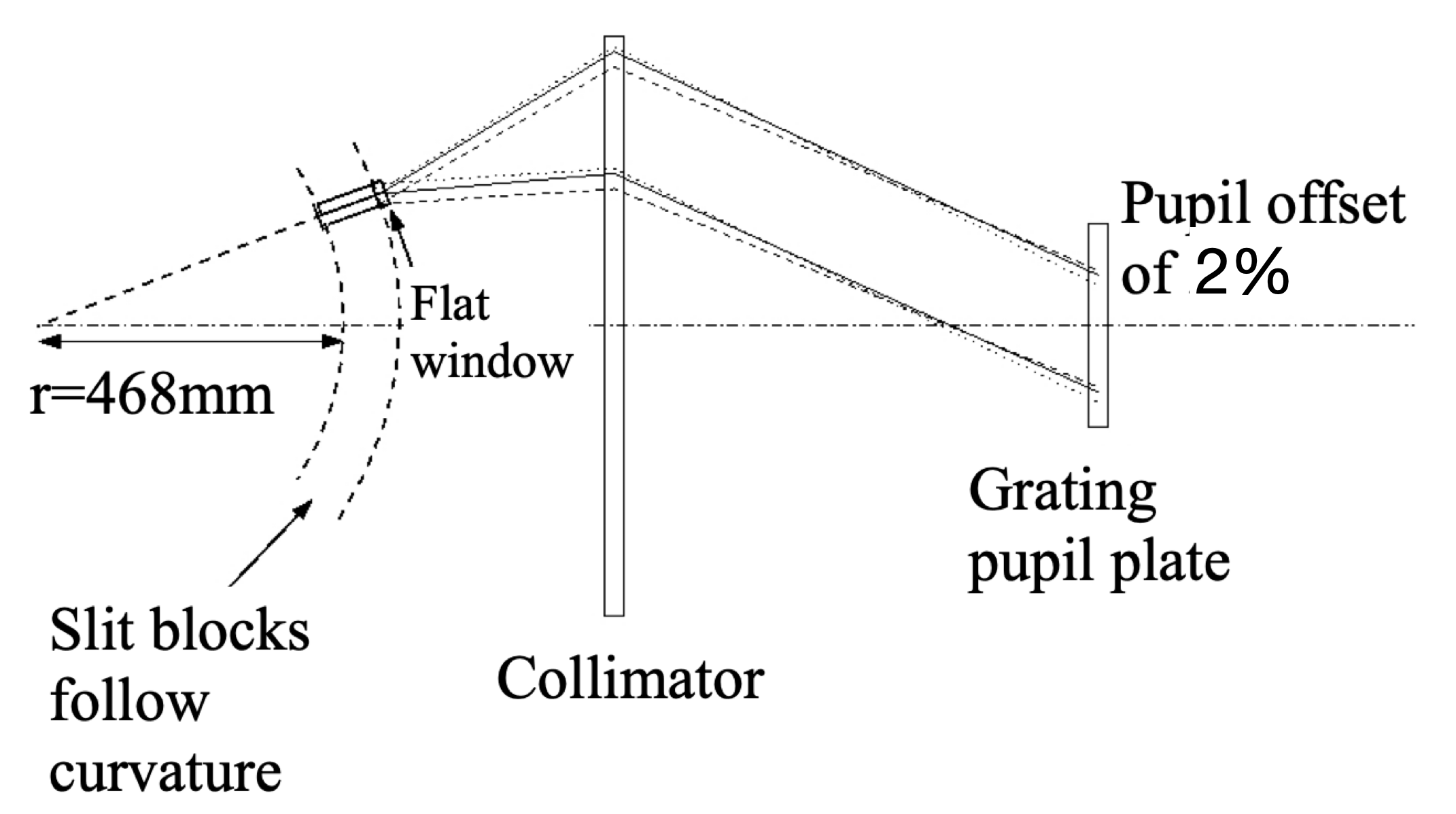}{0.48\textwidth}{(b)}}

\caption{Two scenarios for the construction of the curved slit assembly. Panel (a) -  The theoretically ideal case - a monolithic, curved slit where the fibers are flared, such that their axis passes through the center of curvature of the slit with a curved optical window in front of it. Panel (b) - The parallel V-groove slit block scenario which introduces a slight shift of the pupil in the disperser plane. }
\label{fig:slit_options}
\end{figure}

The left panel of Figure~\ref{fig:slit_options} shows the configuration of a curved slit where all fibres arise from a point on the optical axis that is 468.3\,mm behind their outputs. The pupils of all fibres meet again on the gratings of each spectrograph channel. The right panel of Figure~\ref{fig:slit_options} shows the design where 20 blocks of 25 fibres each are separately fed from the fibre cable breakout to the slit. Each block is about 6\,mm wide and comprises a glass V-groove array with the fibres bonded into parallel grooves. A glass lid is attached above the fibers as they sit in their respective V-grooves. The slit block has an angled `knee' along one edge to allow fitting to the curved slit plate and an AR-coated window is bonded to the front, optically polished surface.

We used Zemax simulations to compare the three slit scenarios: 1) the `ideal' scenario with a slit that is curved with 486.3\,mm radius; 2) a scenario with 20 blocks of 25 fibres, each running in parallel grooves, and that approximate the correct radius; 3) a scenario where the grooves of the blocks are flared to enable the angles to follow the ideal values along each slit block. Compared to scenario 1, both other systems showed no significant deviations with regard to spot shapes or RMS spot radius, and only a slight defocus introduced by the straight blocks. With 6\,mm wide blocks, the sag difference between a flat and a curved surfaces of $r = 468.3$\,mm is $10\,\mu$m, which translates to a defocus related image size increase of $2.6\,\mu$m when using an $\mathrm{NA}=0.13$ fibre. With a core diameter of $107\,\mu$m, this is an increase of 2.4\% and is negligible after taking into account the demagnification and the sampling by the detector pixels. By balancing the focus, this value can be decreased by another factor of two.

\begin{figure}
\centering
\includegraphics[width=0.7\textwidth]{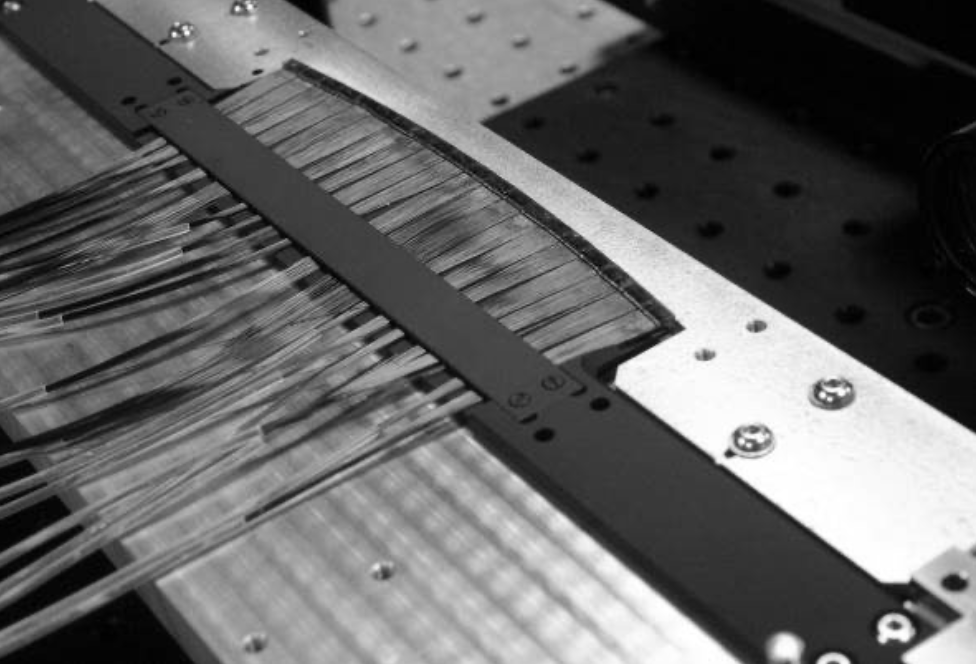}
\caption{Twenty V-blocks bonded with 25 fibers each bonded along the length of the slit backing plate, approximating the slit curvature of $486.3\pm3$\,mm (convex). The fibers are sub-furcated in to PTFE tubing bundles. Each V-block has an AR-coated window attached on the front face.}
\centering
\label{fig:populated_slit_plate}
\end{figure}

The angular misfit of the parallel groove block for scenario 2 would cause a slight shift of the pupil in the disperser plane when comparing fibres along a monolithic, curved block. The fact that all 25 fibres are parallel give rise to an angular misalignment that increases from 0° at the slit block centre to 0.34$^\circ$ at the slit edges. The consequence of this misalignment is a pupil displacement of up to 2.4 mm in the plane of the grating (2\% of the pupil diameter). Assuming a top-hat function of the fibre output (pessimistic approach) this translates into 2.7\% light loss unless the spectrograph optics can still accept these marginal beams.  Whilst this effect could be compensated by flaring the grooves, it is a small effect. In addition, our geometrical throughput analysis showed that the angular misfits caused by parallel grooves along a block have no geometrical vignetting effect on the current spectrograph model.   Hence we decided to use slit blocks with parallel V-grooves as they simplified the manufacture and reduced costs. Figure~\ref{fig:populated_slit_plate} shows 20 slit blocks bonded to form a curved slit.

\section{Production Schedule and performance}
\label{sec:Production}
The production of the DESI fiber system began in August 2016 and was completed in April of 2019 (with an additional two spare petals with complete fiber cables and slitheads as required by the project). As discussed in Section \ref{sec:RandD}, the fiber system was built in two discrete units in order to ease production flow and reduce risk, where one unit was the PFAs and the other unit was the cable and slit assembly. The PFAs were built at Lawrence Berkeley National Laboratory (LBNL) before being installed into robotic fiber positioners\citep{10.1117/12.2311573} that were then installed into a focal plane petal. The cable and slit assemblies were built by a team at Durham University (DU). The fully populated focal plane petals are then connected to the cable and slit assemblies via fusing splicing. We now refer to the input of the fiber system as the focal plane rather than PFAs. The fiber system was subject to acceptance testing at each of these discrete stages, and then again when it was fully integrated. The fiber system was installed at the 4m-Mayall telescope  with $ \geq 99\%$ of fibers intact, $ \geq 90\%$ fibers with collimated FRD $\leq$ 1.8$^\circ$, and $ \geq 90\%$ throughput delivered from the prime focus corrector to the spectrograph. 

The production of the PFAs is described in Section \ref{sec:ProdPFA}, the production of the cables is described in Section \ref{sec:prodcables}, the production of the slits is described in Section \ref{sec:prodslits}, and the fusion splicing is described in Section \ref{sec:prod_splicing}.

The production of the fiber system was mostly performed by highly skilled technicians at LBNL, SSL, and DU. Although quality assurance was built into the design and production of the system, the success of the project is in large part due to the diligence and commitment of these technicians to whom we are eternally grateful. 

 \subsection{Positioner Fiber Assemblies (PFAs)}\label{sec:ProdPFA}
All 5,000 PFAs were built at Lawrence Berkeley National Laboratory (LBNL) over the course of 15 months  \citep{Poppett2018}. Throughout this period, an average of 24 passing assemblies were made per day with a maximum of 62 per day. The pass rate based on FRD analysis and focus inspection was 97\%.

The production of the PFAs required a number of steps. First a fiber was cleaved and installed into a ferrule that had been plasma cleaned. The fiber was then aligned in focus with the front of the ferrule. Unlike traditional polishing techniques whereby the fiber is automatically coplanar with the front surface of the ferrule, when cleaving and bonding care must be taken to place the fiber at the same focus as the ferrule. In order to achieve this result with a cleaved fiber, the ferrule is held in a custom clamp at the focus of a telecentric microscope with a short depth-of-focus. The fiber is inserted from the back of the ferrule on a motorised stage, back illuminated, and then independently translated until it is brought to the same focus as the ferrule. When both the ferrule and the fiber are at the same focus position, the fiber was bonded in place with Dymax 431 as described in Section \ref{sec:PFA_RD}. Since both the fiber and ferrule are clamped when the glue is applied and cured, the fiber must be well aligned in both $x$ and $y$ so as not to bond in stress. Two cameras that are perpendicular to each other were used to verify this alignment.

The next stage in PFA manufacture is the installation of a 320mm length strain relief tube that protects the fiber from the internal mechanics of the fiber positioner. 

The strain relief tube is a polyimid tube with ID=343 $\mu$m and wall thickness=38$\mu
$m. This tube was laser cut in order to provide a clean bond to the back of the ferrule. 

\begin{figure}[ht]
	\centering
\gridline{\fig{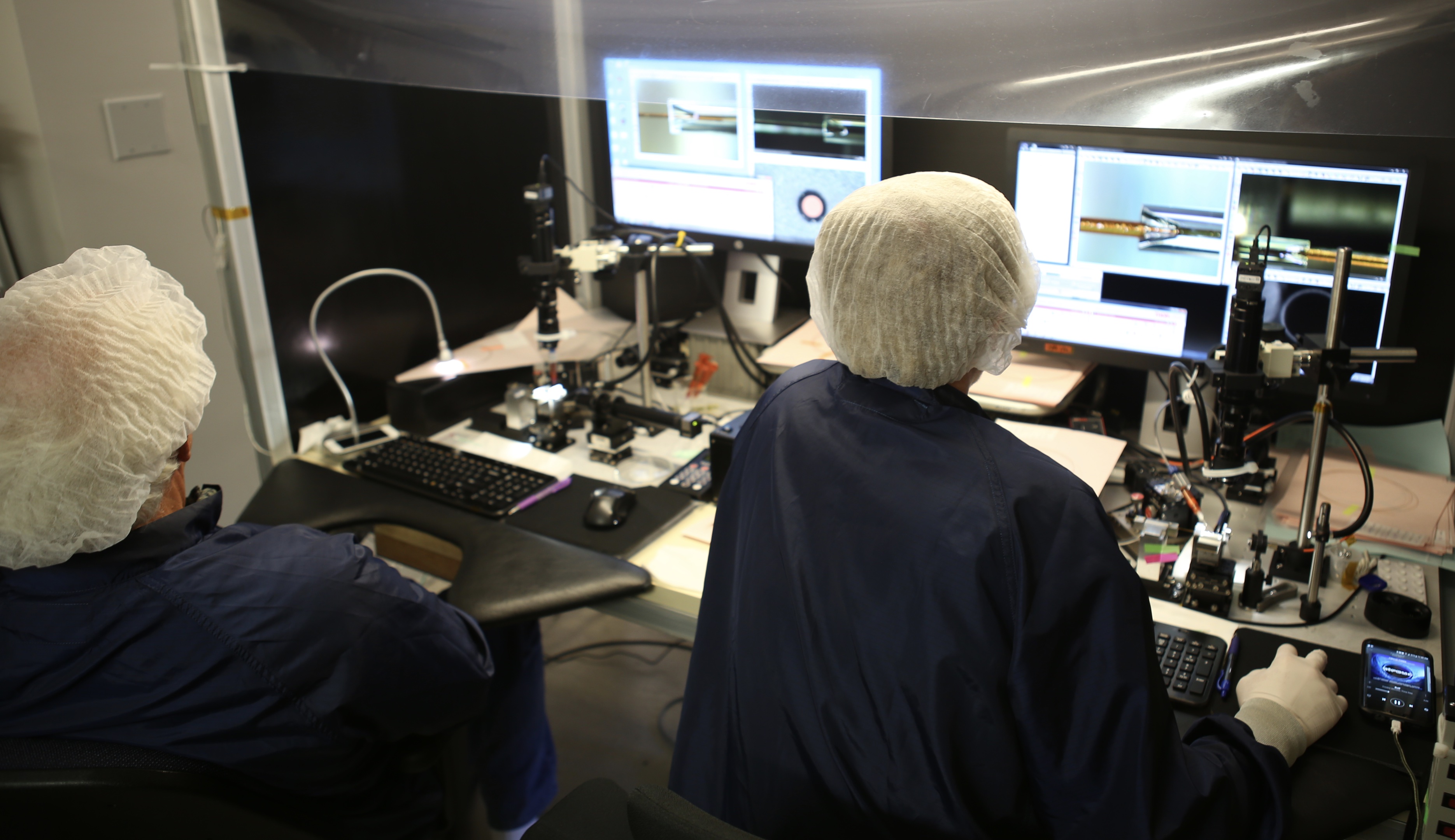}{0.48\textwidth}{(a)}
\fig{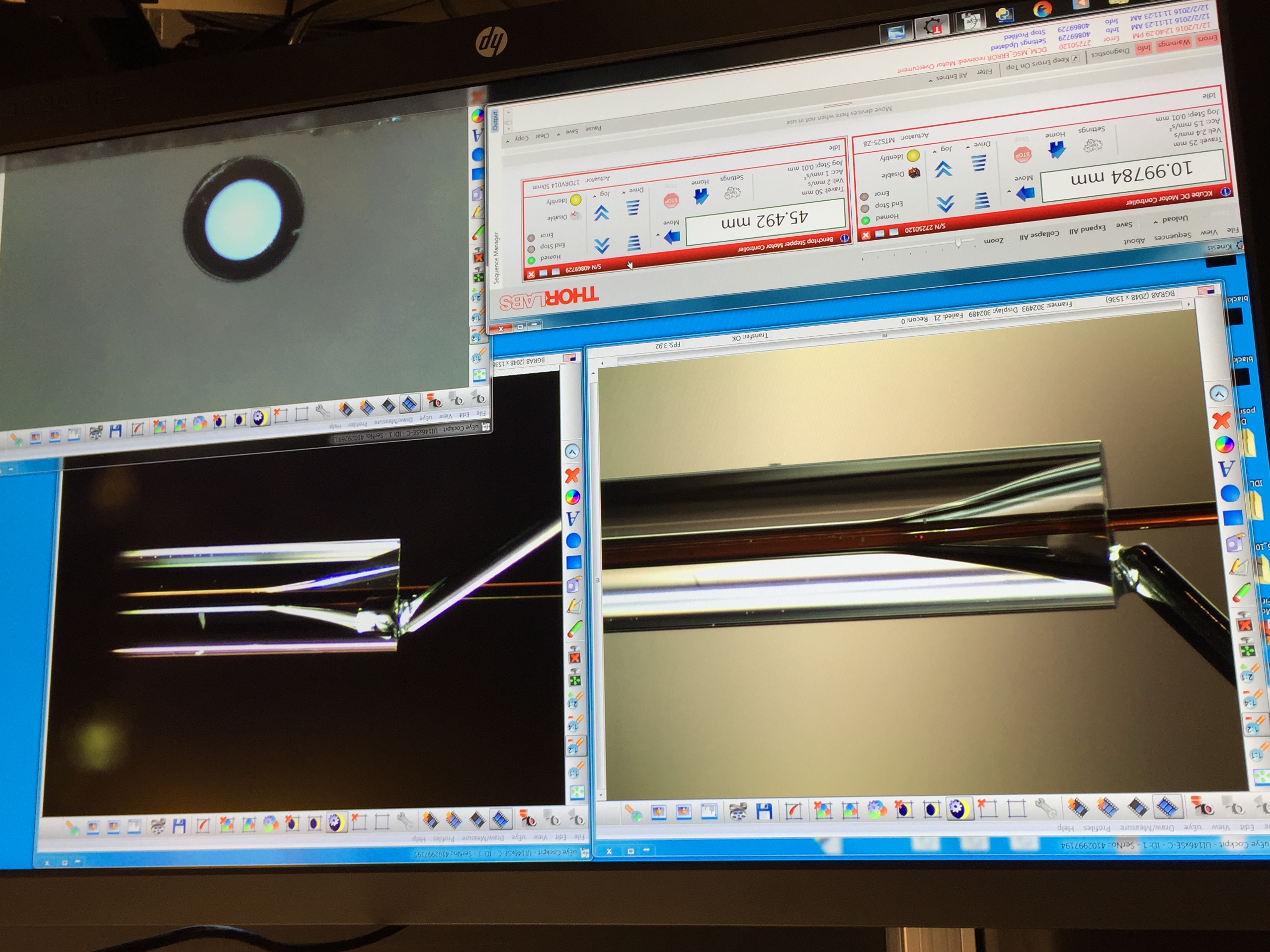}{0.4\textwidth}{(b)}}
\caption{The fibers were aligned and bonded into the ferrules whilst carefully monitoring from three different angles. This screen shot shows the rear of the ferrule as the glue is added, in addition to the fiber which is back illuminated in order to verify that it remains at the correct focus.}
\label{fig:bonding}
\end{figure}

A custom, ergonomic bonding rig was implemented due to the scale of the PFA production. This rig provided two stations, one to bond the fiber into the ferrule, and a further station to bond the strain relief onto the ferrule and measure the collimated FRD of the completed unit. This bonding station is shown in Figure \ref{fig:bonding}.

\begin{figure}[ht]
	\centering
\gridline{\fig{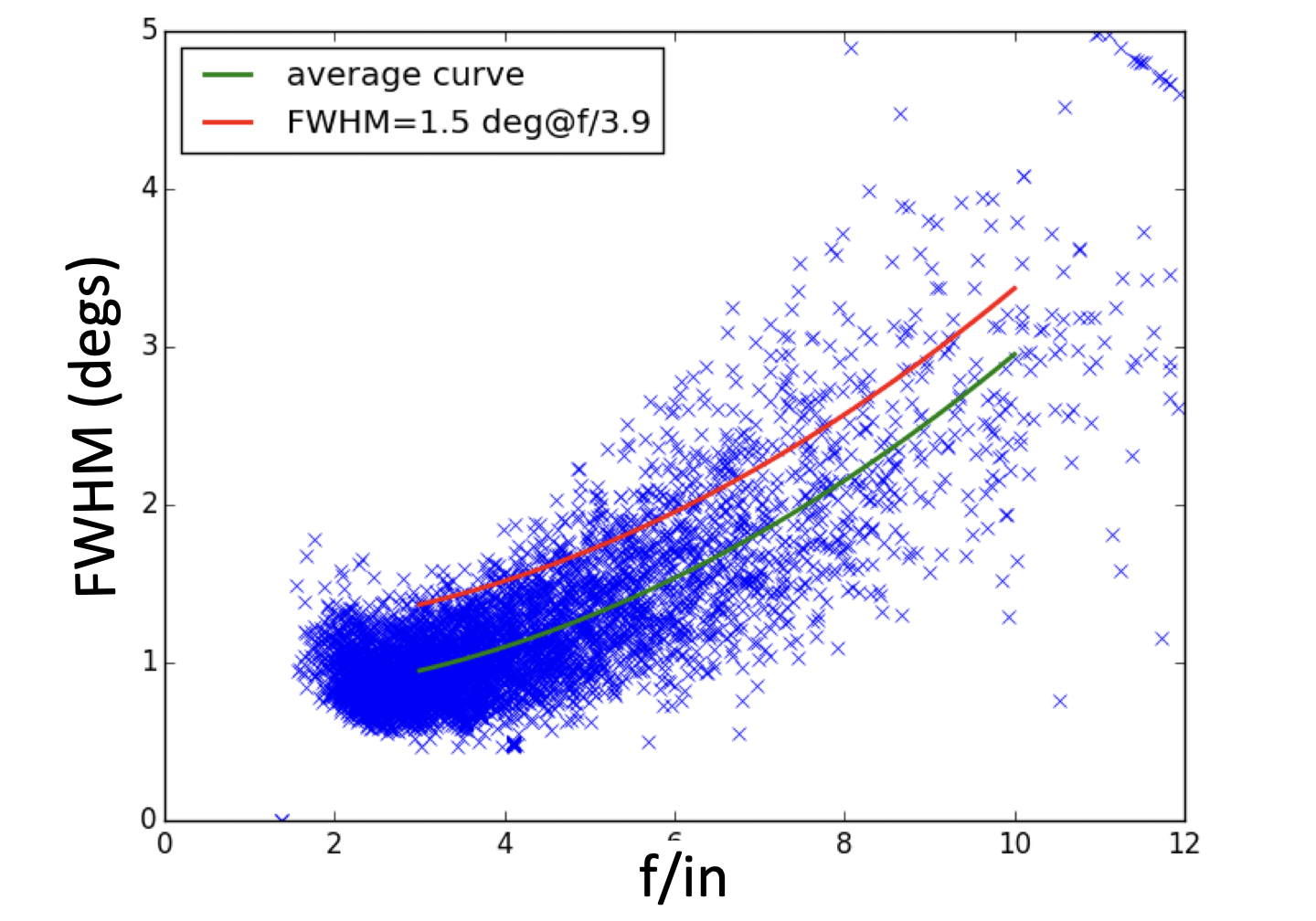}{0.4\textwidth}{(a)}
 \fig{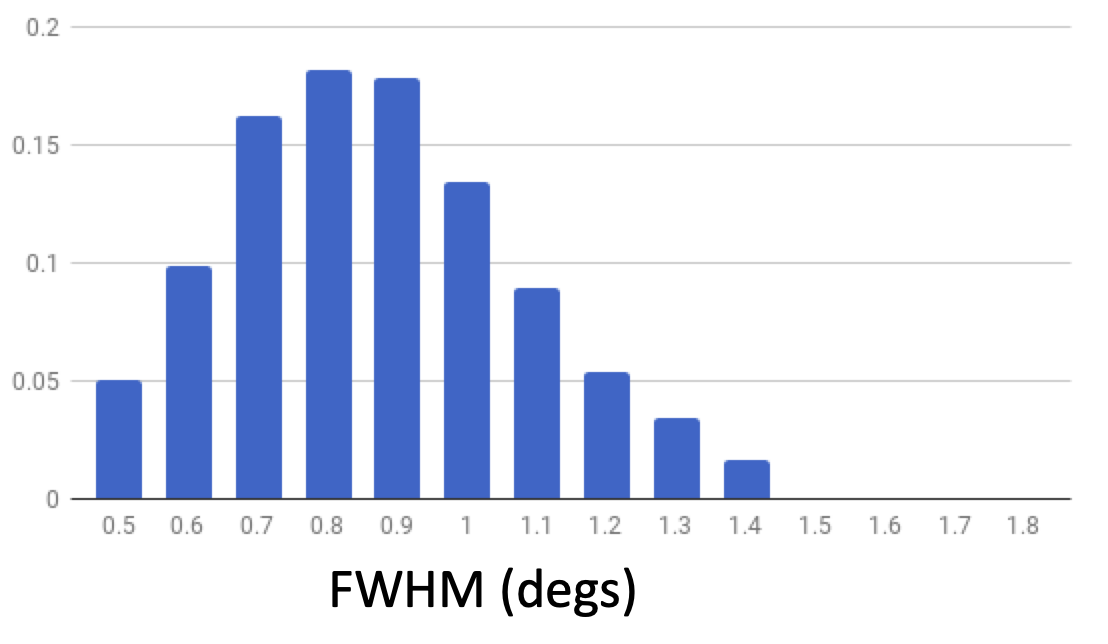}{0.48\textwidth}{(b)}}
\caption{Collimated FRD performance at the DESI input focal ratio of $f$/3.9 for a sample of Positioner Fiber Assemblies. Panel (a) shows the FWHM as a function of input focal ratio for 546 PFAs with a $\mathrm{FWHM} \leq 1.5^\circ$ at $f$/3.9 (note that slower focal ratios are also shown but the requirement is set at$f$/3.9). Panel (b) shows the as built collimated FRD performance of DESI PFAs at $f$/3.9. If the FRD performance exceeded 1.5$^\circ$ the fiber was discarded and rebuilt. }
\label{fig:PFA_perf}
\end{figure}

The as-built FRD performance at an input focal ratio of $f$/3.9 is shown in panel (b) of Figure \ref{fig:PFA_perf}. This dataset includes all PFAs built between October 2016 and November 2017, including those that were classified as `fails’ (FWHM at $f$/3.9 $\geq$ 1.5). The median value is 0.9 with a standard deviation of 0.26.

\begin{figure}
\centering
\includegraphics[width=\textwidth]{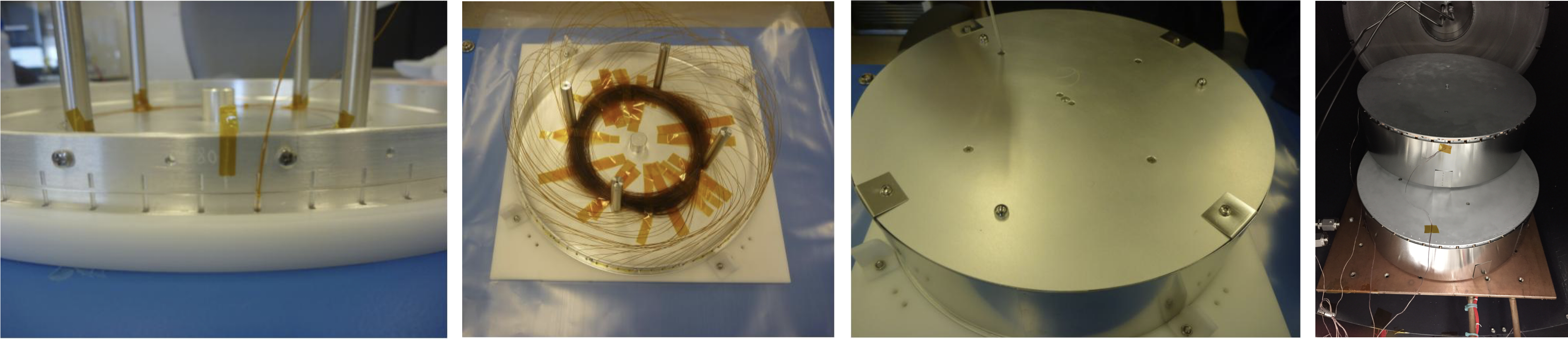}
\caption{In order to reduce risk, the PFAs were delivered to the vendor using custom coating canisters loaded at LBNL. The panel on the left shows the first ferrule being loaded into the outer ring of the coating canister. The second panel shows all 80 fibers loaded into the canister. The third panel shows the canister with the protective cover installed. The final panel shows four canisters loading into a vacuum chamber prior to bakeout which was required to ensure cleanliness.}
\label{fig:AR_loading}
\end{figure}

\begin{figure}[ht]
	\centering
\includegraphics[width=\textwidth]{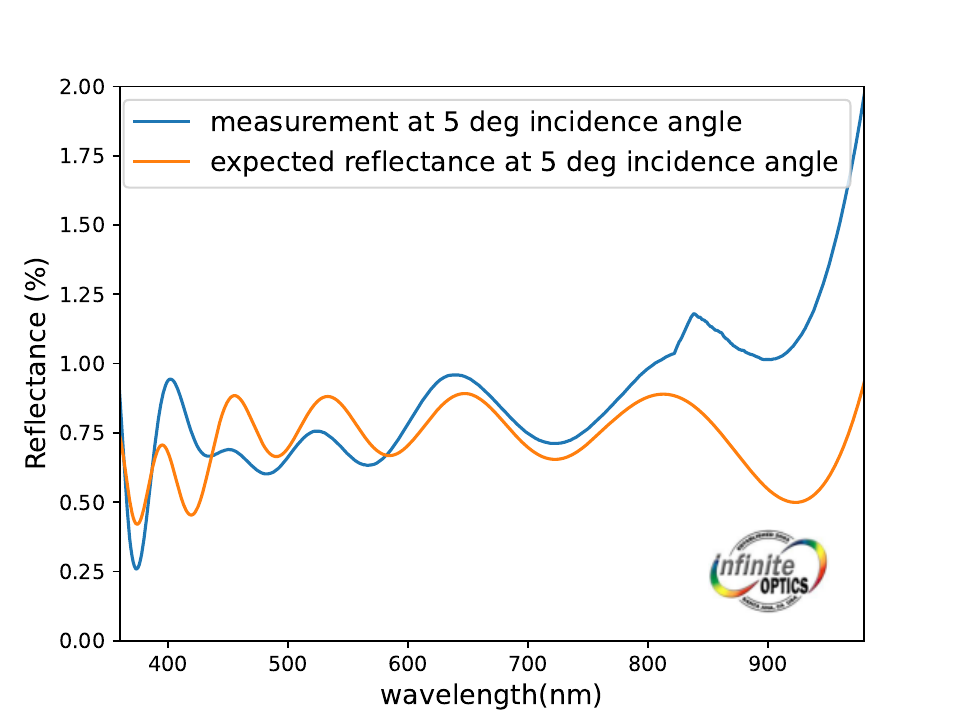}
\caption{
Reflectance achieved by the fiber coating for a typical run. The requirement was for R$\leq1.75$ \% average over  360 to 400nm per surface,  and R$\leq1$ \% average over 400 to 980nm per surface. The angle of incidence is 0 to 8 degrees with an average incident angle = 5.7$^\circ$.}
\label{fig:AR_perf}
\end{figure}

The final stage of PFA production was the application of an Anti-Reflection (AR) coating to the face of the fiber in order to reduce reflective losses at the input of the fiber from $\sim$4\% to $\leq$1.5\%. The coating was
applied by Infinite Optics via a low temperature Ion-assisted-deposition in order to densify the coating film
for environmental stability. In order to reduce risk, the PFAs were delivered to the vendor using custom coating canisters loaded at LBNL. The coating canisters are loaded with 80 fibers at LBNL in order to reduce the risk of damage to the end face of the fibers as is shown in Figure \ref{fig:AR_loading}. The loaded canisters are vacuum baked at 50C in order to ensure cleanliness. The canister design allows the entire length of fiber to be protected
within the unit and a protective cover was used during shipping .The performance achieved in a single coating run is shown in Figure \ref{fig:AR_perf}. 

Prior to positioner installation all PFAs were visually inspected for inadequate bonds, failed alignment, and Focal Ratio Degradation (FRD) tests are performed to verify optical performance. Although extensive R\&D was performed, many challenges were faced when scaling up the production in order to produce 5,000 units. The main challenge was technician fatigue and ergonomic issues. These challenges were overcome and, at completion, 97\% of all PFAs manufactured passed all performance requirements.\\

\subsection{Cables}\label{sec:prodcables}
The 10 fiber cables were manufactured by DU. This activity began in May 2016 and the final cable and slit assembly was delivered to LBNL for splicing in October 2019. The following sections describe the production of the fiber cable and slit assemblies.
\subsubsection{Fiber Bundling}
\label{sec:fiber_bundling}

38.1m of fiber is sheathed inside a strong steel-reinforced Adpataflex external tubing. This rugged tubing is composed of a corrugated steel spiral with a polyvinyl chloride (PVC) outer jacket. The Adaptaflex protects the fiber bundles  yet also must be flexible enough to pass through the 4m-Mayall telescope's two e-chain systems in the Hour Angle and Declination bearings. This section of cable is subject to large motion during telescope slews and acquisitions, as well as constant smaller motion when target tracking during an observation.

Inside the aforementioned Adaptaflex conduit, a central Kevlar strength member acts as an internal core for the bundled fibers, serving as a strain-relief mechanism to support the weight of the fibre bundle. This Parafil Kevlar cable is 11mm in outer diameter. Coated on the exterior with polymer, the Kevlar cable is capable of withstanding 6000\,kg loads.

Adaptaflex cables are further sub-furcated inside the larger conduit into eleven smaller plastic Miniflex tubes. Each plastic tube is 4\,mm outer diameter and 2.6\,mm inner diameter. One of the eleven Miniflex tubes is used to carry three sky fibers, later terminated with individual SMA connectors. The remaining ten Miniflex tubes contain science target fibers to be fusion spliced to positioners, with 50 fibers per tube (plus spares, usually 53 fibers in total) making for 500+ fibers per petal/spectrograph.

Since any bending of the fiber bundles acts to stretch the outer fibers whilst compressing inner fibers (due to the differences in fiber path length); the Miniflex sub-bundles are helically wound about the central Kevlar Parafil strength member cable. This results in a 2\% increase in the fiber lengths required to span the focal plane to spectrograph distance to order to accommodate the additional length in spiral stranding. More information on fiber cable winding is available in Section \ref{sec:fiber_winding} below.

\begin{figure}[ht]
\centering
\includegraphics[width=0.7\textwidth]{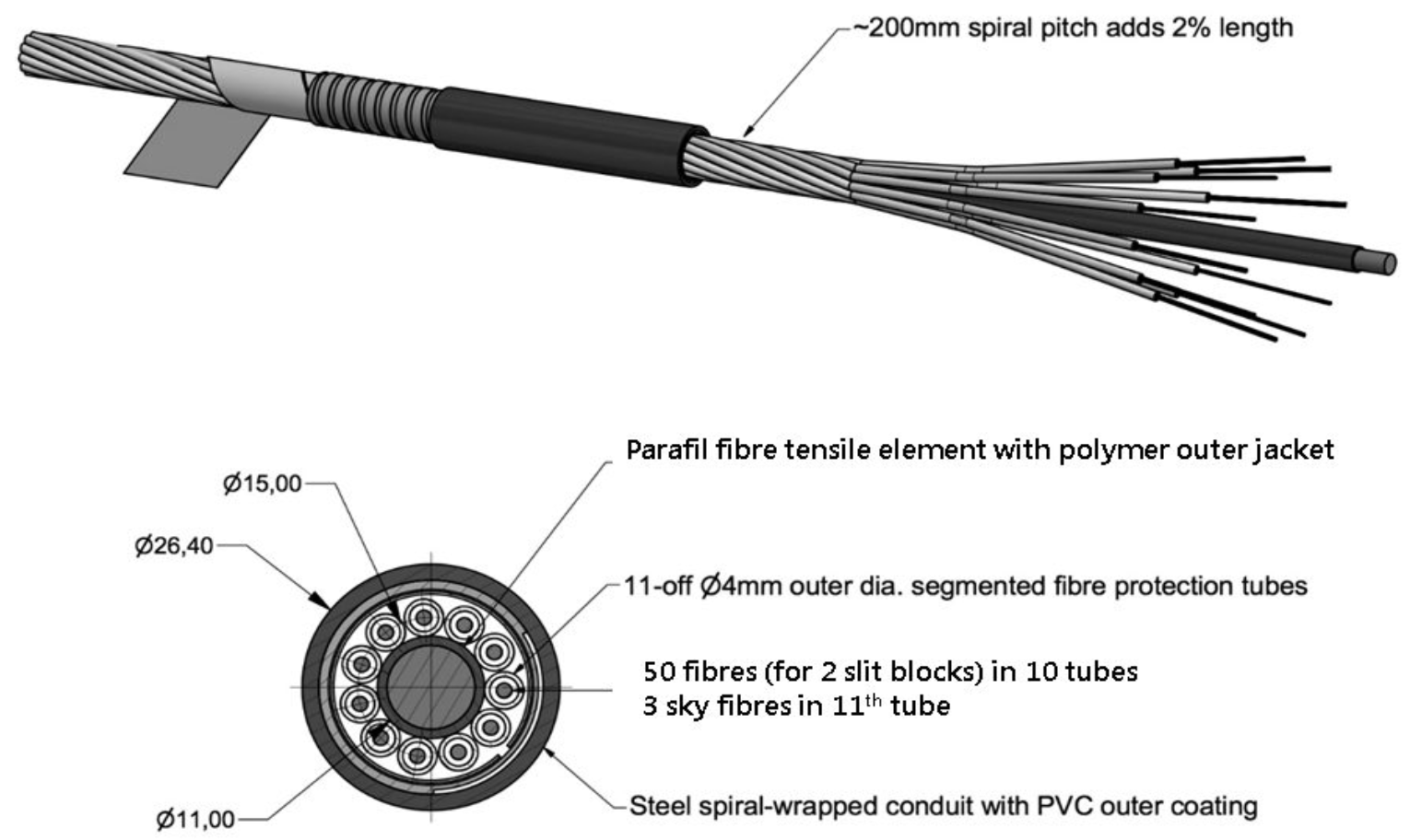}
\caption{Eleven Miniflex conduits carry either sky fibers or fifty (plus spare) PFA science fibers each. These Miniflex conduits are helically wound about a central Parafil Kevlar strength member with polymer coating for strain relief. The bundled spiral is taped in an opposing direction to that of the Miniflex wind with hygroscopic tape, before insertion inside a larger PVC-coated steel Adaptaflex tube.}
\centering
\label{fig:fiber_bundle_diagram}
\end{figure}

A hygroscopic tape is applied to the outside of the helix of Miniflex conduits in an opposing direction to that of the Miniflex wind. This protects the conduits from the steel inner of the Adaptaflex conduit, prevents moisture ingress and allows for a lower friction bundle to be more easily pulled through the Adaptaflex after helical stranding. Figure \ref{fig:fiber_bundle_diagram} shows a diagram of the fiber bundling, both a breakdown of the internal structure (top) and cross-section of the bundle (bottom).

Initial experiments to produce the fiber cables `by hand' were met with a variety of challenges. Foremost is of course an issue of scale, given that the DESI cables require in excess of 200\,km of fiber in total. Further, at roughly 50\,m in straight length, few buildings could accommodate manufacture, and certainly not in a controlled clean-room environment. Finally, a commitment to reducing fiber stress by all means possible proved difficult: although the fiber bundle design described above was optimized to alleviate stress; the mechanism of creating the bundle itself without inducing initial stress in the system was extremely difficult. DU collaborated with PPC Miniflex Ltd (the manufacturer of Miniflex conduit) to industrialize the fiber cable process.

\begin{figure}
\centering
\includegraphics[width=\textwidth]{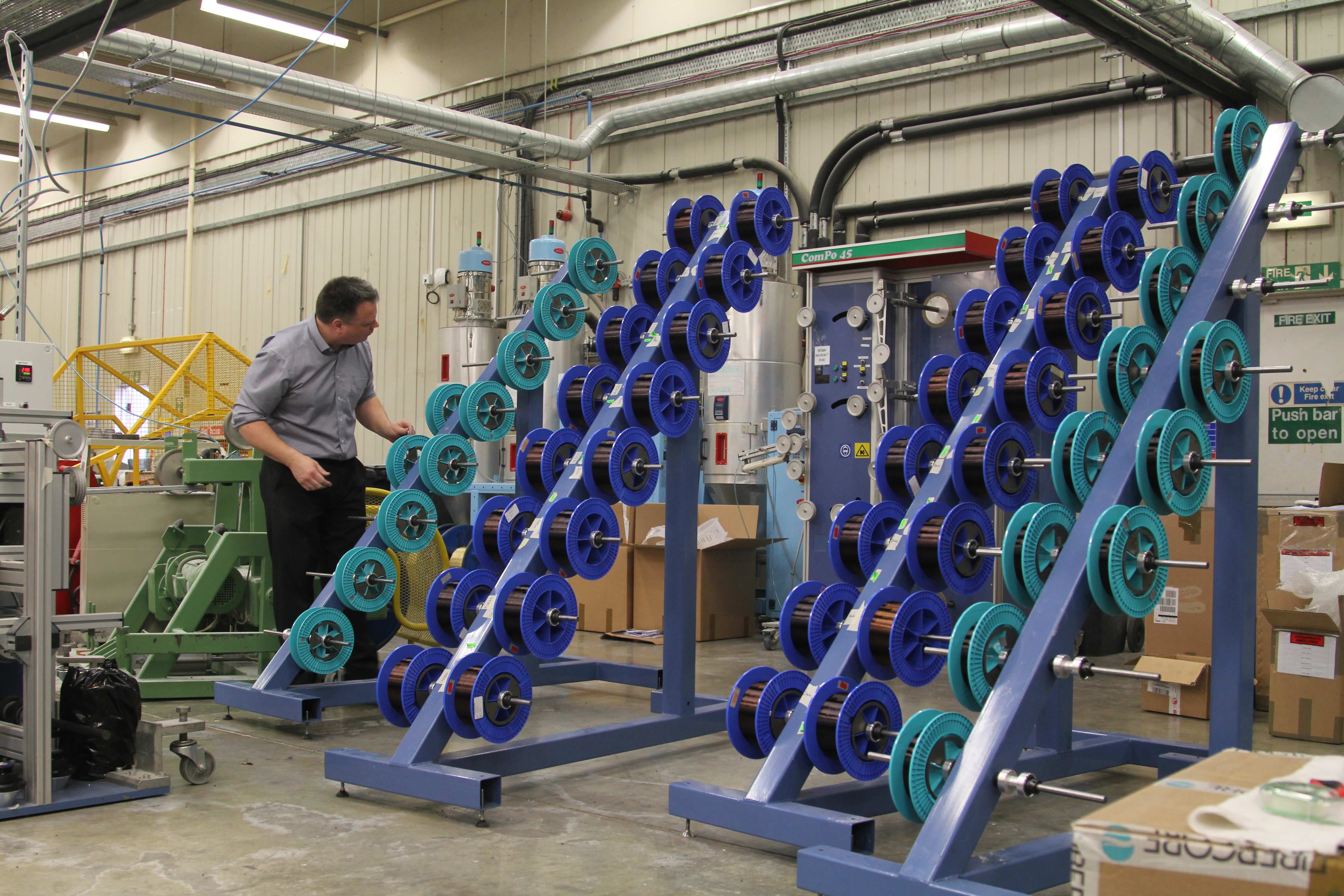}
\caption{Manufacturing bundles of science fibers inside Miniflex conduit in a collaboration between Durham University and PCC Broadband Fiber Ltd.}
\label{fig:fiber_m2fx_photo}
\end{figure}

First the fiber was fed deterministically in to a Miniflex aperture using a mandrel with the addition of Boron Nitride powder as a lubricant. The conduit was then extruded around the fiber bundle. Figure~\ref{fig:fiber_m2fx_photo} shows large staggered racks fabricated to holster the 53 fiber spools (left), each fitted with a clutch mechanism on its axle. Rather than using a mandrel composed of small diameter Polytetrafluoroethylene (PTFE) tubing, a metal form fitted with custom-made ceramic eyelets gathered the individual fibers with minimal friction and static (right). The bunched fibers then passed through a vibrating hopper of finely powdered Boron Nitride. Miniflex conduit is extruded around the bundled fibers allowing for continuous production of fiber. Individual lengths of Miniflex jacketed fiber bundle can then be cut down to length as required from a single $\sim$1.2\,km spool.

\subsubsection{Fiber winding}
\label{sec:fiber_winding}

\begin{figure}
\centering
\includegraphics[width=\textwidth]{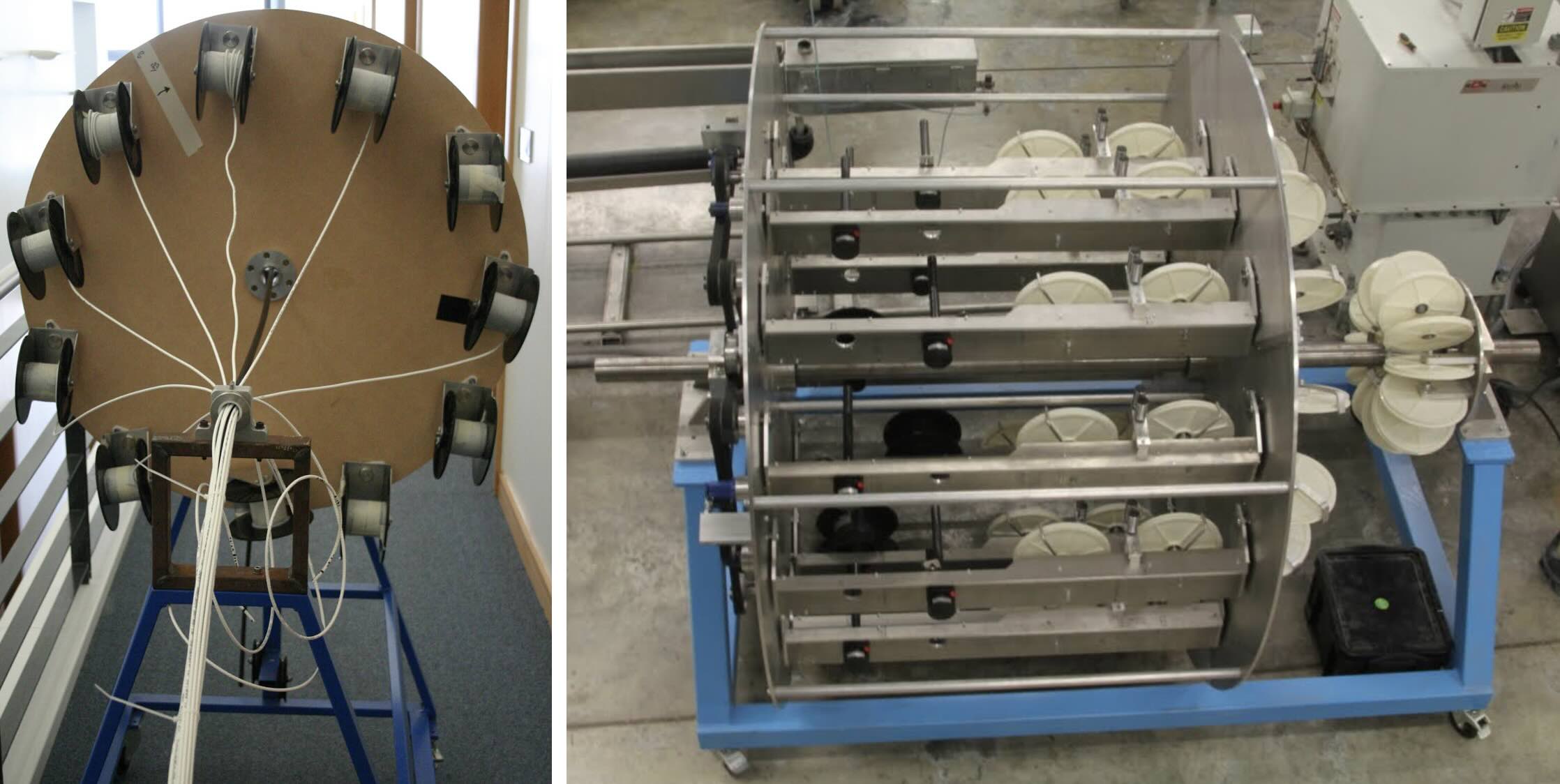}
\caption{A prototype fiber bundle winding cart, which is propelled to drive the rotating wooden disk upon which
fiber-populated Miniflex bundles are spooled. The axis of rotation is about the central Kevlar strength member.}
\centering
\label{fig:fiber_winding_cart_photo}
\end{figure}

The fiber winding method is based on a process previously optimized for Fiber Multi-Object Spectrograph (FMOS) and Subaru Prime Focus Spectrograph (PFS) \citep{murray17}. Prototype and sparse slit test bundles were wound using an in-house rig whilst science fibers installed in DESI used a larger mechanized system at PCC Broadband. Figure \ref{fig:fiber_winding_cart_photo} (left) shows the in-house winding cart. Eleven Miniflex fiber bundles are spooled individually and mounted at the circumference of the rotating disk. The large wooden disk rotates about the central Parafil Kevlar strength member. A steel wire is strung under tension and runs through a pulley system under the cart and up to the axis of rotation of the wooden disk. Pushing the cart forward along said steel wire therefore causes the wooden disk to rotate at a speed proportional to the speed of the cart, ensuring a constant wind pitch. The gearing between the steel wire pulley and the rotation of the wooden disk (and its diameter) sets the pitch. Each Miniflex spool has its own bearing ensuring that as the wooden disk rotates, the spool remains oriented perpendicular to the floor. This ensures no torsion is applied to the Miniflex bundle.

Figure \ref{fig:fiber_winding_cart_photo} (right) shows an industrialized version of the winding machine, built by PPC Broadband in conjunction with the PFS team and made available for DESI use. This machine operates in the same manner as the proof-of-concept cart, but also includes two tape guns to wrap the wound Miniflex conduits.

Finally, the helically wound and taped bundle of Miniflex tubes and Kevlar strength member was pulled through the outer Adaptaflex conduit. This consists of clamping the conduit in a straight line between two vices on the floor. A wire can be threaded through the conduit and attached to a drawstring that is used to pull the taped bundle back through the Adaptaflex. The Adpataflex terminates in an off-the-shelf threaded fitting, to which the Kevlar member is bonded. This fitting attaches to the spool box (see Figure \ref{fig:fiber_spool_photo}). Miniflex is then trimmed from each of the eleven fiber bundles. The sky fibers terminate in SMA connectors whilst the now bare fiber is looped in the spool boxes. The bare fiber leaving the spool box is inserted in to short runs of black Miniflex tubing for routing through the spectrograph slit assembly. Since each Miniflex bundle of 50 (plus spare) fibers terminates in two V-block arrays of 25 fibers each, two smaller clear PTFE tubes further subdivide the fibers in to two V-blocks for routing through the slit `race track' (see Figure \ref{fig:polishedslit}). These PTFE tubes are simply slid inside the larger Miniflex tubing.

\subsubsection{Fiber spectrograph spool boxes} \label{sec:spool}
Once the fibers were wound into cables, the fiber spectrograph spool boxes needed to be installed. These spool boxes allow for loops of fiber to variously expand or contract in order to take up any excess in the system due to motion or fiber creep. This also allows for quality control and production issues such as cutting off and re-bonding a damaged slit V-block.

\begin{figure}
\centering
\includegraphics[width=.35\textwidth,angle=90]{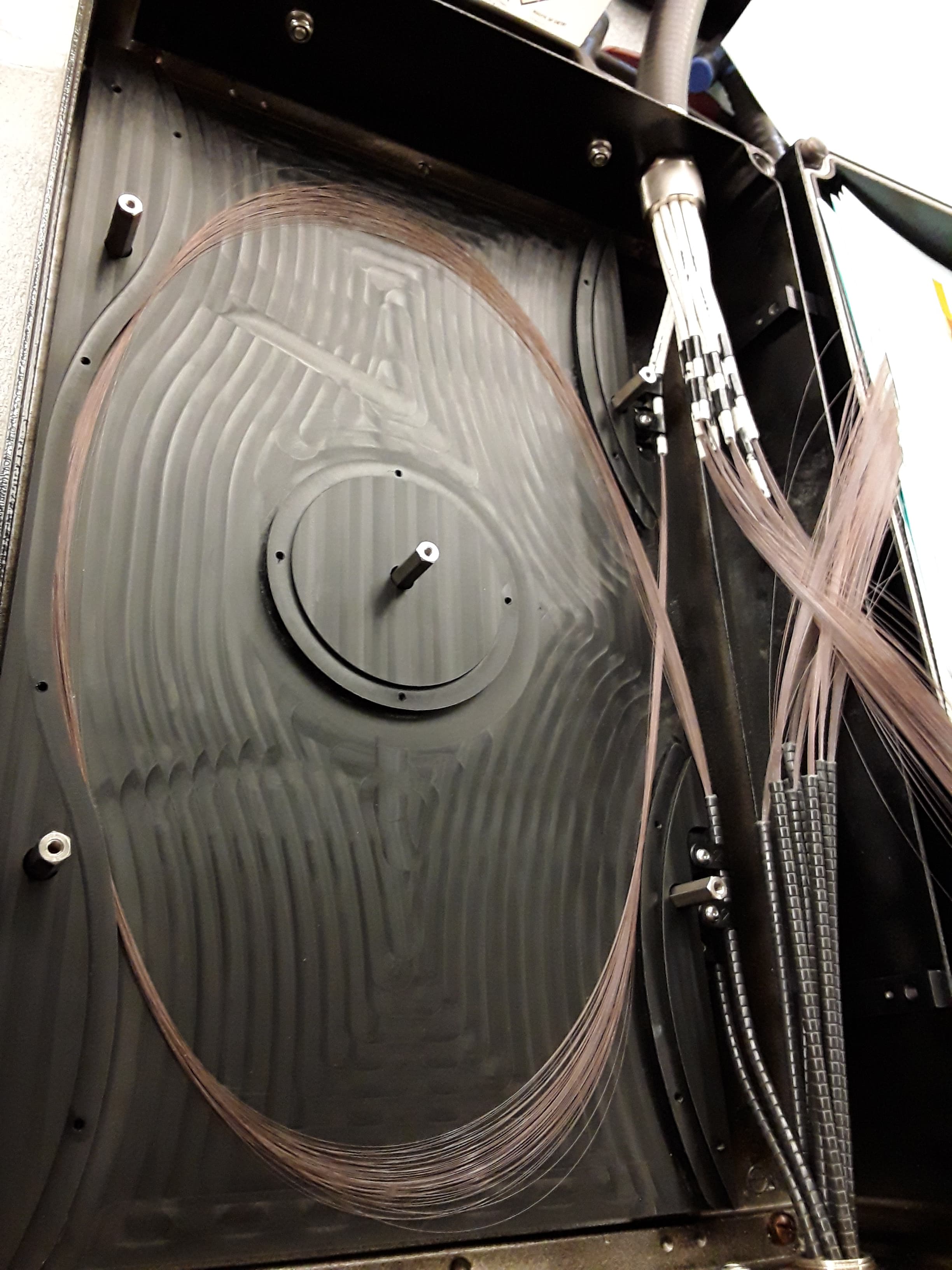}
\includegraphics[width=.35\textwidth,angle=90]{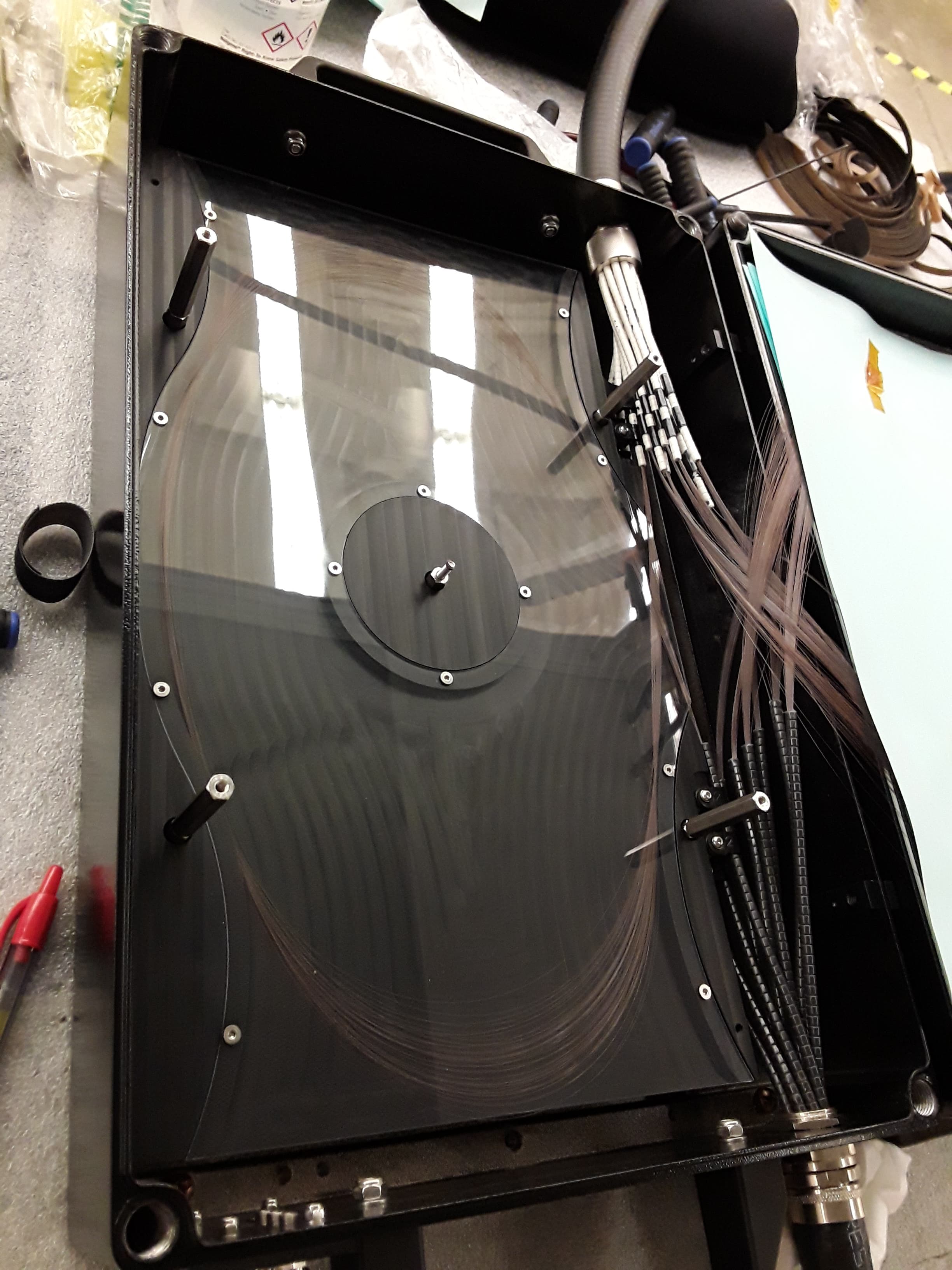}\\
\caption{A spectrograph spool box tray populated with bare looped fiber. A clear Perspex lid is installed which allows the fiber technician to ensure that all fibers are able to move freely inside the cassette before building the next bundle.}
\centering
\label{fig:fiber_spool_photo}
\end{figure}

Figure \ref{fig:fiber_spool_photo} shows the internal trays of the spool box populated with bare fiber. Guides within the trays are used to ensure the minimum bend radius is not exceeded. Each bundle of 50 fibers is separated with a clear Perspex lid which allows the fiber technician to ensure that all fibers are able to move freely inside the cassette. On the left side of the image, the long fiber bundle terminates. On the right side of the image, the bare fiber re-enters larger 5mm diameter Miniflex tubing before termination at the spectrograph slit. Since this Miniflex is not taped or inside Adaptaflex, an opaque black tubing is chosen. One spool box tray carries the sky fibers that terminate separately on the right side of the spool box with individual SMA connectors.

\subsection{Slits}\label{sec:prodslits}
\begin{figure}[ht]
	\centering
\includegraphics[height=.65\textwidth]{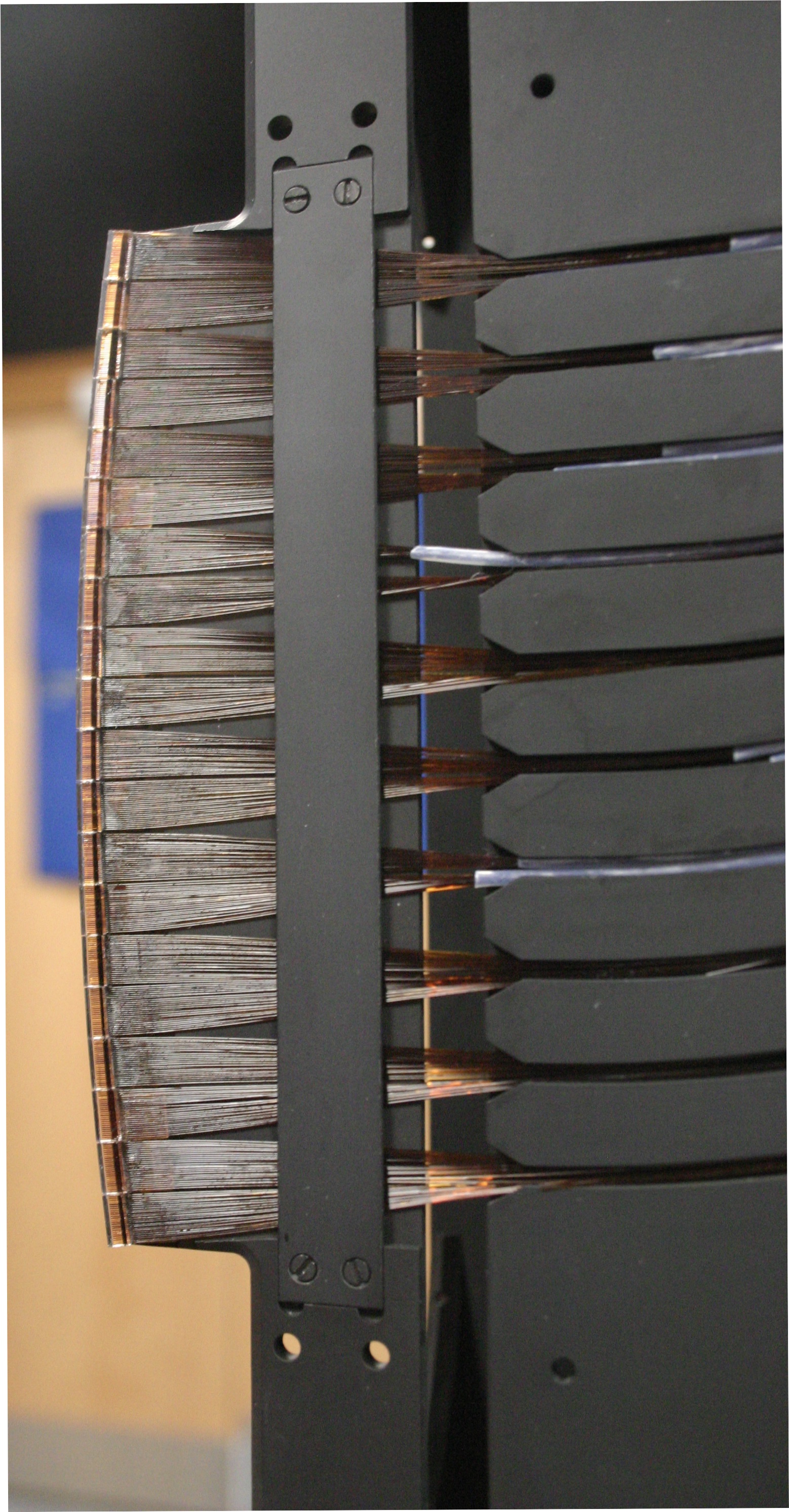}
\includegraphics[height=.65\textwidth]{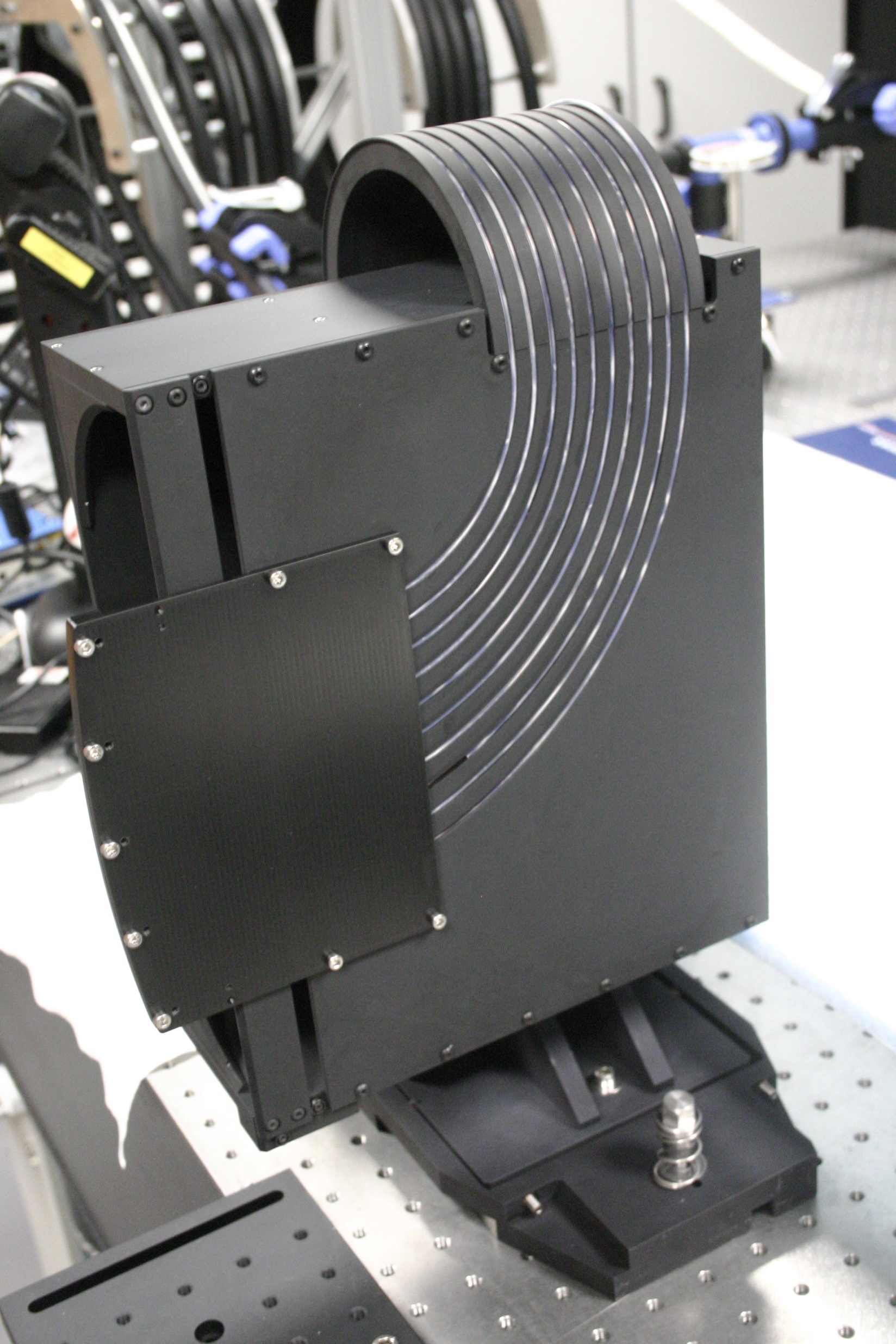}\\
\caption{$(Left)$ A close-up image of a curved slit plate to which 20 V-blocks (fibers inside 20 PTFE tubes and 10 Miniflex tubes) are bonded. Fibers, cover lids, V-blocks and AR windows are visible, with the 20 V-blocks appoximating the curvature of the slit plate below. $(Right)$ An image of the entire slit assembly featuring the `race track' of fiber bundles retained in `T-cut' slots.}
\label{fig:polishedslit}
\end{figure}

The final stage to the production of the cable assembly is the addition of the spectrograph slits. The fibers exit the spectrograph spool box in their Miniflex conduit and each bundle (containing 50 fibers plus spares) is sub-divided in to two smaller PTFE tubes containing 25 fibers each. These PTFE bundles pass through T-shaped slots in a `race track' slit assembly, designed to reduce vignetting. A slit V-block is bonded onto the end of each of these PTFE bundles of 25 fibers, before being bonded in turn to the curved slit plate (see Figure \ref{fig:polishedslit}).

\begin{figure}
\centering
\includegraphics[width=\textwidth]{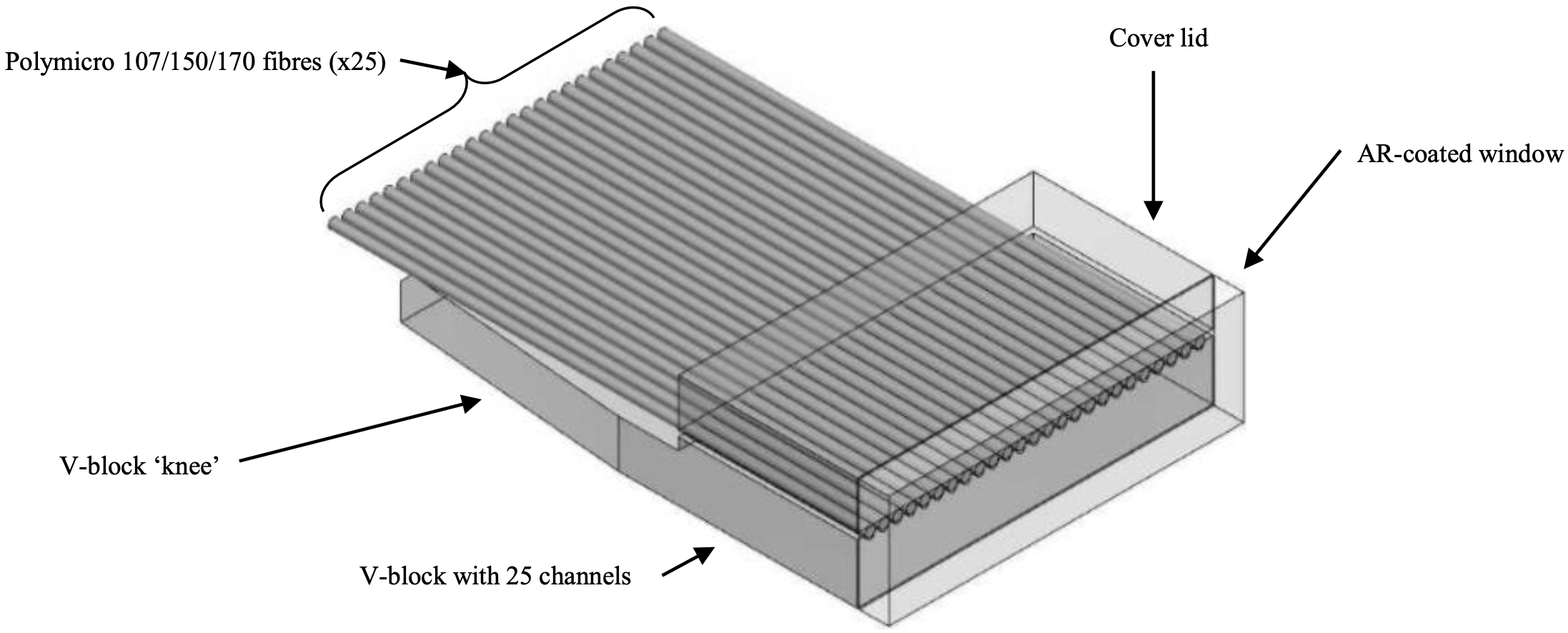}
\caption{A bare fiber is inserted from the rear of the V-block in to its respective V-block channel. The cover lid is held on top of the fibers. Once bonded, the front face of block/fibers/lid is optically polished, before an AR-coated window is attached.}
\centering
\label{fig:vblock diagram}
\end{figure}

The slit block production follows a number of steps. First, the components are cleaned. Next, the fibers must be laid into the 25 V-groove channels in the V-block of the slits. Fibers are inserted by hand in to the channels, above which a fused silica cover lid is overlaid for retention. Epoxy is wicked in to the channels via capillary forces from the front fiber output face, bonding the fibers and cover lid in place. Each block is then individually hand polished to a 0.1$\mu$m optical finish. An anti-reflection coated window is bonded to the polished surface to increase system throughput. Finally, each V-block is bonded to the slit plate. Figure \ref{fig:vblock diagram} shows a diagram of a V-block fitted with fibers, cover lid and AR-coated front window. The knee in the block allows 20 blocks to be fitted to the slit plate side-by-side whilst matching the radius of curvature of the slit.

The following sections give detailed descriptions of the manufacturing process for each step in the slit assembly manufacture and test.

\subsubsection{V-block bonding}
Given that the inter-fiber separation (and hence inter-spectra separation on the CCD) is largely governed by the V-groove pitch; and inter-block spacing is highly constrained to fit the slit plate, a variety of manufacturers were approached to produce V-blocks with PLC Connections\footnote{PLC Connections, 673 North Wilson Road, Columbus, Ohio, 43204, United States of America} the selected vendor. A significant sub-sample of V-blocks were measured visually using a calibrated Keyence VHX-600 digital microscope for block dimensions, V-groove perpendicularity and pitch, and `knee' angle.

The entire block bonding process takes place inside a clean tent. V-blocks and cover lids are individually cleaned on a clean room cloth inside an ultrasonic bath of semiconductor-grade Isopropyl Alcohol (IPA). The components are then dried in a vacuum oven for 15\,minutes at room temperature and 200\,mbar pressure to ensure no remaining solvent can interfere with the epoxy cure. Fibers are cleaned using clean room wipes and semiconductor-grade acetone before insertion under the cover lid.

Table \ref{tab:v_block_table} summarizes the tolerance requirements for individual slit blocks. Variance in AR window glue layer thickness manifest as focus (\textit{z}) error, whilst overhang of the window or cover lid in the \textit{x} direction along the slit prevents suitable spacing of the densely packed slit blocks when bonded to the slit plate. Cover lid underhang can lead to fibers being improperly bonded whilst AR window underhang can lead to cropping of a fiber output. Since the slit blocks fan outwards from the rear on the slit plate, the overhang tolerance for cover lids is stricter than for AR windows. AR window position in \textit{y} (the dispersion direction) is comparatively loose since the window overhangs the slit plate, the main constraint being that fiber outputs are not obstructed. The requirement on pointing accuracy of fibers from the nominal is actually 0.5$^{\circ}$ but since there can be fiber-to-fiber deviations, potential for a wedge in AR window glue thickness and slit plate bonding placement error, $\pm$0.25$^{\circ}$ is adopted for the polishing tolerance.

\begin{table}
\centering
\begin{tabular}{|c|c|} 
 \hline
 \textbf{Requirement} & \textbf{Tolerance} \\
 \hline
 Cover Lid Overhang (wrt. V-Block) in \textit{x} & +$25\mu$m each side \\ 
 AR Window Short Edge (wrt. V-Block) in \textit{x} & $\pm50\mu$m each side \\
 AR Window Long Edge (wrt. V-Block) in \textit{y} & $\pm 300\mu$m each side \\
 AR Window Glue Thickness Variation (between slit blocks) & $\pm10\mu$m \\
 Polish Quality & No observable scratches over fiber faces \\
 Polish Perpendicularity\footnote{polished face with respect to optical
axis of leftmost, rightmost and
central fibers, and the mean of the three} & 90$^{\circ}$ $\pm$ 0.25$^{\circ}$ \\
 \hline
\end{tabular}
\caption{Requirements for Slit Block Manufacture}
\label{tab:v_block_table}
\end{table}

\begin{figure}
\centering
\includegraphics[width=\textwidth]{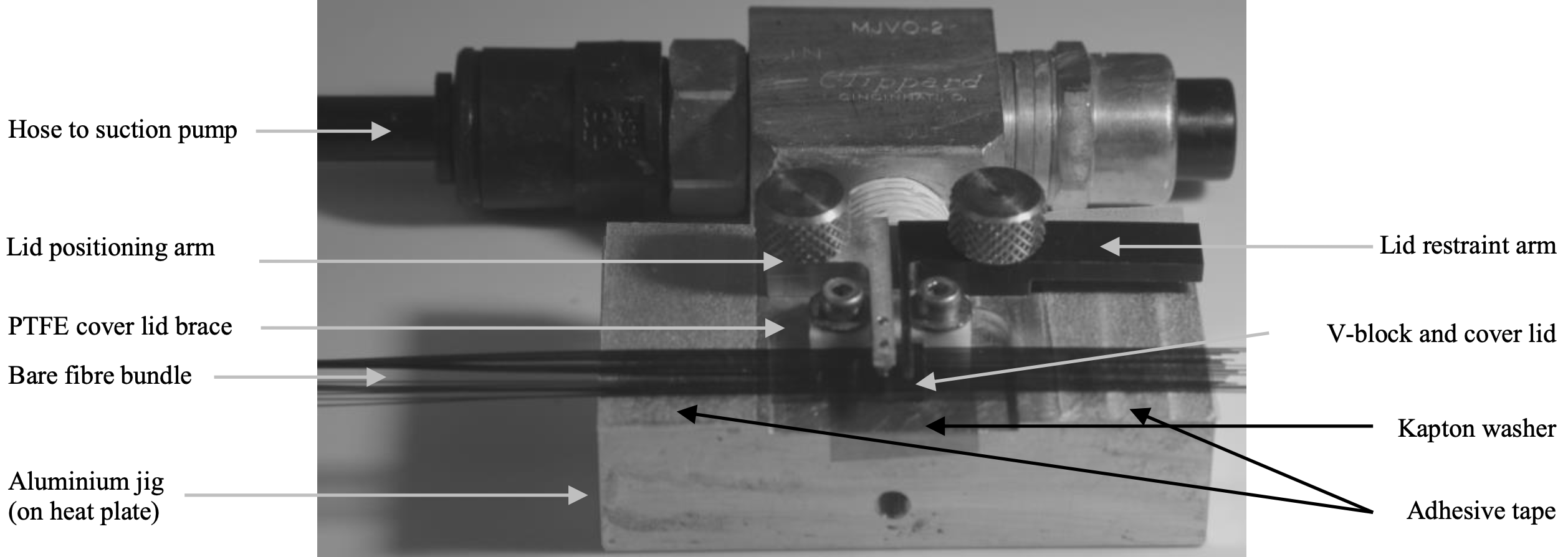}
\caption{V-blocks are held in place by vacuum. The cover lid is positioned and retained with two removable arms. Fibers are insterted in to the V-grooves beneath the cover lid. Double-sided adhesive tape restrains the fibers. The vacuum can be terminated and PTFE brace and cover lid arms removed once the block is bonded and cured.}
\centering
\label{fig:revised_jig_photo}
\end{figure}

The jig used to manufacture populated fiber slit blocks is shown in Figure \ref{fig:revised_jig_photo}. The V-block is retained accurately and securely using a vacuum. The jig is attached to an open/close valve and a Venturi effect pump to create suction, in turn attached to the building compressor line. A thin Kapton washer is placed around the jig vacuum hole between jig and V-block, in case any adhesive is drawn towards the vacuum hole. The cover lid is retained using a removable Nylon arm from above, whilst a metal removable arm to the rear features two pins to set the lid position in \textit{z}. The left and rightmost fibers are populated before the cover lid is installed and held in place with the addition of the Nylon retaining arm. Remaining fibers are then inserted in to the V-grooves beneath the cover lid.

The initial plan was to populate a V-block with fibers before moving the block aside to cure at room temperature (a two day process), simultaneously beginning assembly of another block. Unfortunately this proved impossible since fibers would easily dislodge with the slightest motion. Since the V-block assembly jig is aluminium, and conductive to heat, blocks could instead be cured at an accelerated rate. V-block assembly is performed directly on top of a heat plate beneath the visual microscope and the block is bonded and cured without any further motion. 3D-printed fixtures hold a test tube of acetone for fiber cleaning, as well as provide strain relief to both the PTFE tube bundle of fibers and the stiff vacuum hose attached to the assembly jig. Prior to adopting this manufacturing process, V-blocks cured at accelerated rate were tested for FRD performance and temperature/pressure survivability. No difference in performance or longevity between room temperature or accelerated cure blocks was found.

Fibers are bonded using Angstrom Bond AB9190A/B epoxy, which is directly mixed inside a dispensing syringe in 0.5 g/0.1 g ratio. The two part adhesive is stirred for 4 minutes before de-gassing at 20 mbar pressure for 15\,minutes in a room temperature vacuum oven. (Lower degassing pressures resulted in a loss of volatile components apparently crucial to the curing process.) Epoxy is applied using a Nordson Performus VI glue gun along the overhanging edge of the cover lid. The gun is fed with 80\,psi dry nitrogen from a cylinder and outputs drops of a controlled size with 5\,psi output pressure and 0.150\,s dispense time. Testing using ultra-precise scales confirmed high repeatability in the glue gun delivery. Sufficient epoxy is added to flush any bubbles in the channels out to the rear of the cover lid. Bubble nucleation is also mitigated by the component cleaning processes outlined above. Once the epoxy is applied, the block is cured on a heat plate for 2.5\,hours at 65$^{\circ}$\,C. A variety of cure time/temperature combinations are possible, however this selection provided a minimum of downtime (since slit blocks are assembled in series); exceeds the glass transition temperature of the epoxy; and assures the cure is still sufficiently slow as to avoid internal stresses which may act to diminish FRD performance.

\subsubsection{V-block polishing}
Each V-block is polished by hand, individually before a digital microscope inspection of the surface polish quality. Early concepts had an entire cable (500 fibers in 20 V-blocks) polished by a commercial polishing machine, simultaneously. However, it was determined that debris from the process (fiber, cover lid, epoxy) contaminated the polish and so a hand polishing approach was adopted. 

The fiber slit blocks were loaded into a polishing puck and polished on a glass plate. The tactile feedback and ability to vary the pressure applied helped in the process. Table \ref{tab:polishing_paper_table} shows the selection of ThorLabs diamond lapping films used for hand polishing V-blocks alongside their suggested polishing duration. New wet and dry paper is used for each individual slit block whilst diamond lapping paper is refreshed on a per science cable basis (every 20 blocks). Polishing was performed in a clean room environment and the glass polishing plate and films were rinsed both sides with distilled water prior to use, as was the polishing puck and dummy blocks. Rinses were also performed in the transition between each stage of polish. A `figure of eight' pattern is used in the hand polish to ensure the direction is constantly changing in an analogue to the machine's planetary motion. Visual checks with a high-magnification eyeglass are performed at each stage of the polish process. Although hand polishing ensured the highest quality final product, adopting this process significantly slowed the rate of production.

\begin{table}
\centering
\begin{tabular}{|c|c|} 
 \hline
 \textbf{Polishing Paper/Lapping Film} & \textbf{Duration of Polish (seconds)
} \\
 \hline
 3M P400 grit Wet and Dry Silicon Carbide Paper & 120 \\ 
 ThorLabs LF30D $\pm30\mu$m Diamond Lapping Film & 120 \\
 ThorLabs LF6D $\pm6\mu$m Diamond Lapping Film & 120 \\
 ThorLabs LF3D $\pm3\mu$m Diamond Lapping Film & 240 \\
 ThorLabs LF1D $\pm1\mu$m Diamond Lapping Film & 240 \\
 ThorLabs LFCF $\pm0.2\mu$m Diamond Lapping Film & 240 \\
 \hline
\end{tabular}
\caption{V-Block polishing papers and duration of polish}
\label{tab:polishing_paper_table}
\end{table}

\begin{figure}
\centering
\includegraphics[width=0.65\textwidth]{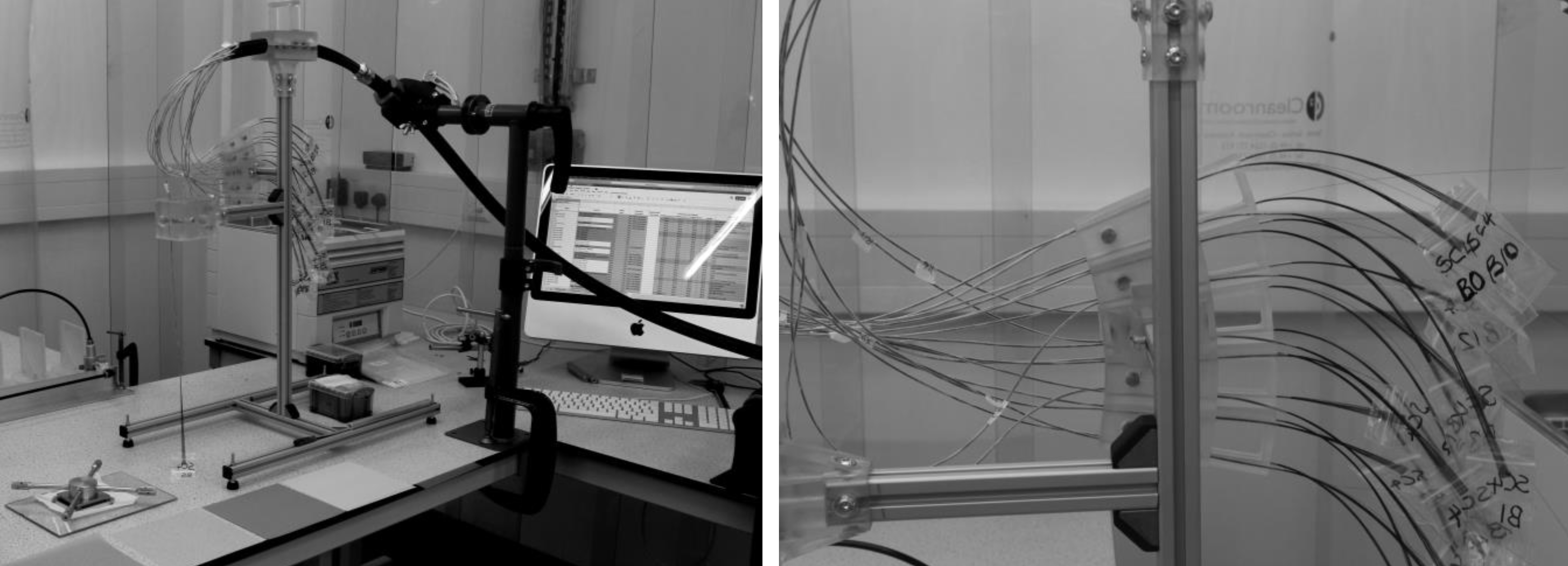}
\caption{$Left$ To the rear of the bench, black tubes connect to the Venturi pumps. The `X-shaped' polishing jig rests upon the glass plate used for polishing single blocks by hand. The Adaptaflex cable is held in black clamps held to the worktop with G-clamps. 3D-printed parts retain and separate the 19 PTFE bundles not in use, and the single bundle to be polished hangs vertically. An array of different colored polishing films sit in front of the plate. A spreadsheet keeps track of production and QA. To the rear, the ultrasonic bath is used for cleaning parts. $Right$ A closer view of the 20 3D-printed guides used to keep PTFE bundles tidy, with clamps held in place with magnets.}
\centering
\label{fig:polishing_restraint}
\end{figure}

Since polishing now proceeded in a serial fashion with one V-block per polish, custom-made parts and a rig were designed to restrain and separate the remaining 19 blocks not being worked upon. Parts which retain bare fiber, PTFE bundles and the Adaptaflex cable are custom 3D-printed from a soft ultraviolet (UV) stereolithography-cured resin. All corners are rounded to prevent snags and the parts clamp together using magnets to avoid latching mechanisms and hinges (see Figure \ref{fig:polishing_restraint}.

\begin{figure}[ht]
	\centering
\gridline{\fig{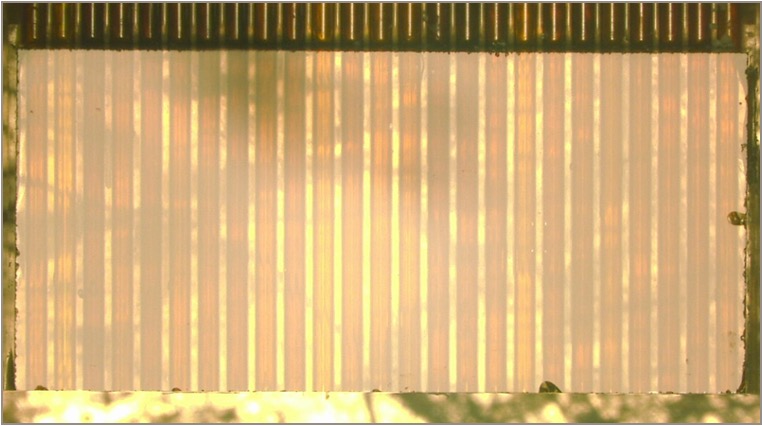}{0.48\textwidth}{(a)}
\fig{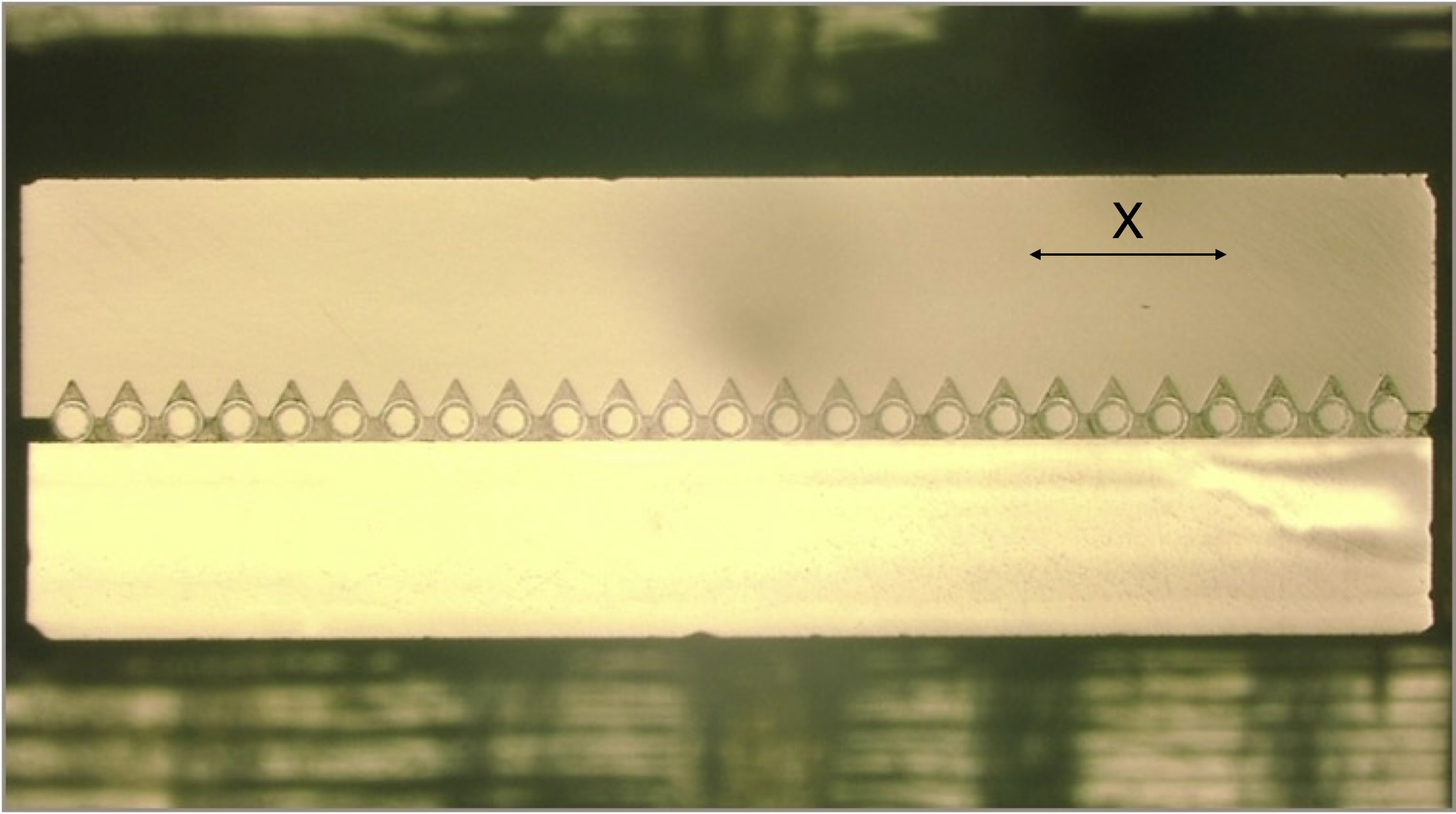}{0.48\textwidth}{(b)}}
\caption{Panel (a) shows a `top down' polished slit block with perpendicular polish (with respect to fibers) and no bubbles in the epoxy channels. Fiber output face is at the bottom of the image. Panel (b) shows a `fiber face on' polished slit block with no bubbles or edge chips, optically flat, scratch-free polish, well-positioned cover lid in \textit{x} and evenly spaced fibres.}
\label{fig:polishing_inspection_photo}
\end{figure}

Following polishing, the V-blocks must be quality control inspected and documented for both polish quality and the dimensional tolerances described previously in Table \ref{tab:v_block_table}. This process involves orienting the V-blocks for a `top down' view (looking through the cover lid at the fibers); and a `fiber face on' view with the polished surface upwards, beneath a calibrated and focused Keyence VHX-600 digital microscope (see Figure~\ref{fig:polishing_inspection_photo}). The former view allows for inspection of the perpendicularity of the polish face with respect to the fiber optical axis (the two extreme fibers, central fiber and the average of these three) and identification of any epoxy bubbles in the V-block grooves. The latter view allows for metrology of the cover lid overhang/underhang and inspection of the quality of polish. V-blocks outside of the 89.75$^\circ{}$ to 90.25$^\circ{}$ perpendicularity requirement are remade, as are blocks with a cover lid overhang/underhang with respect to the V-block greater than +$25\mu$m or $-50\mu$m respectively. Polish quality is a subjective technician judgement: whilst small scratches away from the fiber faces are allowed, scratches on fiber faces are re-polished. Inter-fiber separation is calculated using a MatLab script with automatic circle detection and centroid fitting. Cover lid position can be measured directly using the microscope software.

\subsubsection{V-block anti-reflection windows} \label{sec:arwin}
The final stage of V-block production is to attach an AR coated window to the polished front face of the fiber-populated V-block in order to reduce losses at the air-glass surface. Initially it was planned to use UV-curing Norland NOA-88 glue for this purpose. This glue cures in 30 seconds with an LED UV lightsource for rapid production speeds. NOA-88 also benefits from being low-outgassing, extremely high transmission and with an index of refraction matched to the AR window glass. Unfortunately during pressure and temperature thermal cycling tests several V-blocks with NOA-88 fitted windows began to delaminate. The exact reason for this is unknown but is assumed to relate to a loss of volatiles during the degassing process or an interaction between NOA-88 and the Angstrom Bond epoxy used in fiber V-block bonding, but this is speculation.

A replacement was found in the form of EpoTek 301-2 two part epoxy, which enjoys all of the same transmission, low-outgassing and refractive index-matching characteristics as NOA-88. There are however three issues that make EpoTek less appealing in production use. Firstly, as a two part epoxy, an additional step (and possible source of human error in production) is introduced to weigh and mix the two part adhesive. Secondly, EpoTek 310-2 is not a rapidly UV-curing bond, but rather a slowly curing 48 hours room temperature bond, leading to a vast increase in production time. Finally, Epo-Tek has an extremely low viscosity and surface tension: given the tiny size of the AR windows, this made windows prone to slippage from their initial placement position.

\begin{figure}
\centering
\includegraphics[width=0.65\textwidth]{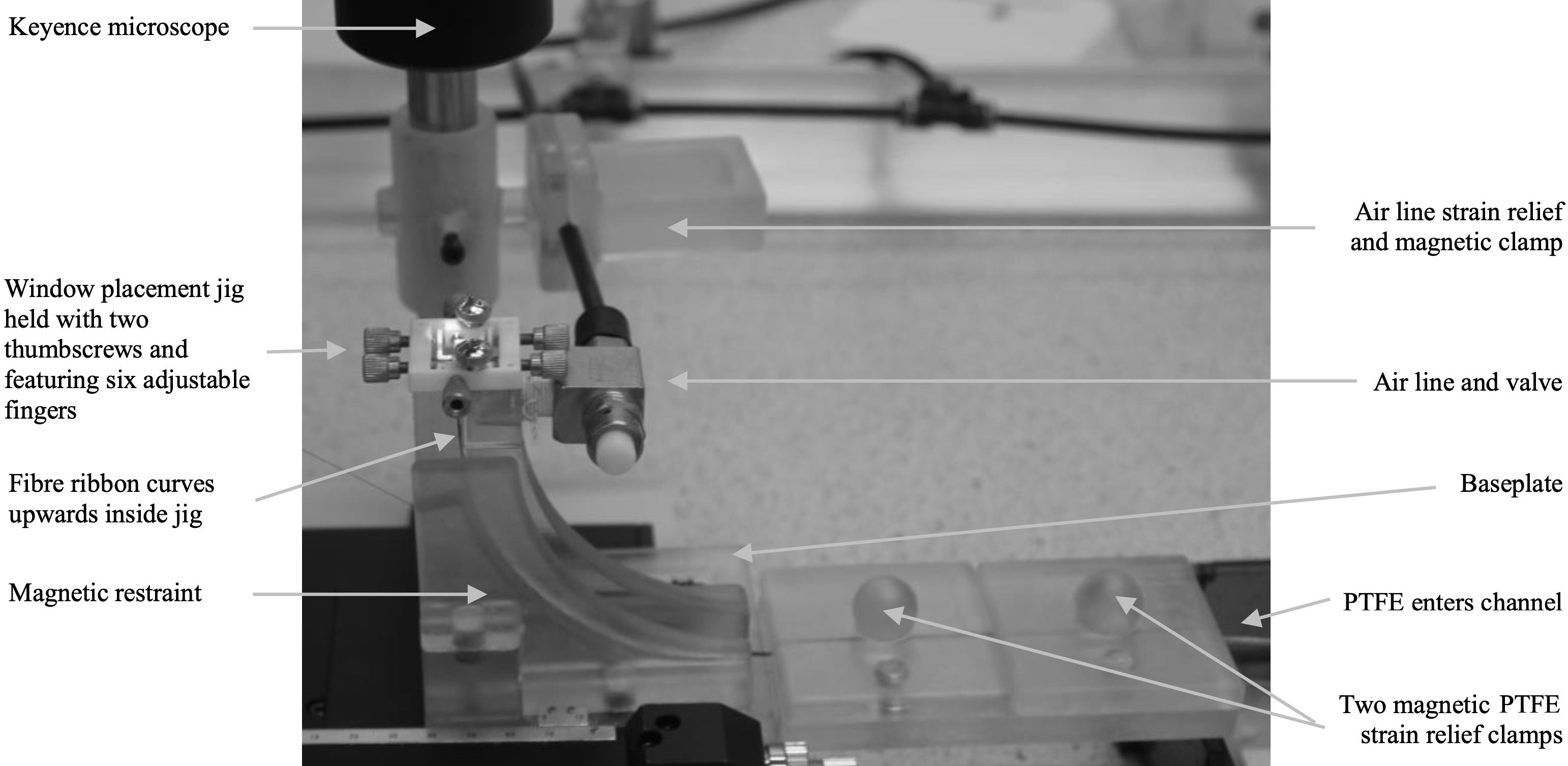}
\caption{A polished V-block is held upright beneath the digital microscope using a 3D-printed vacuum jig. Strain relief for the PTFE fiber bundles and Venturi pump air line is provided. Six fingers adjusted with fine-thread screws can place (and, crucially, retain during the epoxy cure time) the AR coated window.}
\centering
\label{fig:ar_window_jig_photo}
\end{figure}

Figure \ref{fig:ar_window_jig_photo} shows the jig used for bonding AR windows. The PTFE bundled fibers enter along image right, where they are restrained by two magnetically attaching clamps. The bare fiber then arcs upwards through 90$^\circ{}$ as a ribbon inside the arc-shaped part of the jig. A second arc-shaped restraint, also clipped in place with magnets, retains the fiber ribbon. At the top of the jig, the V-block is held in place, again using suction provided from a Venturi pump attached to the laboratory main compressor line. The aforementioned parts are 3D-printed from a soft resin on a UV SLA 3D-printer. A further magnet-clamped strain relief part is attached to the video microscope stand and set at the desired height using grub screws, to prevent the stiff vacuum tubing from twisting the AR window jig. 

\begin{figure}
\centering
\includegraphics[width=0.65\textwidth]{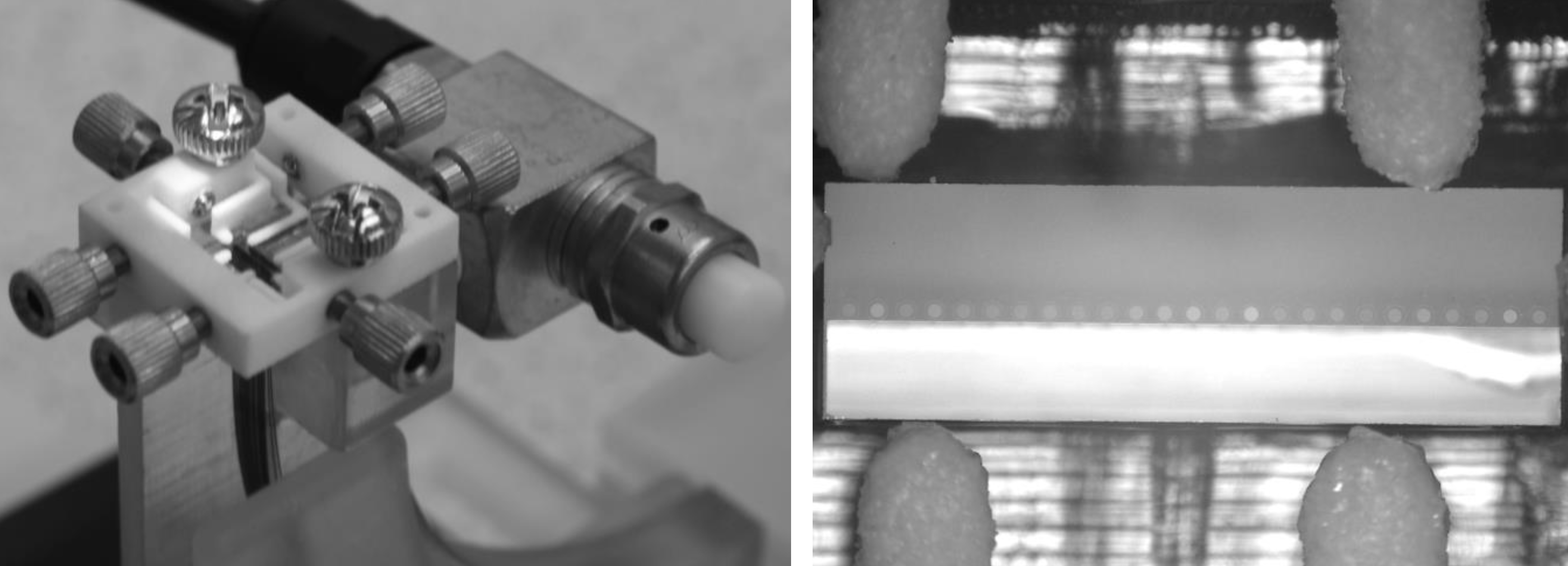}
\caption{$(Left)$ A sintered Nylon part can be placed around the AR window and held in place with screws. Fine threaded screws adjust the six fingers on this jig. $(Right)$ Adjusting the fingers positions (and retains during the slow curing process) the window in place in $x$, $y$ and tilt.}
\centering
\label{fig:ar_window_jig_top}
\end{figure}

On the top of the jig, which sits beneath the Keyence digital microscope, a white sintered 3D-printed Nylon part is attached (See Figure \ref{fig:ar_window_jig_top} left), fitted with six `fingers' akin to a leaf spring. Two fingers are located on each of the two long edges of the V-block, with a single finger at each of the shorter block edges. These fingers are fitted with extremely fine-threaded screws that can bend them inwards as necessary to locate the AR window in to the desired position in $x$, $y$ and tilt (see Figure \ref{fig:ar_window_jig_top} right). The sintered tool is removable using the two button-head top mounted screws. This allows the V-block to be placed in the jig, Epotek 301-2 droplets and the AR window to be placed on top, and the finger adjustment part to be attached on top afterwards. The Nordson Performus VI glue gun is again used using the same settings utilized in V-block bonding, with twenty drops of EpoTek used for V-block bonding. Consistency in glue layer thickness between V-blocks is important since, when bonding blocks to the slit plate, mechanical datums are used against the front of the AR window: a thicker glue layer would therefore put fibers set back in focus ($z$) relative to other blocks. The AR window is visually inspected by a technician before bonding, to determine which side of the window is AR coated. This consists of holding the window with soft-tipped tweezers beneath a white LED ring light and magnifying lens, with the incident light at a grazing angle. Reflections become visible on the non-coated side, but not on the AR coated side. The non-coated side is bonded to the polished V-block.

\subsection{Slit plate bonding} \label{sec:slit_bonding}
As mentioned in Section \ref{sec:slit_RD}, 20 V-blocks are used to approximate the curvature of the slit. Each of the 20 V-blocks is fitted to a black anodized aluminium slit backing plate for mounting in the slit assembly (see Figure \ref{fig:polishedslit}). The requirements for the slit plate are seen in Table \ref{tab:slit_requirements}. 

\begin{table}
\centering
\begin{tabular}{|c|c|c|} 
 \hline
 \textbf{Requirement} & \textbf{Value
} & \textbf{Comments}\\
 \hline
 Slit radius of curvature & 486 $\pm 3$ mm convex & \\ 
 \hline
 Fiber aperture separation & 230 $\pm 10 \mu$m between fibers & \\
  & 556 $\pm 10 \mu$m between blocks & \\
 \hline
 Pointing accuracy & 0.1$^{\circ}$ (dispersion) & Cross dispersion tolerance considers \\
  & 0.5$^{\circ}$ (cross-dispersion) & slit approximation by parallel fibers \\
  \hline
 Lateral tolerance along slit inside each block & $\pm 10 \mu$m & Affects extraction of adjacent spectra \\
 Lateral tolerance between adjacent blocks & $\pm 23 \mu$m & \\
 \hline
 Lateral tolerance in dispersion direction & $\pm 50 \mu$m & Compensated by wavelength calibration \\
 \hline
 Diversion from curvature along slit & $\pm 20 \mu$m & Includes systematic deviation \\
  & & caused by flat V-blocks \\
 \hline
\end{tabular}
\caption{Science slit requirements}
\label{tab:slit_requirements}
\end{table}

Each of the 20 V-blocks is bonded individually, and sequentially, to the slit backing plate. Prior to this bonding process, each plate is measured along the optically active area using a TalySurf aspheric contact profilimeter to check for the radius of curvature; as well as any bowing in the thin part along its full length (which could be caused by manufacturing processes, handling, mounting, shipping or the anodization process). Only plates within $\pm50\mu$m of perfectly flat will be bonded.

\begin{figure}
\centering
\includegraphics[width=0.65\textwidth]{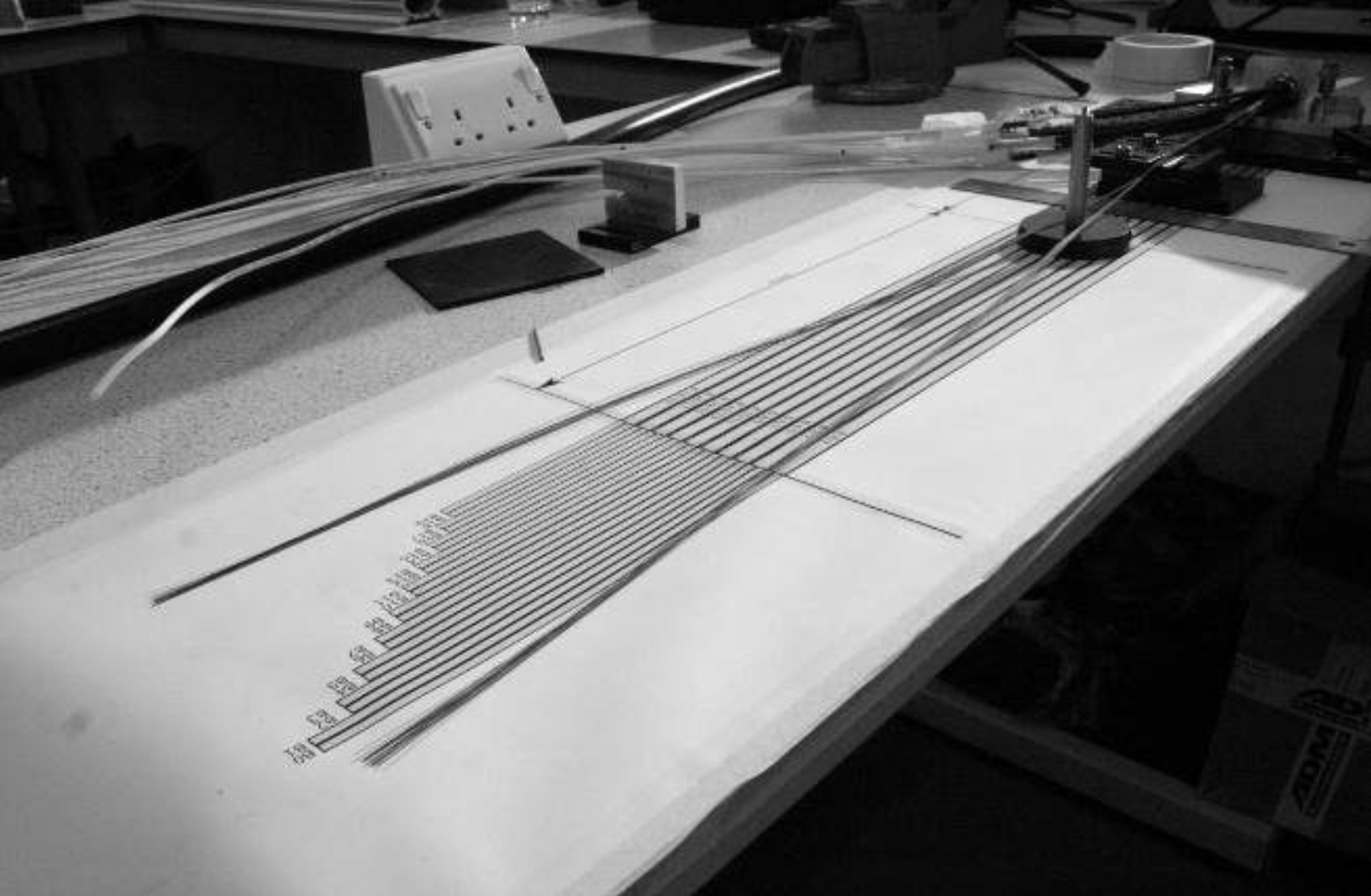}
\caption{A to-scale paper template is used to set the appropriate lengths to the bare and PTFE-clad fiber bundles. Each bundle had a unique length to confirm to the `race track' slots of the slit assembly.}
\centering
\label{fig:Paper_Template}
\end{figure}

Before bonding each V-block, the PTFE and bare fiber sections are adjusted to the appropriate length, using a to-scale paper template attached to the optical bench. Small amounts of fiber can be taken up or withdrawn from the loops in the spool boxes (see Section \ref{sec:spool}). The bench and other parts are wrapped in closed-cell foam to prevent damage to bare fibers (see Figure \ref{fig:Paper_Template}.

A wire-eroded jig is mounted in front of the slit plate with two round `dimples' acting as a datum point for the front face of the AR window on each V-block. Before use as a reference, the jig was measured on a Keyence shadowgraph to ensure the `dimples' were accurate.

In order to confirm the pointing accuracy of each block, the central fiber is illuminated and the fiber output ring is visually matched to a printed target. The target has twenty printed rings (one for each V-block pointing) and is parallel to the slit plate at a distance of 300mm. At this distance, the maximum fiber pointing deviation of $\pm$ 0.5$^{\circ}$ is 2 mm, that is visually discernible on the targets. The central fiber of each block is back-illuminated from the AR-window output side with a laser for identification. The corresponding input end of the fiber is then cleaved to ensure an optical finish for laser injection. The bare fiber is held in a grooved jig with a small magnetic button.

The underside of the V-block and region of slit backing plate used in bonding are cleaned with optical wipes and semi-conductor grade IPA. Norland NOA88 UV-curing glue is applied to the plate and the V-block placed on top. A second Nordson Performus glue gun driven by dry nitrogren feed is used to apply 22 drops of glue per V-block. Tight control of the glue layer thickness ensures uniformity of fiber height (dispersion direction) along the length of the slit plate. The first block, B0, is bonded first, and has three reference `dimples' (two pins at the AR window and on the non-`knee' V-block edge). The block can be manipulated in to place with a clean plastic cocktail stick placed behind the block cover lid. Once located, a soft-tipped M20 bolt provides consistent and replicable downward pressure to retain the V-block during glue curing. The bolt tip is applied centrally to the cover lid to avoid applying any tilt, and the soft tip prevents any damage to fibers and also minimizes stresses (which could increase FRD, and negate the visual ring laser output light check). A ThorLabs CS2010 LED UV curing gun at maximum output is used to cure the Norland glue for 10 seconds.

During the assembly and curing processes, a pair of Dinolite digital microscopes provide real-time placement inspection. A downward facing microscope ensures the AR-window front of each V-block is placed flush against the wire-eroded datums. As mentioned in Section \ref{sec:arwin}, consistency in the thickness of the EpoTek glue layer between the polished fiber faces and the bonded AR window are important to maintain focus between V-blocks, since the jig pins contact the front surface of the AR window. A second orthogonal microscope faces the fiber outputs from the front. Cross-hairs on the microscope confirm that there is no tilt between the V-block and slit base plate.

The combination of visual target `sanity check', twin microscopes and the highly toleranced jig datums ensures accurate and consistent V-block placement and curing. A further 180s of curing with the ThorLabs UV light source permanently fixes the block in place. The first few slit assemblies shipped to LBNL from DU suffered from a small number of V-blocks detaching from the plate, with DU staff travelling to LBNL to remedy the issue. Increasing the cure time with the UV light source remedied the problem, in conjunction with adding Bosch accelerometers and shock-meters to the shipping crates with real-time tracking, to ensure safe handling in freight.

The remaining 19 V-blocks are bonded in a similar manner, in turn, from B1 to B19. The inter-block spacing is set with a 70$\mu$m spacer wire that faces downwards on to the plane of the slit plate. Here, the fiber-facing microscope can also be used to confirm proper block-to-block spacing. Since the wire is so fine and fragile, it is retained lightly from above within a stainless steel hypodermic needle to provide stiffness. The wire must be placed behind the cover lid (in case of any overhang, relative to the V-block beneath, which would add to the diameter of the wire and increase inter-block spacing), but ahead of the V-block `knee' feature (which would reduce inter-block spacing).

\subsection{Slit plate metrology} \label{sec:slit_metrology}

\begin{figure}
\centering
\includegraphics[width=0.65\textwidth]{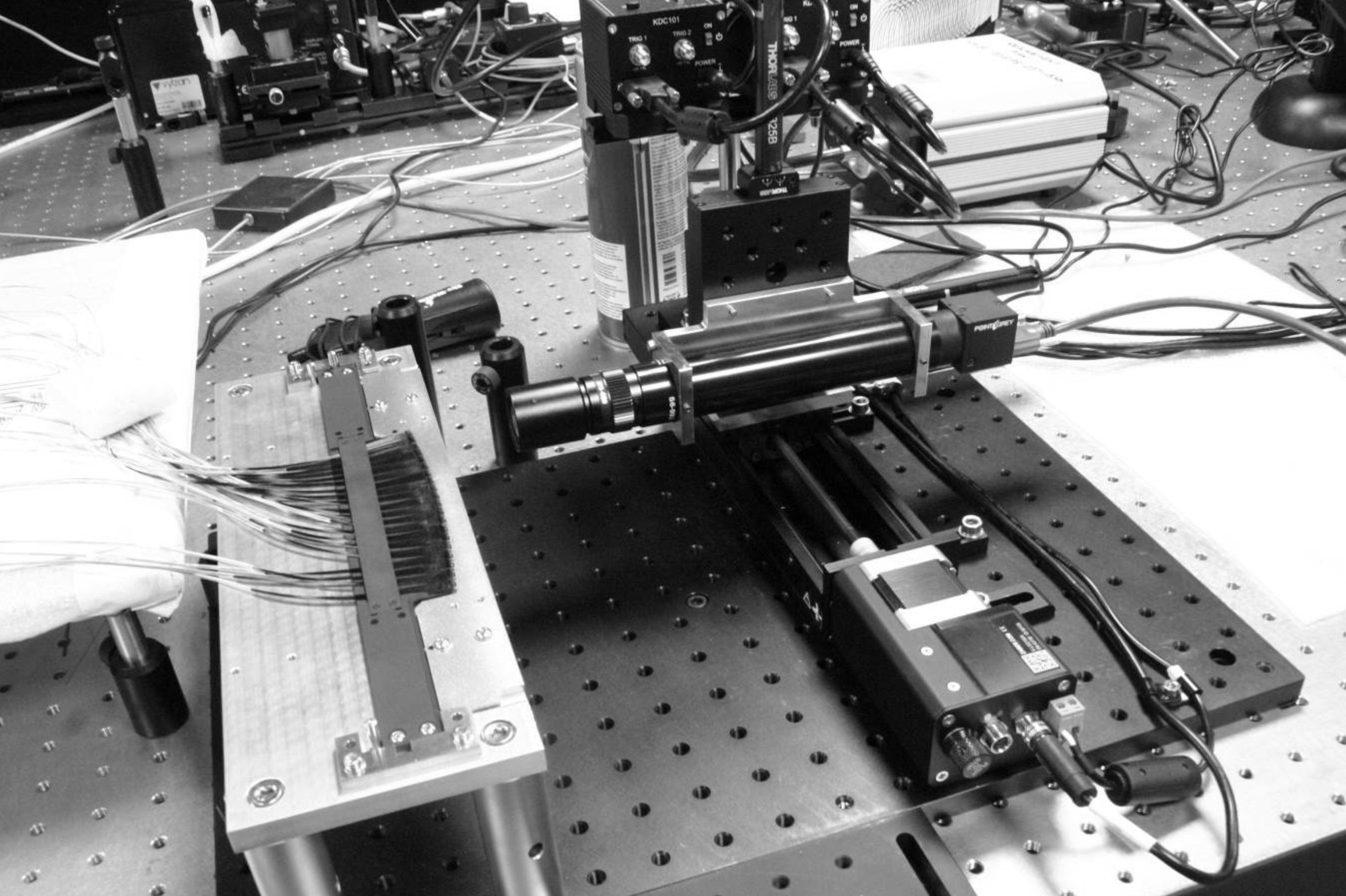}
\caption{The wire-eroded datum jig is removed and the paper target exchanged for a three-axis stage carrying a microscope and camera. The camera scans the slit determining the $x$ and $y$ locations of the illuminated fiber output faces, as well as their focus in $z$, using centroiding and image sharpness algorithms. Each scan is repeated several times and takes around 4 hours to complete.}
\centering
\label{fig:slit_microscope}
\end{figure}

Once all 20 blocks have been fitted to the slit plate, and cured in place, the wire-eroded datum jig can be removed. A three axis metrology jig replaces the paper target screen. The jig, as per Figure \ref{fig:slit_microscope}, comprises three linear translation stages: two ThorLabs PT1/Z-8 25mm stages, and one Zaber LSQ150A 150mm stage. A FLIR BlackFly 5.0M Pixel camera fitted to a 10x Mitutoyo microscope objective lens is carried by the motorized stages. This system has an extremely shallow 5um depth of field coupled with small 3.45um pixels. 

\begin{figure}
\centering
\includegraphics[width=\textwidth]{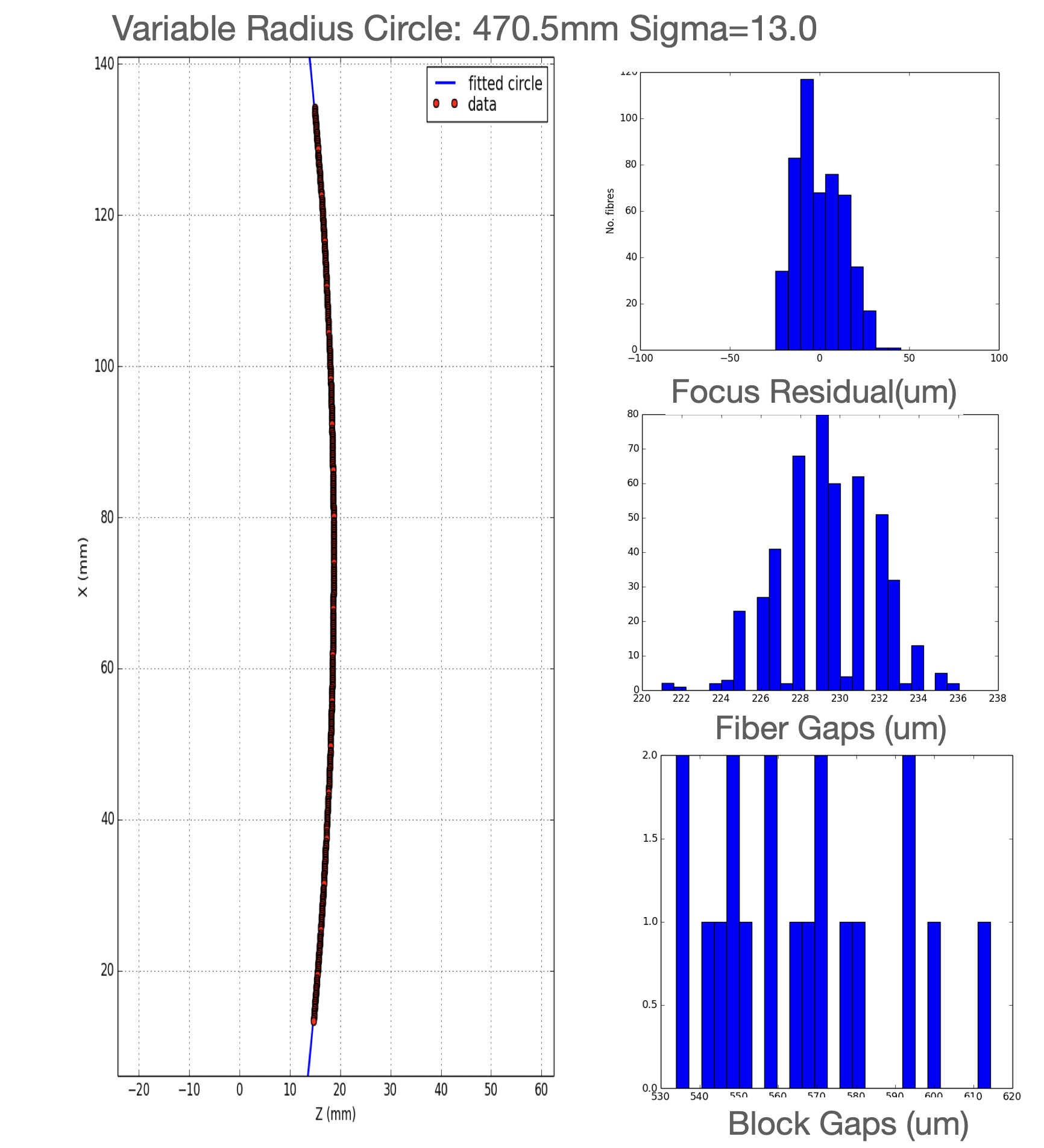}
\caption{$(Left)$ The left side plot shows the positions of each fiber output face determined using travelling microscopes on translation stages, fit to the nominal slit radius. $(Right)$ Three histograms (top to bottom) show the measured focus residual in $z$ (determined by block position on the datum pins and AR window glue layer thickness), the gaps between fibers (determined by the V-groove spacing) and the inter-block spacing (determined by the plate bonding spacer wire).}
\centering
\label{fig:slit_metrology}
\end{figure}

All 500 fibers in the slit are then illuminated with a diffuse continuum light source (cleaving the fiber inputs is not necessary) for metrology. The microscope camera system is scanned across the slit plate to locate all 500 fiber outputs in three dimensional space. The $x$ and $y$ position of each fiber face is determined with a circle-finding and centroiding algorithm. The focus ($z$) position is optimized using an image sharpness criterion. A full scan takes around 4 hours and is repeated several times in a darkened lab to average results. Figure \ref{fig:slit_metrology} shows the results of a scan on one slit plate.

\subsection{Splicing} \label{sec:prod_splicing}
\begin{figure}[ht]
	\centering
\gridline{\fig{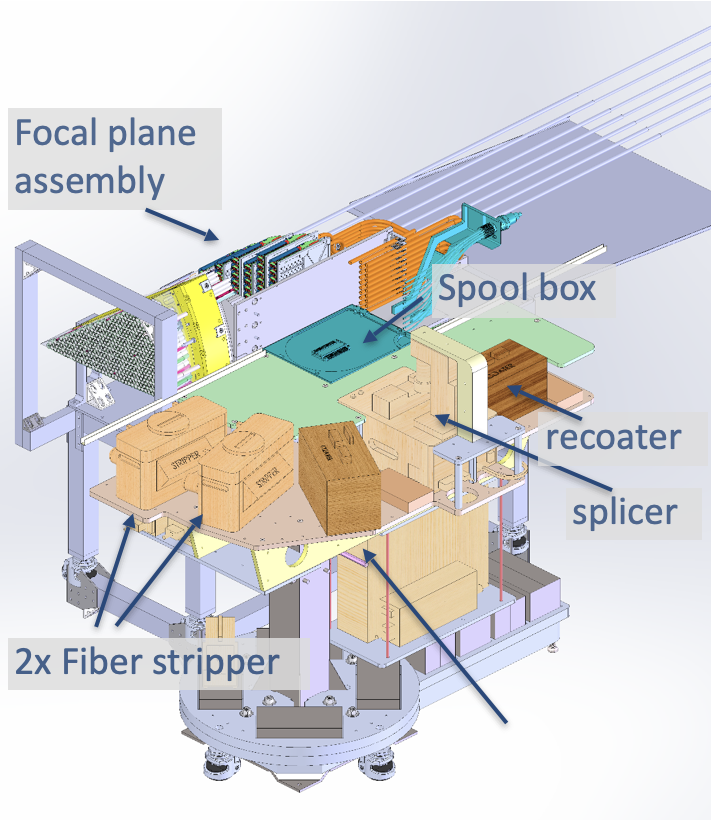}{0.32\textwidth}{(a)}
\fig{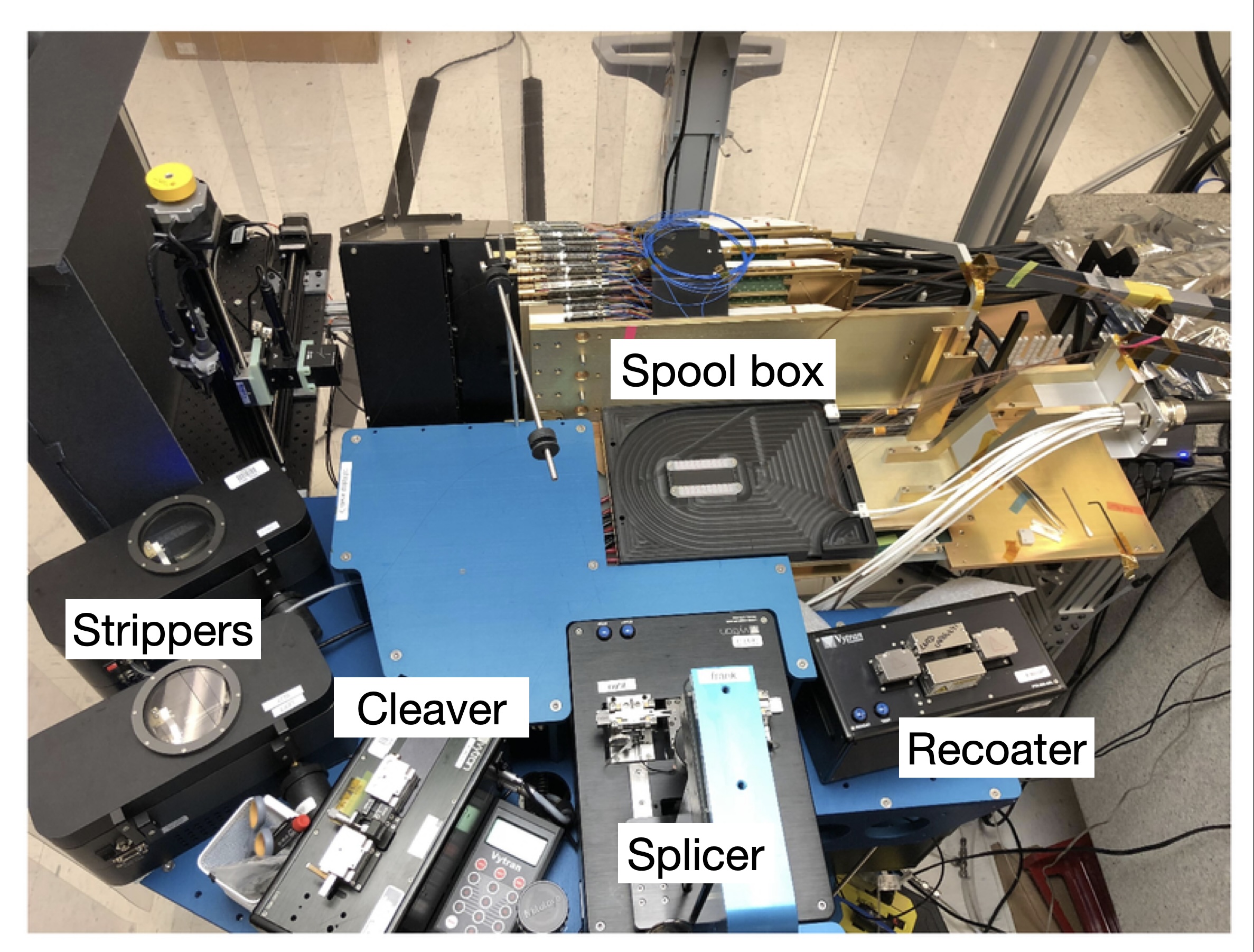}{0.48\textwidth}{(b)}}
\caption{Panel (a) shows a schematic of the custom, ergonomic splicing station. Panel (b) shows the custom ergonomic splicing station in use during production.}
\label{fig:splice_station}
\end{figure}

Due to the repetitive nature of splicing and the large number of splices to be completed, a custom, ergonomic splicing station was designed as shown in Figure \ref{fig:splice_station}. The four key pieces of splicing equipment (buffer stripper, fiber cleaver, fiber splicer, buffer re-coater) were arranged on a table at a comfortable working height for the technicians and which allowed a $\sim1\,$m length of fiber to reach from the spool box to each of the pieces of equipment. The station was designed with no sharp edges where fibers could snag, and the height could be adjusted in order to accommodate the stack of fusion splice cassettes. 

In addition to the splicing equipment, a source of collimated illumination was installed at the input of the focal plane petal and a camera was installed at the spectrograph slit. This equipment is shown in Figure \ref{fig:ps_iloveu} and described in more detail in Section \ref{sec:prod_spliceQA}.

\begin{figure}[ht]
	\centering
\includegraphics[height=8cm]{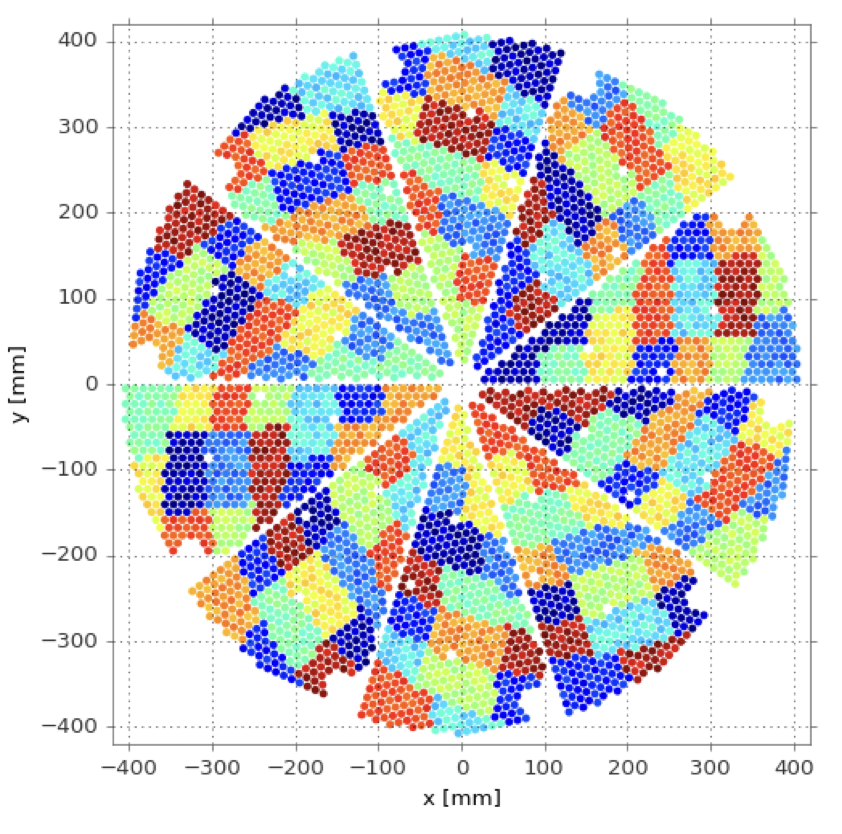}
\caption{The fibers at the focal plane are routed to different areas on the spectrograph for each petal. This randomized mapping allows analyses to break degeneracies between effects which are correlated on the focal plane versus effects which are correlated on the spectrograph slit (and thus on the spectrograph CCDs). The mapping has the following distribution of fibers on the focal plane, color-coded by fiber location on the slithead (first column on the detector = dark blue through last column on the detector = dark red). }
\label{fig:mapping}
\end{figure}

The mapping between the focal plane to the slit was randomised in units of 50 (the number of fibers in a sub-bundle in the fiber cable) and was unique to each petal as is shown by Figure \ref{fig:mapping}. A bundle of 50 fibers from the focal plane was routed to a section on the detector, but adjacent fibers on the focal plane were not placed onto adjacent columns on the detector. This randomized mapping allows analyses to break degeneracy's between effects that are correlated on the focal plane versus effects which are correlated on the spectrograph slit (and thus on the spectrograph CCDs).

\begin{figure}
\centering
\gridline{\fig{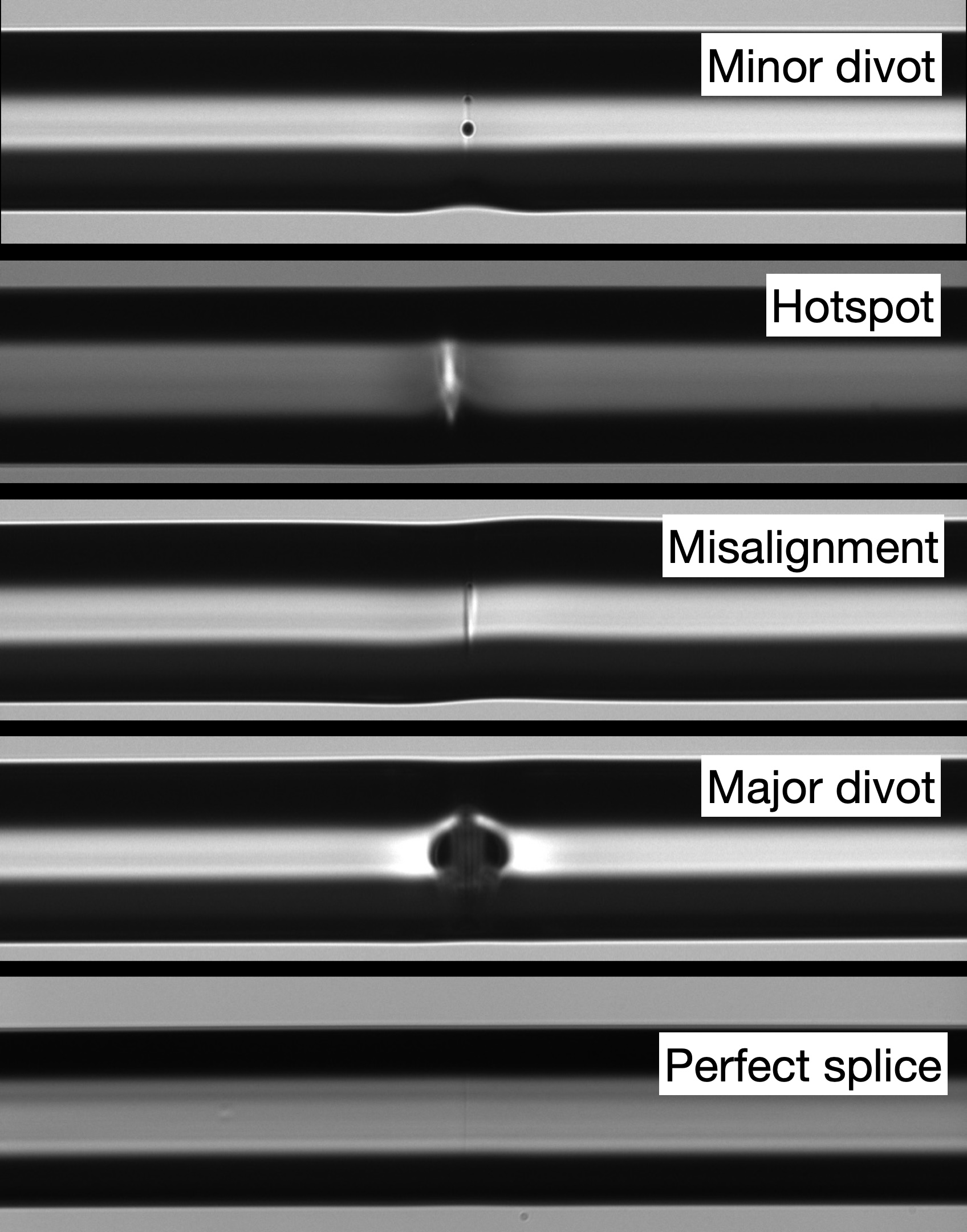}{0.3\textwidth}{(a)}
\fig{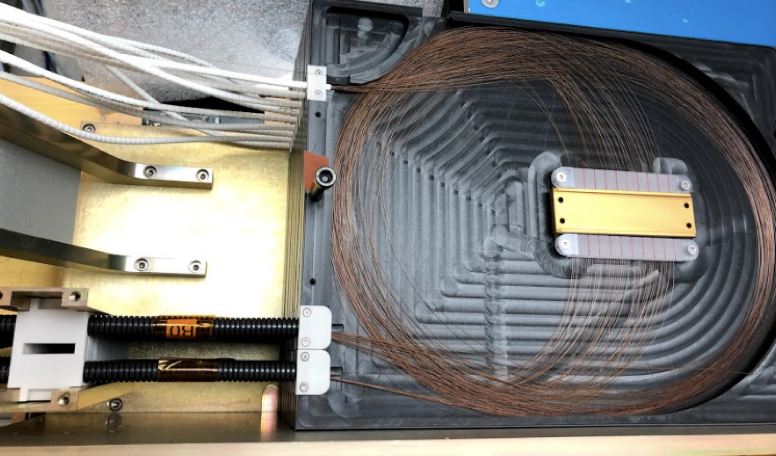}{0.65\textwidth}{(b)}}

\caption{Panel (a) - The splicing proved to be the most unreliable step of the fusion process. Failure modes included divots, misalignment's, and hotspots. A good splice is shown at the bottom of this panel. Panel (b) - One cassette of a splice box. The cassette holds 50 fibers and ensures the spliced area remains straight whilst guaranteeing that the minimum bend radius of the extra fiber is not violated. The spliced area is held between two pieces of foam to reduce stress. }
\label{fig:splice_production}
\end{figure}

The first stage to splicing was to determine which cable sub-bundle should be spliced to which focal plane unit. Once the sub bundle was selected, the Miniflex was removed and the Boron Nitride lubricant was cleaned from the fibers by wiping with Isopropanol. Next the conduit from the appropriate focal plane bundle was indentified and trimmed to align the holder on the splice cassette. Within these two bundles a random fiber was chosen from both the focal plane and the slit and these fibers were cut to the appropriate length and each loaded into the polyimid stripper. When the polyimid was removed the fibers were cleaved and spliced. Splicing proved to be the most unreliable step of the process and so the length of each fiber made it possible to repeat the splice up to three times. Examples of splice failures are shown in the left panel of Figure \ref{fig:splice_production} and include, divots, misalignments, and hotspots. The splice machines were calibrated after every 200 splices in an attempt to reduce failures.

When a successful splice was achieved, the spliced region was re-coated in acrylate to protect from moisture, and pull tested to check integrity. We then loaded the splice into the splice cassette that was capable of holding 50 fibers. We further protected the splice by ensuring the spliced area remained straight by holding it between two fiber combs and clamping the area between foam. The fully loaded splice box is shown in the right panel of Figure \ref{fig:splice_production}. 

The first fiber was successfully spliced on 9th May 2018, and the last fiber was successfully spliced on 3rd April 2019. The average daily rate during this period was 17 successful splices per splice station, and it should also be noted that the average daily rate increased over the duration of the project. The rate-limiting step was the polyimid stripping. Two strippers were installed onto each table to allow us to strip both the focal plane, and spectrograph fibers simultaneously. 

The baseline rate used in our planning was 18 splices per day. In order to increase our production we built a second splice station, and also had a full set of `active' spare equipment that could be swapped in to reduce downtime. The rate of production was largely driven by the rate at which petals were populated with fiber positioners. 

\subsection{Splicing quality and assurance} \label{sec:prod_spliceQA}
During splicing the performance of every splice was measured before the fiber was loaded into the splice box and the next fiber was fused. Both the FRD performance and the focal plane-to slit-position mapping was recorded.

\begin{figure}[ht]
	\centering
\gridline{\fig{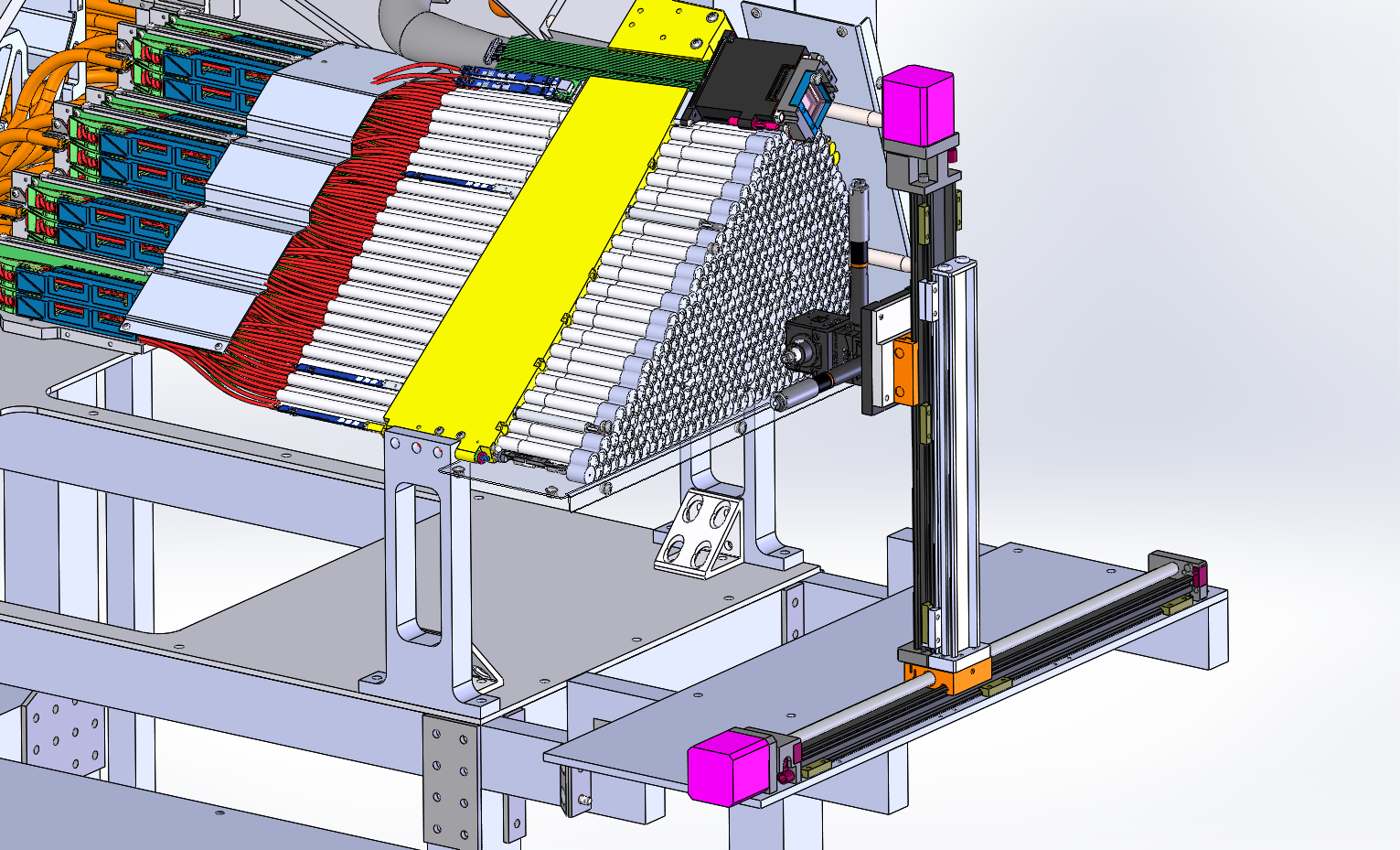}{0.6\textwidth}{(a)}
\fig{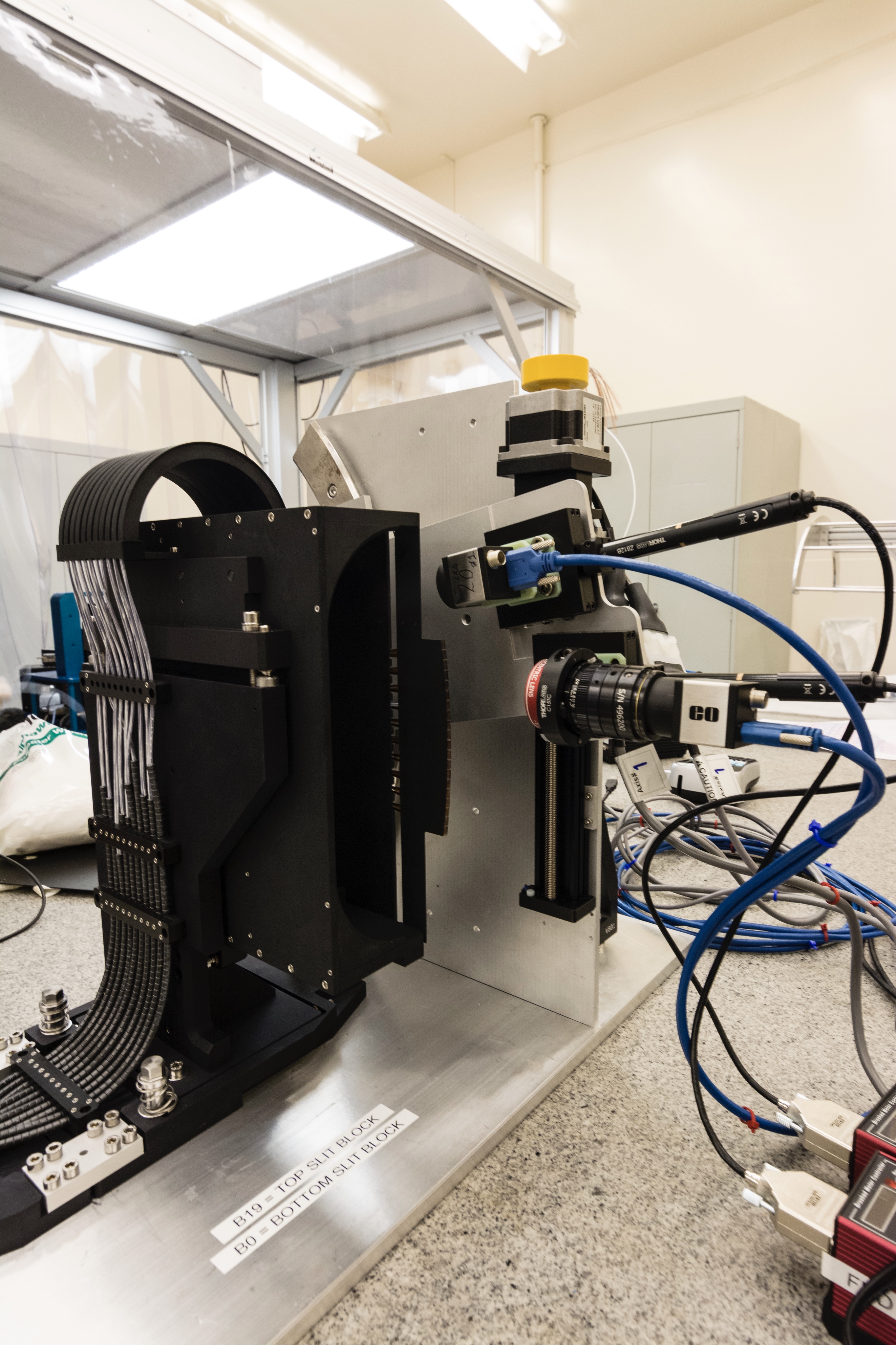}{0.3\textwidth}{(b)}}
\caption{Panel (a) - The petalbot is used to inject a collimated beam mounted on motorized $x$-$y$ stages into focal plane fibers. Panel (b) - The slitbot has two cameras mounted on motorised stages that follow the curve of the spectrograph slit. One camera is used to determine the mapping between the focal plane and the slit. The second camera is used to measure the FRD performance of the fiber.}
\label{fig:ps_iloveu}
\end{figure}

This measurement was performed by injecting a collimated beam mounted onto an automated $x$-$y$ stage (via a robot that we named the petalbot) at the focal plane, and then measuring both the slit position and FRD performance at the fiber slit (via a robot that we named the slitbot). Both the petalbot and slitbot are shown in Figure \ref{fig:ps_iloveu}. The slitbot has two cameras that mount onto a stage that follows the arc of the slit. One camera focuses on the fiber slit in order to map the fiber between the focal plane and the slit, and the other camera captures the far field output of the slit fiber in order to measure the FRD performance. 

\subsection{Final testing and performance}
The final stage to production was a final FRD and throughput measurement. This was performed when splicing was complete and the petal was still in the splice station.

The splice robot (petalbot + slitbot) was used to measure the FRD and throughput of all of the fibers, and the throughput was compared to a pre-calibrated fiber in order to measure the absolute throughput of the full fiber system. The absolute throughput of each fiber was used in addition to the FRD performance in order to calculate the final throughput of the fiber system. The FRD performance had a median value of 2 degrees FWHM  for $f$/3.9 input (less than the budget of 2.3 degrees as described in Section \ref{sec:FRD_budget} and the absolute throughput including these contributions was measured as 91.1\% meeting the requirement imposed by DESI described in section \ref{sec:Reqs} and shown in the requirements table \ref{tab:fibreq}.

\begin{figure}
\centering
\gridline{\fig{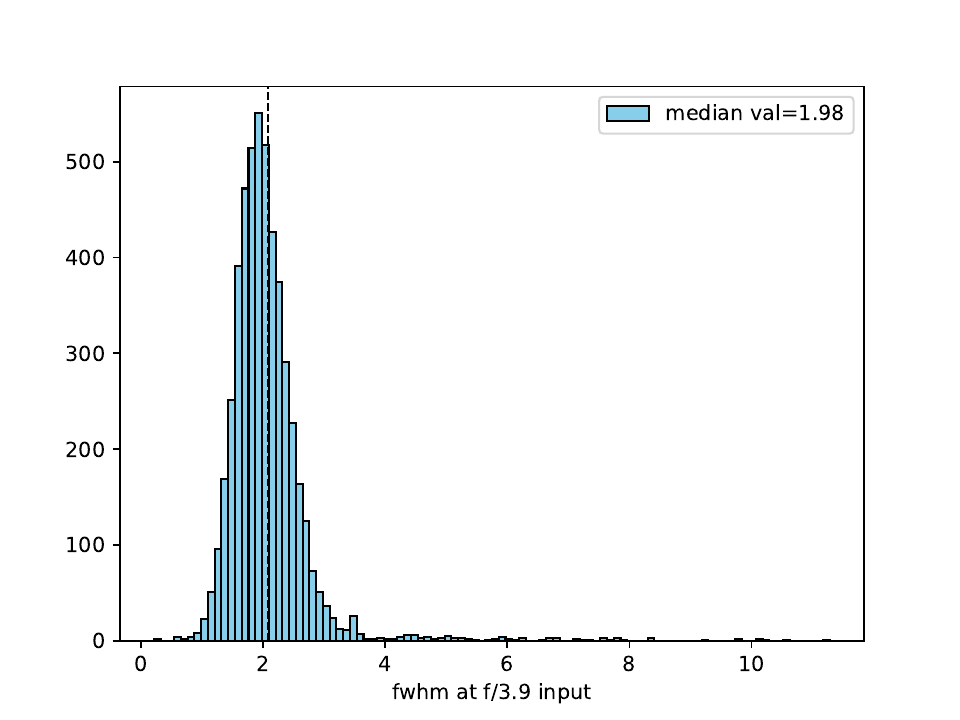}{0.48\textwidth}{(a)}
\fig{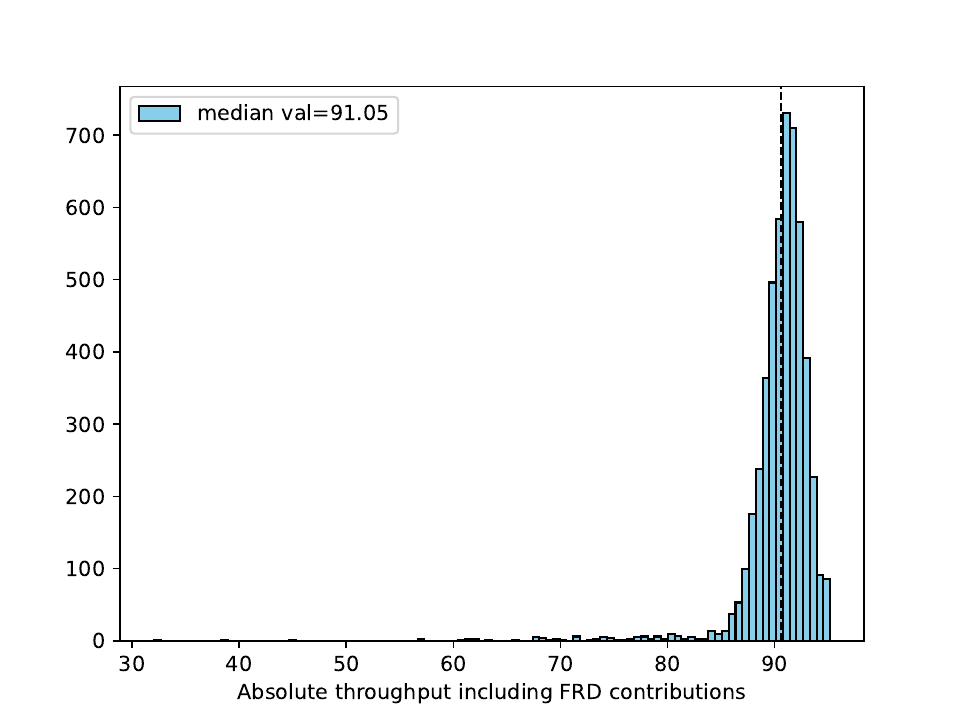}{0.48\textwidth}{(b)}}
\caption{Final measurements of all 5,000 fibers in the fully integrated fiber system prior to shipment to Kitt Peak. The FRD performance had a median value of 2 degrees FWHM  for $f$/3.9 input and the absolute throughput including these contributions was measured as 91.1\%, meeting the requirement imposed by DESI.}
\label{fig:FRD_final}
\end{figure}


\section{Installation at the 4m-Mayall}
\label{sec:installation}
The fully integrated fiber system was installed onto the 4m-Mayall in September 2019. 

All ten petals of the focal plane system were installed in the focal plane, followed by routing the fiber cables down the telescope, through the cable wraps, and to the room where the spectrographs are located. 

\begin{figure}[ht]
	\centering
\includegraphics[width=\textwidth]{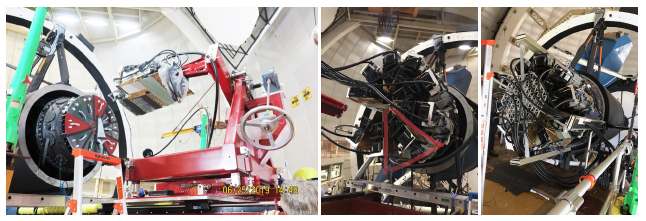}
\caption{Ten optical fiber cables from the focal plane petals are strain relieved to a frame, that will later be covered within the FPE. The routing of each cable was carefully engineered to ensure the bend radius of the cables was everywhere greater than 200 mm. The panel on the left shows the first petal loaded onto the sled prior to insertion. The central panel shows the back of the focal plane when 8/10 were loaded. The panel on the right shows the focal plane cable management system.
}
\label{fig:focal_plane_install}
\end{figure}

To install the focal plane a custom installation sled was used. This sled has an arm with 6 degrees of freedom that holds a single petal and allows it to be safely installed into the focal plane enclosure (FPE), as shown in the left panel of Figure \ref{fig:focal_plane_install}. The petals are installed from the back of the FPE, so petals are installed with the fibers leading into the back of the corrector. The fiber cable and the spectrograph spool box and slithead  remain on their  shipping carts during the focal plane installation.

In order to provide strain relief to the petals, a frame was installed around the focal plane structure. This frame routes the fiber cables in such a way as to ensure that the minimum bend radius is not exceeded, but also allows any unit to be removed without affecting another cables and is shown in the right panel of Figure \ref{fig:focal_plane_install}.

\begin{figure}[ht]
	\centering
\includegraphics[width=\textwidth]{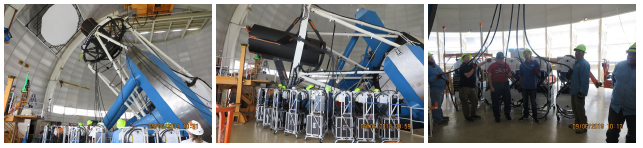}
\caption{The fiber cables and slits were stored on their shipping carts during the focal plane installation and handled by an expert team of NOIRlab engineers during cable routing. The middle and right panels show the NOIRLab team, each person managing one fiber cable and spectrograph spoolbox with slitheads, as the fiber cables are installed to the telescope. }
\label{fig:cable_management}
\end{figure}

The fiber cables and slits were stored on their shipping carts during the focal plane installation and handled by an expert team of NOIRlab engineers during cable routing as is shown in Figure \ref{fig:cable_management}.

After all 10 petals were installed it was time to route the cables down the telescope. 

The fiber cables go from the FPE to the upper ring, down the side of the FPE and over the Southeast and Northeast upper spider vanes, stacked on each other, to minimize blockage of light. From the upper ring, the two groups of five cables are constrained by custom hardware designed,
built, and installed by NOIRLab. The two sets of cables converge into a single group of ten cables about halfway down the Serrurier truss.

\begin{figure}[ht]
	\centering
\includegraphics[width=\textwidth]{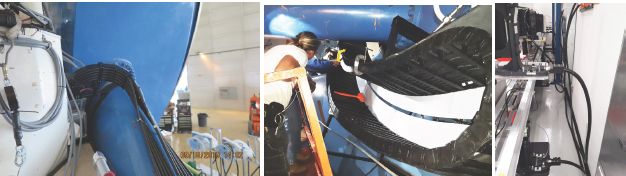}
\caption{Routing of the cables through the Dec and Polar Axis before they penetrate the wall of the spectrograph shack. This process was shown schematically in figure \ref{fig:cables_RD}.}
\label{fig:cable_wraps}
\end{figure}

To ensure the fibers allow telescope rotation about the declination and hour angle axes, without twisting the cables or violating their minimum bend radii of 200 mm, they were installed in articulated cable carriers with reversible bend directions, for the arc-shaped path the rotation requires (as is shown in Figure \ref{fig:cable_wraps}). The bundle of ten cables coming down the telescope truss was loaded into the declination wrap carrier on a bench near the East declination bearing, which was then craned into its service position in its custom guides in the gap between the telescope center section and the hour angle “horseshoe”. Due to the limited width of this gap, the declination wrap has the cables arrayed five wide and two high in the cable carrier, with dividers between the layers of cables. The bundle of ten cables was strain relieved to the oval tube of the horseshoe, then into the hour angle cable wrap in one layer, ten cables wide. The fixed end of the hour angle cable wrap is attached to the fixed telescope mount, and from there, the cables pass straight through penetrations in the wall of the Large Coudé Room over the M floor to the East of the telescope mount. The cables continue from that wall through penetrations in the DESI spectrograph clean room shack.

The penetrations through the walls of the Large Coudé Room and the shack were sized to allow the passage of the slitheads whilst contained in their protective shipping boxes, and insulated panels were then installed to provide environmental separation between the interior and exterior volumes.

All ten units of the fully integrated fiber system (focal plane petals + cable and slit assemblies) were successfully installed within one month, and there was no damage to the fiber system.

\section{On Sky Performance and Validation}
\label{sec:on_sky}

Although throughput and FRD performance were measured pre-shipment at LBNL, it was not possible to send light from the DESI corrector through the fiber system and into the spectrograph before installation at Kitt Peak. Whilst it was relatively straightforward to measure the efficiency of the fiber system in the lab, it was not possible to fully characterise the stability of either the fiber system, or the full instrument. 

The observing strategy for DESI requires that the spectrograph does not require calibrations between observations of each field. This is only possible if the fiber system provides a sufficiently stable illumination pattern into the spectrographs such that the PSF on the detector does not vary. If the PSF is not well characterised, the sky subtraction will be poorer resulting in a decrease in the redshift efficiency. The PSF of the system is composed of
the fiber and spectrograph, such that the PSF is the
convolution of the image of the fiber tip (near field)
convolved with the optical blur from the spectrograph
cameras. The diameter of the fiber image on the CCD
is of 51 $\mu$m, or 3.4 CCD pixels, after accounting for the camera demagnification of 0.48. The PSF appears as a blurred disk presenting a central plateau because the fiber image is partially resolved.

In this section we first  present a series of test results that demonstrate how fiber system requirements are met on sky during the commissioning period. Then we provide two examples of other observations that demonstrate the uniformity and stability of the complete system. These are the Emission Line Galaxy (ELG) redshift efficiency and the radial velocity errors.

\subsection{Validation during commissioning}
The stability of the fiber system has a direct impact on the stability of the PSF, which impacts the performance of the sky subtraction algorithm and optimal spectral extraction with spectroperfectionism \citep{2010PASP..122..248B}. Using
a precise model of the 2D point spread function instead
of a projected 1D cross-dispersion profile, the PSF model is fit on the arc-lamp calibration images. These calibration images are taken in the afternoon
preceding the night of observation, so the spectroscopic PSF has to be stable for a duration of about 12 hours, and should not vary with the fiber flexure induced by either the positioner moves or the change of telescope pointing. These variations in PSF can manifest as a shift in the position on the detector

A quantitative criterion for the PSF stability has to
be defined as there are several scalar numbers one could
use: full width half maximum, second moments, 
flux measurement biases (for a continuum spectrum or emission lines), line fit bias. The most stringent scientific driver is the sky subtraction accuracy because it affects the false identification rate of emission line galaxies, which is a level 2 requirement of the project. This sky subtraction bias is affected by the relative error on the PSF shape from fiber to fiber, and not by a global systematic PSF error. Indeed any average PSF error which is shared among all fibers is compensated by a change of the average sky spectral shape and intensity during the sky subtraction procedure.The most stringent criterion on PSF stability is  the emission line flux bias induced by a change of PSF shape. The effect on this multiplicative bias is maximum when the CCD pixel noise is dominated by readout noise and not Poisson noise.

During commissioning, we measured the stability of the PSF on the detector by studying the emission lines in arc calibrations over a period of 3 days under a range of conditions, and using that as a measure of the fiber system. The first test was the stability of the fiber system with time, since this determines how often calibrations are required. The second test was the stability of the fiber system as the telescope moves, since this determines if the calibrations need to be repeated at different telescope pointings. Finally, we characterized the stability of the fiber system as the fiber positioners move, since we would not be able to perform wavelength calibrations for every target location. If these tests had shown that the PSF was not stable over 12 hours, a new calibration strategy would be required and the efficiency of the survey might be reduced. 

\begin{figure}
\centering
\gridline{\fig{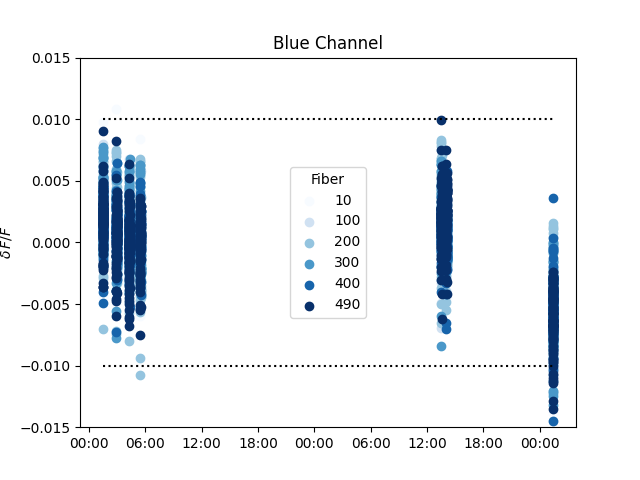}{0.3\textwidth}{(a)}
\fig{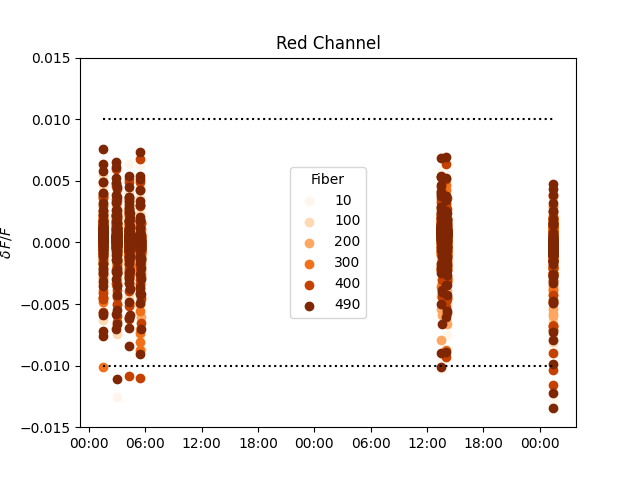}{0.3\textwidth}{(b)}
\fig{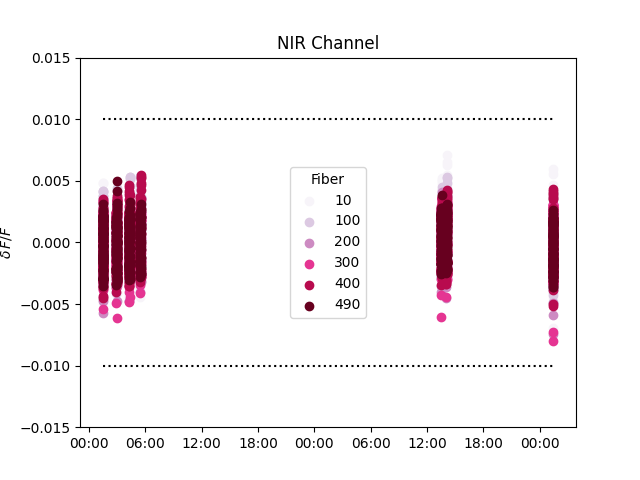}{0.3\textwidth}{(c)}}

\caption{Stability of the PSF as measured during two nights of instrument commissioning. The fractional change in the PSF is measured from three sets of arc calibrations, and the fractional change $\delta$F/F is the change in the PSF relative to the average computed over all exposures for a selection of fibers at a selection of wavelengths. There was a night of observations between each calibration set, and the fiber positioners are in different locations for each calibration. The variation is consequently indicative of the level of variation per night. The fibers were chosen to evenly cover the active region of the CCD. The dashed lines represent the requirement.}
\label{fig:PSF_stabilty}
\end{figure}

The PSF stability over time is shown in Figure \ref{fig:PSF_stabilty} measured as the position on the detector. During these nights the positioners, telescope, and dome were all moved. DESI Requirements state that the PSF bias shall not exceed 1\%. This dataset shows this requirement is marginally exceeded in a few fibers for this particular set of observations. We observed in the analysis a significant drift of the PSF center along the cross-dispersion axis, in other words from fiber to fiber, which physically corresponds to a vertical axis as the pseudo slit is standing vertically on the spectrograph optical bench but this can be corrected in the data analysis. 

In figure 16 of \citep{guy23} the largest contributions to the PSF bias were studied. By looking at data from both moving and non-moving positioners it was determined that the excess scatter caused by the moves was found to be only of 0.002 (0.2\%). The contributions from fiber cable movements were measured by moving the telescope to different dome pointings to take the calibration data. When this data was taken with azimuthal angles of the telescope dome at 253, 359, 107 and 180 degrees, the variation of the PSF shape was $\leq$ 0.6\%.

\subsection{Throughput estimation}
\begin{figure}
\centering
  \includegraphics[height=.5\textwidth]{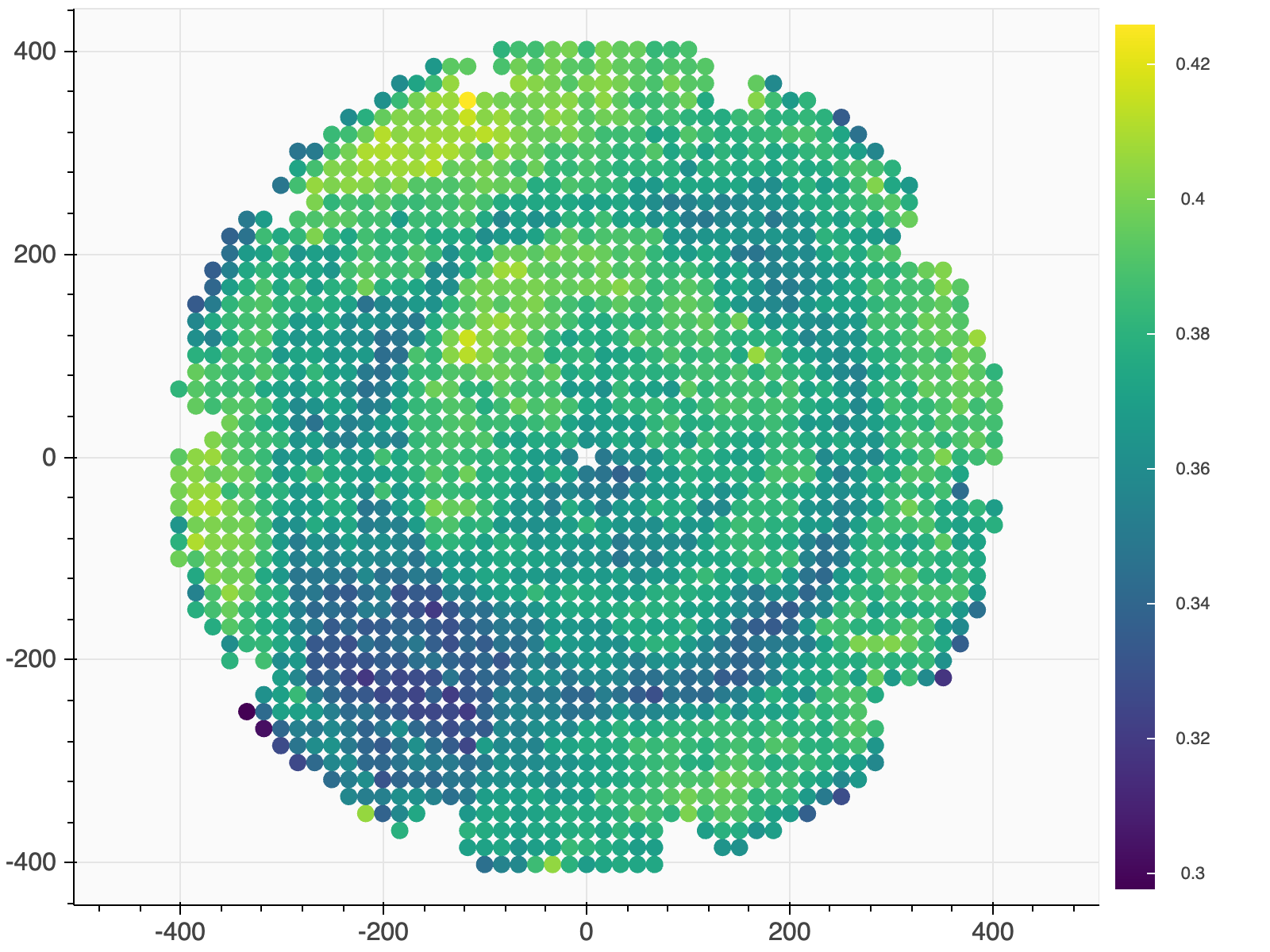}
\caption{Absolute throughput of the DESI instrument at 670nm for every fiber in the focal plane. The relative throughput was measured by using the average ratio of the spectroscopic to imaging flux of standard stars in the red (6000-7300 \r{A}). The absolute throughput was measured during Survey Validation by comparing the light in the spectrographs with the  light in the GFAs for standard stars at airmass one. The residual pattern is due to the image quality  and one imperfection of the DESI corrector as shown in figure 18 of \citep{miller2023optical}.}
\label{fig:spectro_fluxes}
\end{figure}

The variations in the throughput from fiber to fiber is low, as shown in Figure \ref{fig:FRD_final}, though it must still be accounted for during spectral analysis. We obtain a series of flat field images as part of afternoon calibrations to perform this calibration. After we apply corrections for each fiber from the flat fielding, we determined that most of the variation in the full system is a result of imperfections in the DESI corrector. Figure \ref{fig:spectro_fluxes} shows a color map of the average ratio of spectroscopic to imaging fluxes for stars at red wavelengths as a function of focal plane coordinates. This figure shows that the distribution is mostly flat in the red, which is where the throughput is optimised. 

Although we have not remeasured the throughput of the fiber system since initial tests during commissioning, we don't see any throughput or PSF stability change that suggests the performance of the fiber system is deteriorating.

\subsection{Emission line galaxies redshift efficiency}
Emission Line Galaxies (ELGs) are star-forming galaxies with strong emission lines. The ELG sample, which comprises about one-third of all DESI tracers, is used to probe the universe over the 0.6 $\leq z \leq$ 1.6 range, and the 1.1 $\leq z \leq$ 1.6 range is expected to provide the tightest constraints \citep{Raichoor_2023}. The ELG redshift success rate over this redshift range drives several of the technical requirements, especially throughput, resolution, wavelength range, and PSF stability.  In addition, DESI has many requirements on the ELG sample in order to reduce systematic errors and optimise the BAO distance errors. These requirements include limits on the density of targets, the random error on individual redshifts, the systematic inaccuracy in the mean redshift, and the number of  catastrophic failures (defined as redshift errors in excess of  $1000\,\mathrm{km\,s}^{-1}$).

\begin{figure}
\includegraphics[width=.9\textwidth]{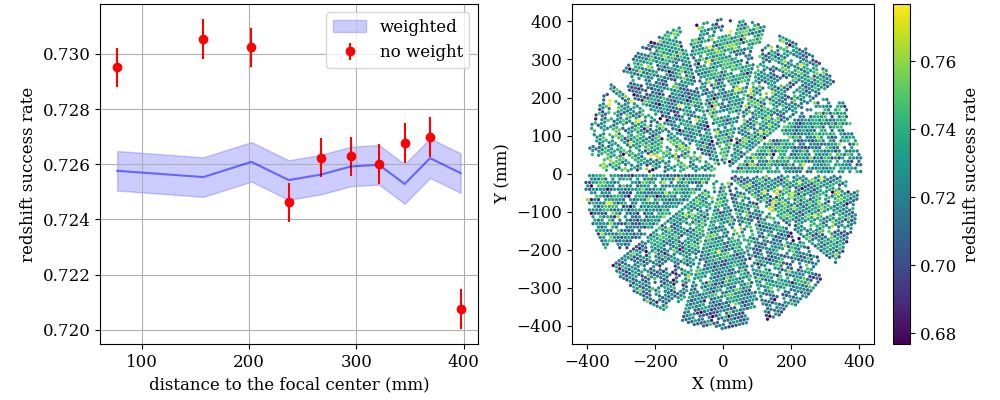}
  \caption{({\it Left}) Total redshift success rate (the number of good redshift measurements divided by the number of ELG targets) as a function of the distance of the fibre to the focal centre. The error bars represent the result without any correction, presenting more than 3-sigma difference in success rate for fibres close to the centre and those in the outskirts. The shaded region is the weight-corrected success rate. ({\it Right}) Per fiber ELG success rate for both the BASS/MzLS survey and DECals survey after the correction for the focal plane location (normalised to one). }
\label{fig:ELG_successFA}
\end{figure}

During the Year 1 data analysis we studied the success rates of the ELG redshift measurement after accounting for the redshift failure rate \citep{Jiaxi_2024}. The redshift failure rate is a function of the effective observing time, and so once this redshift failure weight is applied, the success rate should be uniform with respect to all observing conditions, if the PSF is stable. There is a small dependence on the position of the fiber in the focal plane since there are fewer fibers at the center of the focal plane, however, there is no effect due to the performance of the fibers as is shown in figure \ref{fig:ELG_successFA}.

\subsection{Radial velocity errors}
Although DESI is optimized for galaxy redshift surveys, it is also well suited to observing large numbers of Milky Way stars  \citep{2023ApJ...947...37C}. However, even at low resolution, if the fiber is not stable in the near field, the PSF at the detector will also not be stable and sky subtraction will not be accurate. For high precision radial velocities, we are dominated by sky subtraction and are not testing PSF/LSF stability. 

During Survey Validation, we observed approximately 500,000 unique stellar targets and compared these redshifts to results from other surveys \citep{Cooper_2023}. This dataset in addition to Early Data Release data showed that the DESI RedRock pipelines can measure radial velocities to $\sim0.93$ km$^{\text{-s}}$ \citep{Sergey2023}.

\section{Lessons Learned} \label{sec:lessons}
The DESI Fiber System has been highly successful; In November 2021 alone, DESI cataloged redshifts from 2.5 million galaxies and by June 2022 almost 8 million galaxies were catalogued making it the largest map of our universe to date.

Some key aspects of the design of the fiber system contributed to the success to the instrument. 

\subsection{Positive aspects}
\subsubsection{Procedural}
\paragraph{Technical and Interface Requirements} Beginning the R\&D phase and then the manufacturing phase with controlled, clearly defined technical and interface requirements enabled success at the fiber system level, and during integration with the other subsystems. Technical requirements help define the component selection, process development and testing. Interface requirements, such as boundary conditions, help define the design.

\paragraph{Extensive R\&D program} The R\&D program for DESI lasted for almost 2 years and resulted in many innovations as described in Section \ref{sec:RandD}. Selection of industry components where possible allowed for availability, and reliable and repeatable of form, fit and function. Developing new fiber handling techniques and optimising the performance resulted in a highly stable and efficient fiber system.

\paragraph{Controlled areas} Each part, assembly or test had a dedicated workspace that was access controlled, accessible by trained personnel only. In these controlled areas, personal protective equipment (PPE), such as safety glasses and gloves, were defined. This protected not only the personnel, but also the fiber system and equipment. Levels of particles were also controlled by using laminar flow hoods/workstations and cleanroom garments.

\paragraph{Value of designated Quality Assurance/Control (QA/QC) and Manufacturing professionals for large production quantities} Post-R\&D, the fiber system assembly began with dedicated QA/QC and Manufacturing professionals, with industry experience, who approached the work from a production line standpoint. Controlled documentation began immediately and lasted through the construction life-cycle, from assembly kick-off to crating the fiber system for shipment to Kitt Peak. For individual piece parts alone, controlled documentation included: incoming inspection of new parts, parts storage, inventory management (incoming and outgoing), training of the above, and training logs. This same documentation methodology was repeated for assemblies. Where possible, each phase of the assembly was tested to track performance issues or non-conformance's. Identifying these issues early allowed us to identify trends and the root cause. This reduced the number failed components, re-work and scrap.

\paragraph{Online tracking of production} Both the PFA manufacture and splicing was monitored via a Google spreadsheet. This extra QA step allowed us to monitor the production in real time and address issues quickly. This monitoring was done by the fiber system scientist and the QA manager who were able to stop production if the work was not meeting requirements.  

\paragraph{Value of developing and adhering to procedures} The production procedure was developed, written and reviewed with help from the technicians who would be performing the work. This ensured that they understood and followed the procedure. 

\paragraph{Distribution of work packages} Splitting work packages for positioners and integration to LBNL, with fiber cable and slit manufacture at DU allowed for parallel manufacture, shared risk and independent/repeated quality assurance procedures.

\subsubsection{System design}
\paragraph{Fiber Termination} The quality of the fiber ends are critical to the science. Industry standard cleavers were used to perform fiber cuts at the focal plane end, splice points and spectrograph ends of the fibers. These cleavers were calibrated, monitored and maintained to ensure the optical performance was optimised through the fiber system. Part of the monitoring included measurement of the FWHM of each PFA as they were built.

\paragraph{Glass ferrules} During manufacturing of the PFAs, transparent glass ferrules allowed for careful visual monitoring of the amount of UV cured glue being applied. Controlling the glue amount and quickly UV curing the parts were vital QA steps that increased the yield of the PFAs.

\paragraph{Fusion splicing} By eliminating a connection in the fiber system the throughput was greatly improved. A typical throughput loss from a connector is $\sim10$\% whereas a fusion splice is $\sim2$\%, when throughput loss due to FRD is considered. Multiple splicing stations were implemented, including a full set of spare equipment. Downtime due to machine calibration or maintenance was reduced, enabling efficient production.  

\paragraph{Routing the fiber system on the outside of the telescope} The fiber cable does not pass through any spaces on the telescope until it enters the spectrograph shack. This means that each focal plane petal and its cable can be removed without removing the spectrograph slit.

\paragraph{Spare cables} Two spare cables (slit assemblies, spool boxes, long fiber cables, petals and positioners) were manufactured between LBNL and DU. This of course provides spares in case of shipping loss or damage between DU, LBNL or Kitt Peak; or installation difficulties; but also allows for the potential of like-for-like replacements mid-survey in case of deterioration in performance (large scale fibre breakage/damage or positioner failures). By making two spare fiber cables we could make two full spare petals: silt assemblies, spool boxes, long fiber cables, and positioners with PFAs installed in spare petals which then allows for extensive testing in the lab during the survey, allowing new procedures, firmware updates, communications protocols etc. to be trialed without risk to the operational instrument or creating survey downtime. Given that the survey operational lifetime has been increased beyond the initial five year scope, these petals + cable and slit assemblies may one day be used on-sky.

\paragraph{Fiber slit v-blocks} The use of V-blocks to set inter-fibre spacing worked well and consistently. Multiple batches of V-blocks from the manufacturer were highly consistent.

The use of vacuum jigs required Venturi pumps, creating constant noise and energy consumption in the clean rooms, but allowed for much tighter tolerances in placing the cover lid. The suction also created small glue deposits underneath the V-block which required cleaning, adding a little to production time.

Moving to reagent-grade solvents for cleaning, ultrasonic baths, drying parts in a vacuum oven prior to assembly and use of a clean room added some manufacturing time and cost but ultimately created a superior product. Degassing glue prior to application and the use of a glue gun with small and replicable deposition also allowed for excellent finish.\\

All projects have their pains, however, and we offer here some of the key things we would do differently in hindsight.

\subsection{Negative aspects}
\subsubsection{Procedural}
\paragraph{Scaling to Mass Production}The R\&D program optimised the production method for small quantities, however, we learned that these processes do not scale well when the order of magnitude is increased. Additional R\&D focused on producing large quantities would have improved production. Subsystems such as the positioner manufacture dud produce first articles to verify the procedure before full production began, however, this was not deemed necessary for the fiber system since although the quantities are large, they were determined to be below threshold for this more formal procedure. With hindsight, we would not have omitted this step. At large quantities it may also have been beneficial to outsource production to industry whilst maintaining oversight on Q\&A. 

\paragraph{Ergonomics} Related to the issue of scaled production, some of the manufacture processes were not optimised for the scale of the project and so ergonomics were added after. In future projects this should be designed into
the production process.

\subsubsection{System design}
\paragraph{Cable Design} The design of the cable was based on heritage design, but it may have been over-engineered for our purposes and over-constrained the manufacture of the spectrograph slits. A higher degree of modularity would have relaxed these constraints. The bulky design also makes it difficult to scale for instruments with more multiplexing. More R\&D may have been useful to simplify the design and and reduce the outer diameter of the fiber cable.

\paragraph{Fusion splicing}
Whilst the performance of the fiber system was massively improved due to fusion splices rather than physical connectors, this method makes maintenance and upgrades more difficult. For future instruments, R\&D should be focused on making physical connectors with performance to rival fusion splicing.

\begin{figure}[ht]
	\centering
\includegraphics[height=4cm]{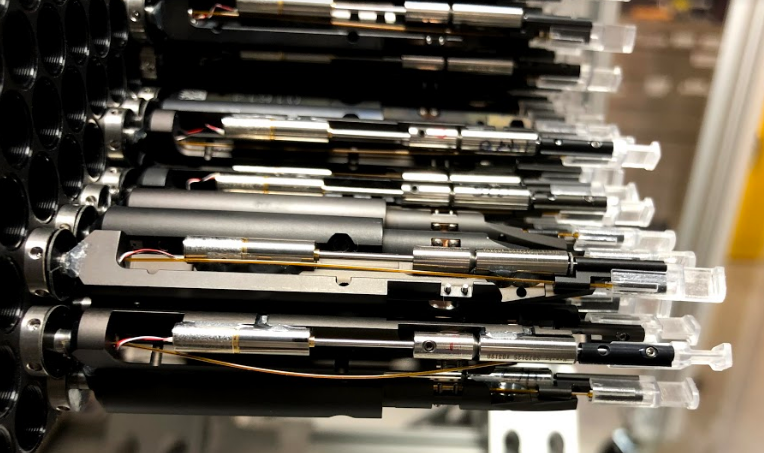}
\caption{A fiber can be seen bowing out of the edge of the positioner. This is due to the ferrule being fixed into the ferrule arm and the Hytrel tubing adding electrostatic stickiness on the fiber and causing it to be pushed, inchworm style, into the positioner. This issue was remedied by shortening the amount of Hytrel tubing on the fiber.}
\label{fig:inchworm}
\end{figure}

\paragraph{Hytrel tubing on PFAs} During the production process it was discovered that the Hytrel tubing protecting the fibers on the PFAs added
electrostatic stickiness on the fiber. Since the ferrule is fixed into the ferrule arm, this caused the fiber to inchworm out of the side of the the positioner as shown in figure \ref{fig:inchworm}. This issue was remedied by shortening the amount of Hytrel tubing on the fiber.

\paragraph{Fiber length} The length of the fiber system was minimised as much as possible for DESI, however, since most fiber attenuation losses appear in the blue wavelength regime (and compound with other losses in the blue within the instrument such as AR-coating performance, diffraction efficiency, transmission losses etc.), future projects might aim to minimize fibre length wherever possible. This could be via placing spectrograph housings/rooms closer to the telescope and/or placing the instrument focal plane at Cassegrain focus.

\paragraph{Slit block polishing} The need to polish V-blocks by hand, one by one, versus the initial plan to machine polish an entire slit plate (20 blocks) in just minutes, dramatically slowed production rate. This required tedious manual labour, the use of disposable dummy blocks, frequent rinsing of polishing papers with distilled water, and regular visual inspection. Many V-blocks required re-polishing due to remaining chips, scratches or a wedged finish.

\paragraph{AR windows} Switching from the initial UV curing glue (which cured in 30s) for bonding AR windows to a slower curing epoxy added significant production time. Four V-blocks (20\% of a slit) took two days, when initial estimates would complete an entire slit in an hour. The need to keep the blocks stationary during this curing period also precluded working on other blocks simultaneously, and required custom jigs to hold the windows in place during the cure. The reason for the delamination was not definitively determined, it may have been caused by surface preparation or perhaps the extreme smoothness of the surfaces being bonded, but there was not sufficient time to investigate since an alternative process was found. 

A fast-curing window adhesive that does not interact with the V-block bonding epoxy, can survive temperature and pressure variations, has high transparency and is index matched to glass would be useful.

\section{Summary} \label{sec:sum}
This paper provides a comprehensive overview of the Fiber System designed for the Dark Energy Spectroscopic Instrument (DESI). As a critical component of DESI, the Fiber System plays a pivotal role in enabling the instrument's ambitious mission to map the large-scale structure of the universe and probe the nature of dark energy. The paper details the key design principles, technological advancements, and operational considerations that underpin the functionality and performance of the Fiber System. 

The DESI fiber system collects light delivered by the DESI prime focus corrector and routes it to the DESI spectrographs. The 5,000 optical fibers are mounted into robotic fiber positioners that patrol the $\sim1$\,m diameter focal plane. The $107\,\mu$m fiber core is chosen to optimize the SNR of the target objects given the plate scale and median seeing at the 4m-Mayall telescope. 

The focal plane fibers are connected to the fiber cable and slit assembly via a fusion splice. A connection in the fiber system was required to ease production and integration flow. This method of connection resulted in a $\sim$2\% throughput loss and a highly stable fiber system compared to the typical loss of ~10\% when using a physical connector \citep{10.1117/12.2629751}.

The ruggardized fiber cables are completely external to the telescope structure, and the 10 bundles route from the focal plane, around both the polar and declination bearings, and terminate in 10 spectrograph slits, each with 500 fibers. The fiber slits are curved but include 20 blocks of parallel V-grooves with an additional AR coated window to reduce Fresnel losses. 

The design, manufacture, and test were an international collaboration between LBNL and DU and is the result of two years of dedicated of R\&D followed by two years of construction. The fiber system was installed at the 4\,m-Mayall telescope with $ \geq 99\% $ of fibers intact, $ \geq 90\%$ fibers with collimated FRD $ \leq$ 1.8$^\circ$, and $ \geq 90\%$ throughput delivered from the prime focus corrector to the spectrograph. 

We completed the installation and commissioning of the complete DESI instrument in 2020. Survey validation began thereafter, and observations for the main survey began on May 14, 2021. By the beginning of 2024 we have obtained 31.8 million spectra over 610 observing nights. This is an average of over 2 million spectra delivered every month. The fiber system has proven to be highly efficient, extremely stable, and robust during operations. No fibers have been lost during operations, and an analysis of sky residuals shows that the sky residuals RMS on emission lines is $\leq$ CCD noise + 1 \% of sky lines.

Analysis of only the first two months of object acquisition has already shown the first detection of the BAO signal \citep{2023MNRAS.525.5406M}. We look forward to further data releases that will further prove the power of this instrument that is enabling new and exciting scientific discoveries.

The DESI fiber system involved extensive R\&D that resulted in many innovations to the design and build. Moving away from the standard method of manufacturing fiber systems (for example, cleaving vs. polishing and fusion splicing vs connectors) and optimizing the performance resulted in a highly stable and efficient fiber system. The performance of the instrument and the quality of the dataset prove that these design choices were good, and it is our hope that the lessons learned will be useful to future instrument builders. 

\acknowledgments

The production of the fiber system was mostly performed by highly skilled technicians at LBNL, SSL, and DU. Although quality assurance was built into the design of the system, the success of the project is in large part due to the diligence and commitment of these technicians. We would like to acknowledge Hanna Ali, Topher Bently, Ann-Marie Dixon, David Gallegos, Ian Heetderks, Yvette Irwin, Sharon Jelinsky, Matthew Kaiser, John Mourelatos, Marco Pelayo, Kyle Perry, Sharon Sutherland, Carly Weaver, Krista Williams and Helen Yuan to whom we are eternally grateful.\\

This research is supported by the Director, Office of Science, Office of High Energy Physics of the U.S. Department of Energy under Contract No. 
DE–AC02–05CH1123, and by the National Energy Research Scientific Computing Center, a DOE Office of Science User Facility under the same 
contract; additional support for DESI is provided by the U.S. National Science Foundation, Division of Astronomical Sciences under Contract No. 
AST-0950945 to the NSF’s National Optical-Infrared Astronomy Research Laboratory; the Science and Technologies Facilities Council of the United Kingdom; the Gordon 
and Betty Moore Foundation; the Heising-Simons Foundation; the French Alternative Energies and Atomic Energy Commission (CEA); 
the National Council of Science and Technology of Mexico; the Ministry of Economy of Spain, and by the DESI 
Member Institutions: Aix-Marseille University; Argonne National Laboratory; Barcelona-Madrid Regional Participation Group; Brookhaven 
National Laboratory; Boston University; Brazil Regional Participation Group; Carnegie Mellon University; CEA-IRFU, Saclay; China Participation Group; Cornell University; Durham University; École 
Polytechnique Fédérale de Lausanne; Eidgenössische Technische Hochschule, Zürich; Fermi National Accelerator Laboratory; Granada-Madrid-Tenerife Regional 
Participation Group; Harvard University; Kansas State University; Korea Astronomy and Space Science Institute; Korea Institute for Advanced Study;
Lawrence Berkeley National Laboratory; Laboratoire de Physique Nucléaire et de Hautes Energies; Max Planck Institute; Mexico Regional Participation Group; New York University;
NSF’s National Optical-Infrared Astronomy Research Laboratory; Ohio University; Perimeter Institute; Shanghai Jiao Tong University; Siena College; SLAC National Accelerator Laboratory; Southern Methodist University; Swinburne University; 
The Ohio State University; Universidad de los Andes; University of Arizona; University of Barcelona; University of California, Berkeley; University of California, Irvine; University of California, 
Santa Cruz; University College London; University of Florida; University of Michigan at Ann Arbor; University of Pennsylvania; University of Pittsburgh; University of Portsmouth; 
University of Queensland; University of Rochester; University of Toronto; University of Utah; University of Waterloo; University of Wyoming; University of Zurich; UK Regional Participation Group; Yale University.  
The authors are honored to be permitted to conduct astronomical research on Iolkam Du’ag (Kitt Peak), a mountain with particular significance to the Tohono O’odham Nation.  For more information, visit desi.lbl.gov.

\vspace{5mm}
\facility{Mayall (DESI)}

\bibliography{desi_fiber}{}
\bibliographystyle{aasjournal}

\end{document}